\documentclass[10pt, journal, letterpaper]{IEEEtran}
\usepackage{url}
\usepackage[utf8]{inputenc}
\usepackage[dvipsnames]{xcolor}
\usepackage{amsmath}
\usepackage{amssymb}
\usepackage{xspace}
\usepackage{epsfig}
\usepackage{balance}

\usepackage[acronyms,nonumberlist,nopostdot,nomain,nogroupskip]{glossaries}

\usepackage{booktabs}
\usepackage{tabularx}
\usepackage{makecell}

\usepackage{tikz}
\usepackage{pgfplots}
\pgfplotsset{compat=newest}
\pgfplotsset{plot coordinates/math parser=false}
\newlength\fheight
\newlength\fwidth
\usetikzlibrary{plotmarks,patterns,decorations.pathreplacing,backgrounds,calc,arrows,arrows.meta,spy,matrix,scopes}
\usepgfplotslibrary{patchplots,groupplots}
\usepackage{tikzscale}
\usepackage[draft]{hyperref}

\newif\ifexttikz
\exttikzfalse

\ifexttikz
	\usetikzlibrary{external}
	\usepackage{fontspec}
\fi

\usepackage[var0]{inconsolata}

\usepackage{multirow}

\usepackage[font=footnotesize]{subcaption}
\usepackage[font=footnotesize]{caption}

\usepackage{mathtools}
\usepackage[numbers,sort&compress]{natbib}

\usepackage{soul}

\newacronym{3gpp}{3GPP}{3rd Generation Partnership Project}
\newacronym{4g}{4G}{4th generation}
\newacronym{5g}{5G}{5th generation}
\newacronym{6g}{6G}{6th generation}
\newacronym{5gc}{5GC}{5G Core}
\newacronym{adc}{ADC}{Analog to Digital Converter}
\newacronym{aerpaw}{AERPAW}{Aerial Experimentation and Research Platform for Advanced Wireless}
\newacronym{ai}{AI}{Artificial Intelligence}
\newacronym{aimd}{AIMD}{Additive Increase Multiplicative Decrease}
\newacronym{am}{AM}{Acknowledged Mode}
\newacronym{amc}{AMC}{Adaptive Modulation and Coding}
\newacronym{amf}{AMF}{Access and Mobility Management Function}
\newacronym{aops}{AOPS}{Adaptive Order Prediction Scheduling}
\newacronym{api}{API}{Application Programming Interface}
\newacronym{apn}{APN}{Access Point Name}
\newacronym{ap}{AP}{Application Protocol}
\newacronym{aqm}{AQM}{Active Queue Management}
\newacronym{ar}{AR}{Augmented Reality}
\newacronym{ausf}{AUSF}{Authentication Server Function}
\newacronym{avc}{AVC}{Advanced Video Coding}
\newacronym{awgn}{AGWN}{Additive White Gaussian Noise}
\newacronym{balia}{BALIA}{Balanced Link Adaptation Algorithm}
\newacronym{bbu}{BBU}{Base Band Unit}
\newacronym{bdp}{BDP}{Bandwidth-Delay Product}
\newacronym{ber}{BER}{Bit Error Rate}
\newacronym{bf}{BF}{Beamforming}
\newacronym{bler}{BLER}{Block Error Rate}
\newacronym{brr}{BRR}{Bayesian Ridge Regressor}
\newacronym{bs}{BS}{Base Station}
\newacronym{bsr}{BSR}{Buffer Status Report}
\newacronym{bss}{BSS}{Business Support System}
\newacronym{ca}{CA}{Carrier Aggregation}
\newacronym{caas}{CaaS}{Connectivity-as-a-Service}
\newacronym{cb}{CB}{Code Block}
\newacronym{cc}{CC}{Congestion Control}
\newacronym{ccid}{CCID}{Congestion Control ID}
\newacronym{cco}{CC}{Carrier Component}
\newacronym{cdd}{CDD}{Cyclic Delay Diversity}
\newacronym{cdf}{CDF}{Cumulative Distribution Function}
\newacronym{cdn}{CDN}{Content Distribution Network}
\newacronym{cn}{CN}{Core Network}
\newacronym{codel}{CoDel}{Controlled Delay Management}
\newacronym{comac}{COMAC}{Converged Multi-Access and Core}
\newacronym{cord}{CORD}{Central Office Re-architected as a Datacenter}
\newacronym{cornet}{CORNET}{COgnitive Radio NETwork}
\newacronym{cosmos}{COSMOS}{Cloud Enhanced Open Software Defined Mobile Wireless Testbed for City-Scale Deployment}
\newacronym{cots}{COTS}{Commercial Off-the-Shelf}
\newacronym{cp}{CP}{Control Plane}
\newacronym{cyp}{CP}{Cyclic Prefix}
\newacronym{up}{UP}{User Plane}
\newacronym{cpu}{CPU}{Central Processing Unit}
\newacronym{cqi}{CQI}{Channel Quality Information}
\newacronym{cr}{CR}{Cognitive Radio}
\newacronym{cran}{C-RAN}{Cloud \gls{ran}}
\newacronym{crs}{CRS}{Cell Reference Signal}
\newacronym{csi}{CSI}{Channel State Information}
\newacronym{csirs}{CSI-RS}{Channel State Information - Reference Signal}
\newacronym{cu}{CU}{Central Unit}
\newacronym{d2tcp}{D$^2$TCP}{Deadline-aware Data center TCP}
\newacronym{d3}{D$^3$}{Deadline-Driven Delivery}
\newacronym{dac}{DAC}{Digital to Analog Converter}
\newacronym{dag}{DAG}{Directed Acyclic Graph}
\newacronym{das}{DAS}{Distributed Antenna System}
\newacronym{dash}{DASH}{Dynamic Adaptive Streaming over HTTP}
\newacronym{dc}{DC}{Dual Connectivity}
\newacronym{dccp}{DCCP}{Datagram Congestion Control Protocol}
\newacronym{dce}{DCE}{Direct Code Execution}
\newacronym{dci}{DCI}{Downlink Control Information}
\newacronym{dctcp}{DCTCP}{Data Center TCP}
\newacronym{dl}{DL}{Downlink}
\newacronym{dmr}{DMR}{Deadline Miss Ratio}
\newacronym{dmrs}{DMRS}{DeModulation Reference Signal}
\newacronym{drlcc}{DRL-CC}{Deep Reinforcement Learning Congestion Control}
\newacronym{drs}{DRS}{Discovery Reference Signal}
\newacronym{du}{DU}{Distributed Unit}
\newacronym{e2e}{E2E}{end-to-end}
\newacronym{ecaas}{ECaaS}{Edge-Cloud-as-a-Service}
\newacronym{ecn}{ECN}{Explicit Congestion Notification}
\newacronym{edf}{EDF}{Earliest Deadline First}
\newacronym{embb}{eMBB}{Enhanced Mobile Broadband}
\newacronym{empower}{EMPOWER}{EMpowering transatlantic PlatfOrms for advanced WirEless Research}
\newacronym{enb}{eNB}{evolved Node Base}
\newacronym{endc}{EN-DC}{E-UTRAN-\gls{nr} \gls{dc}}
\newacronym{epc}{EPC}{Evolved Packet Core}
\newacronym{eps}{EPS}{Evolved Packet System}
\newacronym{es}{ES}{Edge Server}
\newacronym{etsi}{ETSI}{European Telecommunications Standards Institute}
\newacronym[firstplural=Estimated Times of Arrival (ETAs)]{eta}{ETA}{Estimated Time of Arrival}
\newacronym{eutran}{E-UTRAN}{Evolved Universal Terrestrial Access Network}
\newacronym{faas}{FaaS}{Function-as-a-Service}
\newacronym{fapi}{FAPI}{Functional Application Platform Interface}
\newacronym{fdd}{FDD}{Frequency Division Duplexing}
\newacronym{fdm}{FDM}{Frequency Division Multiplexing}
\newacronym{fdma}{FDMA}{Frequency Division Multiple Access}
\newacronym{fed4fire}{FED4FIRE+}{Federation 4 Future Internet Research and Experimentation Plus}
\newacronym{fir}{FIR}{Finite Impulse Response}
\newacronym{fit}{FIT}{Future \acrlong{iot}}
\newacronym{fpga}{FPGA}{Field Programmable Gate Array}
\newacronym{fr2}{FR2}{Frequency Range 2}
\newacronym{fs}{FS}{Fast Switching}
\newacronym{fscc}{FSCC}{Flow Sharing Congestion Control}
\newacronym{ftp}{FTP}{File Transfer Protocol}
\newacronym{fw}{FW}{Flow Window}
\newacronym{ge}{GE}{Gaussian Elimination}
\newacronym{gnb}{gNB}{Next Generation Node Base}
\newacronym{gop}{GOP}{Group of Pictures}
\newacronym{gpr}{GPR}{Gaussian Process Regressor}
\newacronym{gpu}{GPU}{Graphics Processing Unit}
\newacronym{gtp}{GTP}{GPRS Tunneling Protocol}
\newacronym{gtpc}{GTP-C}{GPRS Tunnelling Protocol Control Plane}
\newacronym{gtpu}{GTP-U}{GPRS Tunnelling Protocol User Plane}
\newacronym{gtpv2c}{GTPv2-C}{\gls{gtp} v2 - Control}
\newacronym{gw}{GW}{Gateway}
\newacronym{harq}{HARQ}{Hybrid Automatic Repeat reQuest}
\newacronym{hetnet}{HetNet}{Heterogeneous Network}
\newacronym{hh}{HH}{Hard Handover}
\newacronym{hol}{HOL}{Head-of-Line}
\newacronym{hqf}{HQF}{Highest-quality-first}
\newacronym{hss}{HSS}{Home Subscription Server}
\newacronym{http}{HTTP}{HyperText Transfer Protocol}
\newacronym{ia}{IA}{Initial Access}
\newacronym{iab}{IAB}{Integrated Access and Backhaul}
\newacronym{ic}{IC}{Incident Command}
\newacronym{ietf}{IETF}{Internet Engineering Task Force}
\newacronym{imsi}{IMSI}{International Mobile Subscriber Identity}
\newacronym{imt}{IMT}{International Mobile Telecommunication}
\newacronym{iot}{IoT}{Internet of Things}
\newacronym{ip}{IP}{Internet Protocol}
\newacronym{itu}{ITU}{International Telecommunication Union}
\newacronym{kpi}{KPI}{Key Performance Indicator}
\newacronym{kpm}{KPM}{Key Performance Measurement}
\newacronym{kvm}{KVM}{Kernel-based Virtual Machine}
\newacronym{los}{LOS}{Line-of-Sight}
\newacronym{lsm}{LSM}{Link-to-System Mapping}
\newacronym{lstm}{LSTM}{Long Short Term Memory}
\newacronym{lte}{LTE}{Long Term Evolution}
\newacronym{lxc}{LXC}{Linux Container}
\newacronym{m2m}{M2M}{Machine to Machine}
\newacronym{mac}{MAC}{Medium Access Control}
\newacronym{manet}{MANET}{Mobile Ad Hoc Network}
\newacronym{mano}{MANO}{Management and Orchestration}
\newacronym{mc}{MC}{Multi-Connectivity}
\newacronym{mcc}{MCC}{Mobile Cloud Computing}
\newacronym{mchem}{MCHEM}{Massive Channel Emulator}
\newacronym{mcs}{MCS}{Modulation and Coding Scheme}
\newacronym{mec}{MEC}{Multi-access Edge Computing}
\newacronym{mec2}{MEC}{Mobile Edge Cloud}
\newacronym{mfc}{MFC}{Mobile Fog Computing}
\newacronym{mgen}{MGEN}{Multi-Generator}
\newacronym{mi}{MI}{Mutual Information}
\newacronym{mib}{MIB}{Master Information Block}
\newacronym{miesm}{MIESM}{Mutual Information Based Effective SINR}
\newacronym{mimo}{MIMO}{Multiple Input, Multiple Output}
\newacronym{ml}{ML}{Machine Learning}
\newacronym{mlr}{MLR}{Maximum-local-rate}
\newacronym[plural=\gls{mme}s,firstplural=Mobility Management Entities (MMEs)]{mme}{MME}{Mobility Management Entity}
\newacronym{mmtc}{mMTC}{Massive Machine-Type Communications}
\newacronym{mmwave}{mmWave}{millimeter wave}
\newacronym{mpdccp}{MP-DCCP}{Multipath Datagram Congestion Control Protocol}
\newacronym{mptcp}{MPTCP}{Multipath TCP}
\newacronym{mr}{MR}{Maximum Rate}
\newacronym{mrdc}{MR-DC}{Multi \gls{rat} \gls{dc}}
\newacronym{mse}{MSE}{Mean Square Error}
\newacronym{mss}{MSS}{Maximum Segment Size}
\newacronym{mt}{MT}{Mobile Termination}
\newacronym{mtd}{MTD}{Machine-Type Device}
\newacronym{mtu}{MTU}{Maximum Transmission Unit}
\newacronym{mumimo}{MU-MIMO}{Multi-user \gls{mimo}}
\newacronym{mvno}{MVNO}{Mobile Virtual Network Operator}
\newacronym{nalu}{NALU}{Network Abstraction Layer Unit}
\newacronym{nas}{NAS}{Non-Access Stratum}
\newacronym{nbiot}{NB-IoT}{Narrow Band IoT}
\newacronym{nfv}{NFV}{Network Function Virtualization}
\newacronym{nfvi}{NFVI}{Network Function Virtualization Infrastructure}
\newacronym{ni}{NI}{Network Interfaces}
\newacronym{nic}{NIC}{Network Interface Card}
\newacronym{nlos}{NLOS}{Non-Line-of-Sight}
\newacronym{now}{NOW}{Non Overlapping Window}
\newacronym{nsm}{NSM}{Network Service Mesh}
\newacronym{nr}{NR}{New Radio}
\newacronym{nrf}{NRF}{Network Repository Function}
\newacronym{nsa}{NSA}{Non Stand Alone}
\newacronym{nse}{NSE}{Network Slicing Engine}
\newacronym{nssf}{NSSF}{Network Slice Selection Function}
\newacronym{o2i}{O2I}{Outdoor to Indoor}
\newacronym{oai}{OAI}{OpenAirInterface}
\newacronym{oaicn}{OAI-CN}{\gls{oai} \acrlong{cn}}
\newacronym{oairan}{OAI-RAN}{\acrlong{oai} \acrlong{ran}}
\newacronym{oam}{OAM}{Operations, Administration and Maintenance}
\newacronym{ofdm}{OFDM}{Orthogonal Frequency Division Multiplexing}
\newacronym{olia}{OLIA}{Opportunistic Linked Increase Algorithm}
\newacronym{omec}{OMEC}{Open Mobile Evolved Core}
\newacronym{onap}{ONAP}{Open Network Automation Platform}
\newacronym{onf}{ONF}{Open Networking Foundation}
\newacronym{onos}{ONOS}{Open Networking Operating System}
\newacronym{oom}{OOM}{\gls{onap} Operations Manager}
\newacronym{opnfv}{OPNFV}{Open Platform for \gls{nfv}}
\newacronym{oran}{O-RAN}{Open \gls{ran}}
\newacronym{orbit}{ORBIT}{Open-Access Research Testbed for Next-Generation Wireless Networks}
\newacronym{os}{OS}{Operating System}
\newacronym{oss}{OSS}{Operations Support System}
\newacronym{pa}{PA}{Position-aware}
\newacronym{pase}{PASE}{Prioritization, Arbitration, and Self-adjusting Endpoints}
\newacronym{pawr}{PAWR}{Platforms for Advanced Wireless Research}
\newacronym{pbch}{PBCH}{Physical Broadcast Channel}
\newacronym{pcef}{PCEF}{Policy and Charging Enforcement Function}
\newacronym{pcfich}{PCFICH}{Physical Control Format Indicator Channel}
\newacronym{pcrf}{PCRF}{Policy and Charging Rules Function}
\newacronym{pdcch}{PDCCH}{Physical Downlink Control Channel}
\newacronym{pdcp}{PDCP}{Packet Data Convergence Protocol}
\newacronym{pdsch}{PDSCH}{Physical Downlink Shared Channel}
\newacronym{pdu}{PDU}{Packet Data Unit}
\newacronym{pf}{PF}{Proportional Fair}
\newacronym{pgw}{PGW}{Packet Gateway}
\newacronym{phich}{PHICH}{Physical Hybrid ARQ Indicator Channel}
\newacronym{phy}{PHY}{Physical}
\newacronym{pmch}{PMCH}{Physical Multicast Channel}
\newacronym{pmi}{PMI}{Precoding Matrix Indicators}
\newacronym{powder}{POWDER}{Platform for Open Wireless Data-driven Experimental Research}
\newacronym{ppo}{PPO}{Proximal Policy Optimization}
\newacronym{ppp}{PPP}{Poisson Point Process}
\newacronym{prach}{PRACH}{Physical Random Access Channel}
\newacronym{prb}{PRB}{Physical Resource Block}
\newacronym{psnr}{PSNR}{Peak Signal to Noise Ratio}
\newacronym{pss}{PSS}{Primary Synchronization Signal}
\newacronym{pucch}{PUCCH}{Physical Uplink Control Channel}
\newacronym{pusch}{PUSCH}{Physical Uplink Shared Channel}
\newacronym{qam}{QAM}{Quadrature Amplitude Modulation}
\newacronym{qci}{QCI}{\gls{qos} Class Identifier}
\newacronym{qoe}{QoE}{Quality of Experience}
\newacronym{qos}{QoS}{Quality of Service}
\newacronym{quic}{QUIC}{Quick UDP Internet Connections}
\newacronym{rach}{RACH}{Random Access Channel}
\newacronym{ran}{RAN}{Radio Access Network}
\newacronym[firstplural=Radio Access Technologies (RATs)]{rat}{RAT}{Radio Access Technology}
\newacronym{rcn}{RCN}{Research Coordination Network}
\newacronym{rc}{RC}{RAN Control}
\newacronym{rec}{REC}{Radio Edge Cloud}
\newacronym{red}{RED}{Random Early Detection}
\newacronym{renew}{RENEW}{Reconfigurable Eco-system for Next-generation End-to-end Wireless}
\newacronym{rf}{RF}{Radio Frequency}
\newacronym{rfc}{RFC}{Request for Comments}
\newacronym{rfr}{RFR}{Random Forest Regressor}
\newacronym{ric}{RIC}{\gls{ran} Intelligent Controller}
\newacronym{rlc}{RLC}{Radio Link Control}
\newacronym{rlf}{RLF}{Radio Link Failure}
\newacronym{rlnc}{RLNC}{Random Linear Network Coding}
\newacronym{rmr}{RMR}{RIC Message Router}
\newacronym{rmse}{RMSE}{Root Mean Squared Error}
\newacronym{rnis}{RNIS}{Radio Network Information Service}
\newacronym{rr}{RR}{Round Robin}
\newacronym{rrc}{RRC}{Radio Resource Control}
\newacronym{rrm}{RRM}{Radio Resource Management}
\newacronym{rru}{RRU}{Remote Radio Unit}
\newacronym{rs}{RS}{Remote Server}
\newacronym{rsrp}{RSRP}{Reference Signal Received Power}
\newacronym{rsrq}{RSRQ}{Reference Signal Received Quality}
\newacronym{rss}{RSS}{Received Signal Strength}
\newacronym{rssi}{RSSI}{Received Signal Strength Indicator}
\newacronym{rtt}{RTT}{Round Trip Time}
\newacronym{ru}{RU}{Radio Unit}
\newacronym{rw}{RW}{Receive Window}
\newacronym{rx}{RX}{Receiver}
\newacronym{s1ap}{S1AP}{S1 Application Protocol}
\newacronym{sa}{SA}{standalone}
\newacronym{sack}{SACK}{Selective Acknowledgment}
\newacronym{sap}{SAP}{Service Access Point}
\newacronym{sc2}{SC2}{Spectrum Collaboration Challenge}
\newacronym{scef}{SCEF}{Service Capability Exposure Function}
\newacronym{sch}{SCH}{Secondary Cell Handover}
\newacronym{scoot}{SCOOT}{Split Cycle Offset Optimization Technique}
\newacronym{sctp}{SCTP}{Stream Control Transmission Protocol}
\newacronym{sdap}{SDAP}{Service Data Adaptation Protocol}
\newacronym{sdk}{SDK}{Software Development Kit}
\newacronym{sdm}{SDM}{Space Division Multiplexing}
\newacronym{sdma}{SDMA}{Spatial Division Multiple Access}
\newacronym{sdn}{SDN}{Software-defined Networking}
\newacronym{sdr}{SDR}{Software-defined Radio}
\newacronym{seba}{SEBA}{SDN-Enabled Broadband Access}
\newacronym{sgsn}{SGSN}{Serving GPRS Support Node}
\newacronym{sgw}{SGW}{Service Gateway}
\newacronym{si}{SI}{Study Item}
\newacronym{sib}{SIB}{Secondary Information Block}
\newacronym{sinr}{SINR}{Signal to Interference plus Noise Ratio}
\newacronym{sip}{SIP}{Session Initiation Protocol}
\newacronym{siso}{SISO}{Single Input, Single Output}
\newacronym{sla}{SLA}{Service Level Agreement}
\newacronym{sm}{SM}{Service Model}
\newacronym{smf}{SMF}{Session Management Function}
\newacronym{smo}{SMO}{Service Management and Orchestration}
\newacronym{sms}{SMS}{Short Message Service}
\newacronym{smsgmsc}{SMS-GMSC}{\gls{sms}-Gateway}
\newacronym{snr}{SNR}{Signal-to-Noise-Ratio}
\newacronym{son}{SON}{Self-Organizing Network}
\newacronym{sptcp}{SPTCP}{Single Path TCP}
\newacronym{srb}{SRB}{Service Radio Bearer}
\newacronym{srn}{SRN}{Standard Radio Node}
\newacronym{srs}{SRS}{Sounding Reference Signal}
\newacronym{ss}{SS}{Synchronization Signal}
\newacronym{sss}{SSS}{Secondary Synchronization Signal}
\newacronym{st}{ST}{Spanning Tree}
\newacronym{svc}{SVC}{Scalable Video Coding}
\newacronym{tb}{TB}{Transport Block}
\newacronym{tcp}{TCP}{Transmission Control Protocol}
\newacronym{tdd}{TDD}{Time Division Duplexing}
\newacronym{tdm}{TDM}{Time Division Multiplexing}
\newacronym{tdma}{TDMA}{Time Division Multiple Access}
\newacronym{tfl}{TfL}{Transport for London}
\newacronym{tfrc}{TFRC}{TCP-Friendly Rate Control}
\newacronym{tft}{TFT}{Traffic Flow Template}
\newacronym{tgen}{TGEN}{Traffic Generator}
\newacronym{tip}{TIP}{Telecom Infra Project}
\newacronym{tm}{TM}{Transparent Mode}
\newacronym{to}{TO}{Telco Operator}
\newacronym{tr}{TR}{Technical Report}
\newacronym{trp}{TRP}{Transmitter Receiver Pair}
\newacronym{ts}{TS}{Technical Specification}
\newacronym{tti}{TTI}{Transmission Time Interval}
\newacronym{ttt}{TTT}{Time-to-Trigger}
\newacronym{tx}{TX}{Transmitter}
\newacronym{uas}{UAS}{Unmanned Aerial System}
\newacronym{uav}{UAV}{Unmanned Aerial Vehicle}
\newacronym{udm}{UDM}{Unified Data Management}
\newacronym{udp}{UDP}{User Datagram Protocol}
\newacronym{udr}{UDR}{Unified Data Repository}
\newacronym{ue}{UE}{User Equipment}
\newacronym{uhd}{UHD}{\gls{usrp} Hardware Driver}
\newacronym{ul}{UL}{Uplink}
\newacronym{um}{UM}{Unacknowledged Mode}
\newacronym{uml}{UML}{Unified Modeling Language}
\newacronym{upa}{UPA}{Uniform Planar Array}
\newacronym{upf}{UPF}{User Plane Function}
\newacronym{urllc}{URLLC}{Ultra Reliable and Low Latency Communications}
\newacronym{usa}{U.S.}{United States}
\newacronym{usim}{USIM}{Universal Subscriber Identity Module}
\newacronym{usrp}{USRP}{Universal Software Radio Peripheral}
\newacronym{utc}{UTC}{Urban Traffic Control}
\newacronym{vim}{VIM}{Virtualization Infrastructure Manager}
\newacronym{vm}{VM}{Virtual Machine}
\newacronym{vnf}{VNF}{Virtual Network Function}
\newacronym{volte}{VoLTE}{Voice over \gls{lte}}
\newacronym{voltha}{VOLTHA}{Virtual OLT HArdware Abstraction}
\newacronym{vr}{VR}{Virtual Reality}
\newacronym{vran}{vRAN}{Virtualized \gls{ran}}
\newacronym{vss}{VSS}{Video Streaming Server}
\newacronym{wbf}{WBF}{Wired Bias Function}
\newacronym{wf}{WF}{Waterfilling}
\newacronym{wg}{WG}{Working Group}
\newacronym{wlan}{WLAN}{Wireless Local Area Network}
\newacronym{osm}{OSM}{Open Source Management and Orchestration}
\newacronym{pnf}{PNF}{Physical Network Function}
\newacronym{drl}{DRL}{Deep Reinforcement Learning}
\newacronym{mtc}{MTC}{Machine-type Communications}
\newacronym{osc}{OSC}{O-RAN Software Community}
\newacronym{mns}{MnS}{Management Services}
\newacronym{ves}{VES}{\gls{vnf} Event Stream}
\newacronym{ei}{EI}{Enrichment Information}
\newacronym{fh}{FH}{Fronthaul}
\newacronym{fft}{FFT}{Fast Fourier Transform}
\newacronym{laa}{LAA}{Licensed-Assisted Access}
\newacronym{plfs}{PLFS}{Physical Layer Frequency Signals}
\newacronym{ptp}{PTP}{Precision Time Protocol}
\newacronym{asic}{ASIC}{Application-specific Integrated Circuit}
\newacronym{aal}{AAL}{Acceleration Abstraction Layer}
\newacronym{fec}{FEC}{Forward Error Correction}
\newacronym{sdl}{SDL}{Shared Data Layer}
\newacronym{nib}{NIB}{Network Information Base}
\newacronym{rnib}{R-NIB}{RAN \gls{nib}}
\newacronym{fcaps}{FCAPS}{Fault, Configuration, Accounting, Performance, Security}
\newacronym{ie}{IE}{Information Element}
\newacronym{fg}{FG}{Focus Group}
\newacronym{osfg}{OSFG}{Open Source Focus Group}
\newacronym{sdfg}{SDFG}{Standard Development Focus Group}
\newacronym{tifg}{TIFG}{Test \& Integration Focus Group}
\newacronym{sfg}{SFG}{Security Focus Group}
\newacronym{swg}{SWG}{Security Work Group}
\newacronym{e2sm}{E2SM}{E2 Service Model}
\newacronym{tsc}{TSC}{Technical Steering Committee}
\tikzstyle{startstop} = [rectangle, rounded corners, minimum width=2cm, minimum height=0.5cm,text centered, draw=black]
\tikzstyle{io} = [trapezium, trapezium left angle=70, trapezium right angle=110, minimum width=3cm, minimum height=1cm, text centered, draw=black]
\tikzstyle{process} = [rectangle, minimum width=2cm, minimum height=0.5cm, text centered, draw=black, alignb=center]
\tikzstyle{decision} = [ellipse, minimum width=2cm, minimum height=1cm, text centered, draw=black]
\tikzstyle{arrow} = [thick,<->,>=stealth]
\tikzstyle{line} = [thick,>=stealth]
\tikzstyle{darrow} = [thick,<->,>=stealth,dashed]
\tikzstyle{sarrow} = [thick,->,>=stealth]
\tikzstyle{larrow} = [line width=0.1mm,dashdotted,->,>=stealth]
\tikzstyle{llarrow} = [line width=0.1mm,->,>=stealth]

\makeatletter
\def\grd@save@target#1{%
  \def\grd@target{#1}}
\def\grd@save@start#1{%
  \def\grd@start{#1}}
\tikzset{
  grid with coordinates/.style={
    to path={%
      \pgfextra{%
        \edef\grd@@target{(\tikztotarget)}%
        \tikz@scan@one@point\grd@save@target\grd@@target\relax
        \edef\grd@@start{(\tikztostart)}%
        \tikz@scan@one@point\grd@save@start\grd@@start\relax
        \draw[minor help lines] (\tikztostart) grid (\tikztotarget);
        \draw[major help lines] (\tikztostart) grid (\tikztotarget);
        \grd@start
        \pgfmathsetmacro{\grd@xa}{\the\pgf@x/1cm}
        \pgfmathsetmacro{\grd@ya}{\the\pgf@y/1cm}
        \grd@target
        \pgfmathsetmacro{\grd@xb}{\the\pgf@x/1cm}
        \pgfmathsetmacro{\grd@yb}{\the\pgf@y/1cm}
        \pgfmathsetmacro{\grd@xc}{\grd@xa + \pgfkeysvalueof{/tikz/grid with coordinates/major step x}}
        \pgfmathsetmacro{\grd@yc}{\grd@ya + \pgfkeysvalueof{/tikz/grid with coordinates/major step y}}
        \foreach \x in {\grd@xa,\grd@xc,...,\grd@xb}
        \node[anchor=north] at (\x,\grd@ya) {\pgfmathprintnumber{\x}};
        \foreach \y in {\grd@ya,\grd@yc,...,\grd@yb}
        \node[anchor=east] at (\grd@xa,\y) {\pgfmathprintnumber{\y}};
      }
    }
  },
  minor help lines/.style={
    help lines,
    gray,
    line cap =round,
    xstep=\pgfkeysvalueof{/tikz/grid with coordinates/minor step x},
    ystep=\pgfkeysvalueof{/tikz/grid with coordinates/minor step y}
  },
  major help lines/.style={
    help lines,
    line cap =round,
    line width=\pgfkeysvalueof{/tikz/grid with coordinates/major line width},
    xstep=\pgfkeysvalueof{/tikz/grid with coordinates/major step x},
    ystep=\pgfkeysvalueof{/tikz/grid with coordinates/major step y}
  },
  grid with coordinates/.cd,
  minor step x/.initial=.5,
  minor step y/.initial=.2,
  major step x/.initial=1,
  major step y/.initial=1,
  major line width/.initial=1pt,
}
\makeatother

\usepackage{dblfloatfix}

\makeglossaries

\usepackage{listings}

\definecolor{codegray}{rgb}{0.25,0.25,0.25}
\definecolor{codepurple}{rgb}{0.58,0,0.82}
\lstdefinestyle{mystyle}{
  commentstyle=\color{PineGreen},
  keywordstyle=\color{MidnightBlue}\bfseries,
  numberstyle=\tiny\color{codegray},
  stringstyle=\color{codepurple},
  basicstyle=\ttfamily\scriptsize,
  breakatwhitespace=true,         
  breaklines=true,                 
  captionpos=b,
  frame=tb,
  keepspaces=true,                 
  numbers=left,                    
  numbersep=5pt,                  
  showspaces=false,                
  showstringspaces=false,
  showtabs=false,                  
  tabsize=2,
  xleftmargin=10pt,
  belowskip=-10pt,
  float=htbp,  
}
 
\lstdefinelanguage{myxml}{%
  language     = XML,
  morekeywords = {E2AP,PDU,criticality,value,id},
}

\begin{document}

\title{Understanding O-RAN: Architecture, Interfaces, Algorithms, Security, and Research Challenges}

\author{\IEEEauthorblockN{Michele Polese, Leonardo Bonati, Salvatore D'Oro, Stefano Basagni, Tommaso Melodia}
\thanks{The authors are with the Institute for the Wireless Internet of Things, Northeastern University, Boston, MA, USA. E-mail: \{m.polese, l.bonati, s.doro, s.basagni, melodia\}@northeastern.edu.}
\thanks{This work was partially supported by the U.S.\ National Science Foundation under Grants CNS-1923789 and CNS-2112471, and by the U.S.\ Office of Naval Research under Grant N00014-20-1-2132.}
}


\flushbottom
\setlength{\parskip}{0ex plus0.1ex}

\maketitle
\glsunset{nr}
\glsunset{lte}
\glsunset{3gpp}
\glsunset{usrp}
\glsunset{sctp}

\begin{abstract}
The Open \gls{ran} and its embodiment through the O-RAN Alliance specifications are poised to revolutionize the telecom ecosystem. O-RAN promotes virtualized 
\glspl{ran} where disaggregated components are connected via open interfaces and optimized by intelligent controllers. The result is a new paradigm for the \gls{ran} design, deployment, and operations: O-RAN networks can be built with multi-vendor, interoperable components, and can be programmatically optimized through a centralized abstraction layer and data-driven closed-loop control.
Therefore, understanding O-RAN, its architecture, its interfaces, and workflows is key for researchers and practitioners in the wireless community. 
In this article, we present the first detailed tutorial on O-RAN. We also discuss the main research challenges and review early research results. 
We provide a deep dive of the O-RAN specifications, describing its architecture, design principles, and the O-RAN interfaces. 
We then describe how the O-RAN \glspl{ric} can be used to effectively control and manage \gls{3gpp}-defined \glspl{ran}. 
Based on this, we discuss innovations and challenges of O-RAN networks, including the \gls{ai} and \gls{ml} workflows that the architecture and interfaces enable,  security and standardization issues. 
%
%
Finally, we review experimental research platforms that can be used to design and test O-RAN networks, along with recent research results, and we outline future directions for O-RAN development. 
\end{abstract}

\begin{picture}(0,0)(10,-435)
\put(0,0){
\put(0,0){\footnotesize This work has been submitted to the IEEE for possible publication.}
\put(0,-10){
\footnotesize Copyright may be transferred without notice, after which this version may no longer be accessible.}}
\end{picture}



\glsresetall
\glsunset{nr}
\glsunset{lte}
\glsunset{3gpp}
\glsunset{usrp}
\glsunset{sctp}

\section{Introduction}
\label{sec:intro}

%
%
The complexity of cellular networks is increasing~\cite{3gpp.38.300,polese2020toward}, with next-generation wireless systems built on a host of heterogeneous technologies and frequency bands. New developments include massive \gls{mimo}~\cite{marzetta2010noncooperative}, millimeter wave and sub-terahertz communications~\cite{AKYILDIZ201416,giordani2020toward}, network-based sensing~\cite{bourdoux20206g}, network slicing~\cite{doro2021coordinated,bonati2020open,doro2020slicing,doroInfocom2019Slicing,doro2020sledge,doro2019tnet,bonati2020cellos}, and \gls{ml}-based digital signal processing~\cite{oshea2017introduction}, among others.
This will impose increasing capital and operational costs for the networks operators, which will have to continuously upgrade and maintain their infrastructure to keep up with new market trends and technology and customer requirements~\cite{deloitte2021oran}.

Managing and optimizing these new network systems require solutions 
that \emph{open the \gls{ran}}. This makes it possible to expose data and analytics and to enable data-driven optimization, closed-loop control, and automation~\cite{klaine2017survey,challita2020when,polese2018machine}.
Current approaches to cellular networking, however, are far from open. Today, \gls{ran} components are monolithic units, \textit{all-in-one solutions} that implement each and every layer of the cellular protocol stack. They are provided by a limited number of vendors and seen by the operators as black-boxes. 
%
%
Reliance on black-box solutions has resulted in:
(i) limited reconfigurability of the \gls{ran}, with equipment whose operations cannot be fine-tuned to support diverse deployments and different traffic profiles; 
(ii) limited coordination among network nodes,  preventing joint optimization and control of \gls{ran} components; and 
(iii) vendor lock-in, with limited options for operators to deploy and interface \gls{ran} equipment from multiple vendors. 
Under these circumstances, optimized radio resource management and efficient spectrum utilization through real-time adaptation become extremely challenging~\cite{bonati2021intelligence}.
%

\setcounter{figure}{1}
\begin{figure*}[b]
    \centering

    \includegraphics[width=\textwidth]{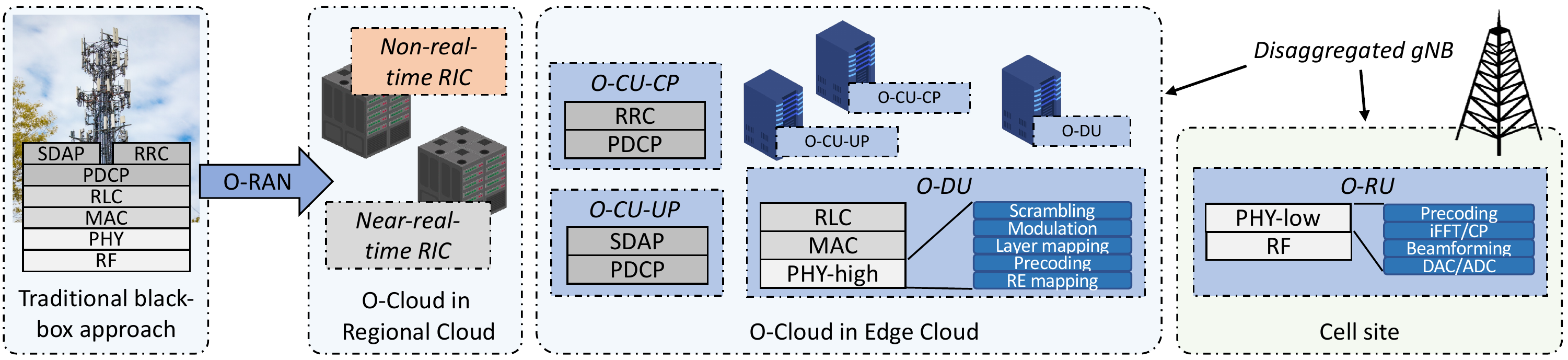}
    \caption{Evolution of the traditional black-box base station architecture (left) toward a virtualized gNB with a functional split (right, including the CU and DU at the edge, and the RU at the cell site). The functional split distributes the higher layers of the stack in the CU, which features RRC, PDCP, and SDAP. The DU features the RLC, MAC, and the higher part of the physical layer. This is distributed according to the 3GPP 7.2x split, which features frequency-domain functionalities in the DU (including scrambling, modulation, layer mapping, part of precoding, and mapping into physical resource blocks), and the time-domain functionalities in the RU (with precoding, \gls{fft} and \gls{cyp} addition/removal, beamforming, and the \gls{rf} components).}
    \label{fig:gnb-split}
\end{figure*}

To overcome these limitations, in the last decade several research and standardization efforts have promoted the Open RAN as the new paradigm for the \gls{ran} of the future. 
Open \gls{ran} deployments are based on disaggregated, virtualized and software-based components, connected through open and standardized interfaces, and interoperable across different vendors~\cite{oran-wg1-arch-spec}. 
Disaggregation and virtualization enable flexible deployments, based on cloud-native principles. This increases the resiliency and reconfigurability of the \gls{ran}. 
Open and standardized interfaces also allow operators to onboard different equipment vendors, which opens the \gls{ran} ecosystem to smaller players. 
%
%
Finally, open interfaces and software-defined protocol stacks enable the integration of intelligent, data-driven closed-loop control for the \gls{ran}~\cite{oran-wg2-ml}.

The O-RAN specifications implement these principles on top of \gls{3gpp} \gls{lte} and NR \glspl{ran}~\cite{3gpp.38.300, 3gpp.36.300}.
Specifically, O-RAN embraces and extends the \gls{3gpp} NR 7.2~split for base stations~\cite{3gpp.38.801}. The latter disaggregates base station functionalities into a \gls{cu}, a \gls{du}, and a \gls{ru}.
Moreover, it connects them to intelligent controllers through open interfaces that can stream telemetry from the \gls{ran} and deploy control actions and policies to it. The O-RAN architecture includes indeed two \glspl{ric} that perform management and control of the network at near-real-time (10\:ms to 1\:s) and non-real-time (more than 1\:s) time scales~\cite{oran-wg2-non-rt-ric-architecture,oran-wg3-ricarch}.
Finally, the O-RAN Alliance is standardizing a virtualization platform for the \gls{ran}, and extending the definition of \gls{3gpp} and eCPRI interfaces to connect \gls{ran} nodes~\cite{oran-wg1-arch-spec}.

\textbf{Contributions.} The Open \gls{ran} paradigm and, specifically, O-RAN networks will drastically change the design, deployment, and operations of the next generations of cellular networks. They will enable, among other things, transformative applications of \gls{ml} for optimization and control of the \gls{ran}~\cite{bonati2021intelligence}.
In this paper, we provide a detailed overview of how O-RAN will revolutionize future cellular networks.
We do so through a comprehensive analysis of the O-RAN technical specifications, architectural components, of the interfaces connecting them, and of the \gls{ml} and closed-loop control workflows that O-RAN enables and is standardizing.
We also discuss the new security challenges and opportunities introduced by O-RAN, as well as the main publicly available experimental platforms that enable research and development of O-RAN components. Finally, we survey recent results on design and optimization of O-RAN, and discuss the issues that need to be addressed to fully realize the O-RAN vision. The goal is to offer the interested reader a clear picture of the state of the art in O-RAN, and a deep understanding of the opportunities that the Open \gls{ran} introduces in the cellular ecosystem. 

Other papers~\cite{bonati2020open,lee2020hosting,bonati2021intelligence,abdalla2021generation,garciasaavedra2021oran,brik2022deep,arnaz2022toward} introduce the O-RAN building blocks and architecture, with use cases mostly related to the application of machine learning to the RAN. The literature on Open RAN also includes several high-level white papers that summarize different elements of the O-RAN architecture~\cite{rimedo2021wp,nolle2021wp,ntt2021wp,comcores2020wp,dell2021wp,deloitte2021oran,viavi2021wp,altiostar2021wp,ericsson2021securitywp,security2022,xran2021story,openranwp-ria,openranwp}. Differently from these, we introduce here a multi-faceted perspective on O-RAN, which starts from the foundational principles, covers in details the architectural components and the interfaces, and then connects these elements to highlight AI/ML use cases, security issues, deployment options, testbeds, and future research. Notably, this is the first paper that describes in detail the full set of O-RAN specifications for the \glspl{ric} and interfaces, including how O-RAN effectively enables control of \gls{3gpp}-defined network elements through custom logic running on the intelligent controllers.

\setcounter{figure}{0}
\begin{figure}[t]
    \centering
    \includegraphics[width=\columnwidth]{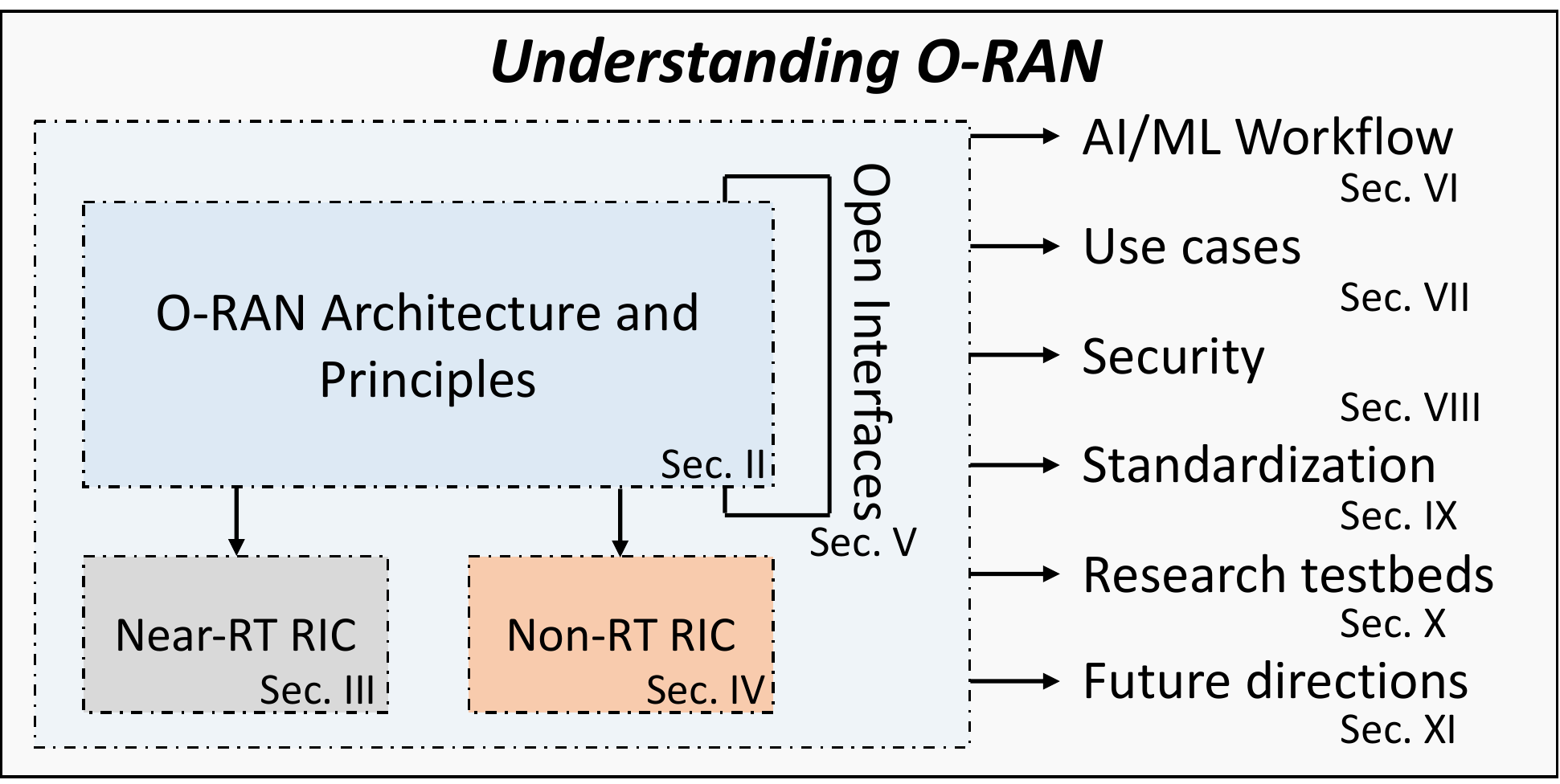}
    \caption{O-RAN components and paper organization. Sections~\ref{sec:arch}---~\ref{sec:orchestration} (left part of the figure) introduce the general architecture of O-RAN, the \glspl{ric}, and the open interfaces connecting them. Sections~\ref{sec:ai-ml-workflow}---~\ref{sec:future} (right part of the figure) discuss topics that relate to the overall Open RAN architecture.}
    \label{fig:paper-structure}
\end{figure}

\textbf{Paper structure.} The rest of this paper is organized as shown in Figure~\ref{fig:paper-structure}. 
Sections~\ref{sec:arch} to~\ref{sec:interfaces} introduce specific components of O-RAN networks; Sections~\ref{sec:ai-ml-workflow} to~\ref{sec:future} discuss topics that are relevant to the overall O-RAN vision and architecture; Section~\ref{sec:conclusions} concludes this work.
In particular, Section~\ref{sec:arch} describes the key principles of the O-RAN architecture, and introduces its components and the control loops that O-RAN enables. The near-real-time \gls{ric} and \gls{ran} control are discussed in Section~\ref{sec:xapps}, while the non-real-time \gls{ric} is presented in Section~\ref{sec:orchestration}. Section~\ref{sec:interfaces} is a deep dive on the O-RAN interfaces that connect the \gls{ran} and the \glspl{ric}.
Section~\ref{sec:ai-ml-workflow} describes the \gls{ai}/\gls{ml} workflow supported in O-RAN networks. Section~\ref{sec:results} summarizes the main O-RAN use cases and related research results.
Section~\ref{sec:security} reviews security challenges in O-RAN, and Section~\ref{sec:standardization} presents the standardization efforts and structure of the O-RAN Alliance. 
Publicly-available research and experimental platforms for O-RAN are discussed in Section~\ref{sec:testbeds}.
Finally, Section~\ref{sec:future} provides an outlook on future directions for the Open \gls{ran}, and Section~\ref{sec:conclusions} concludes the paper. We also include examples of O-RAN messages and a list of acronyms at the end of the paper.


\setcounter{figure}{2}
\begin{figure*}[b]
    \centering
    \includegraphics[width=\textwidth]{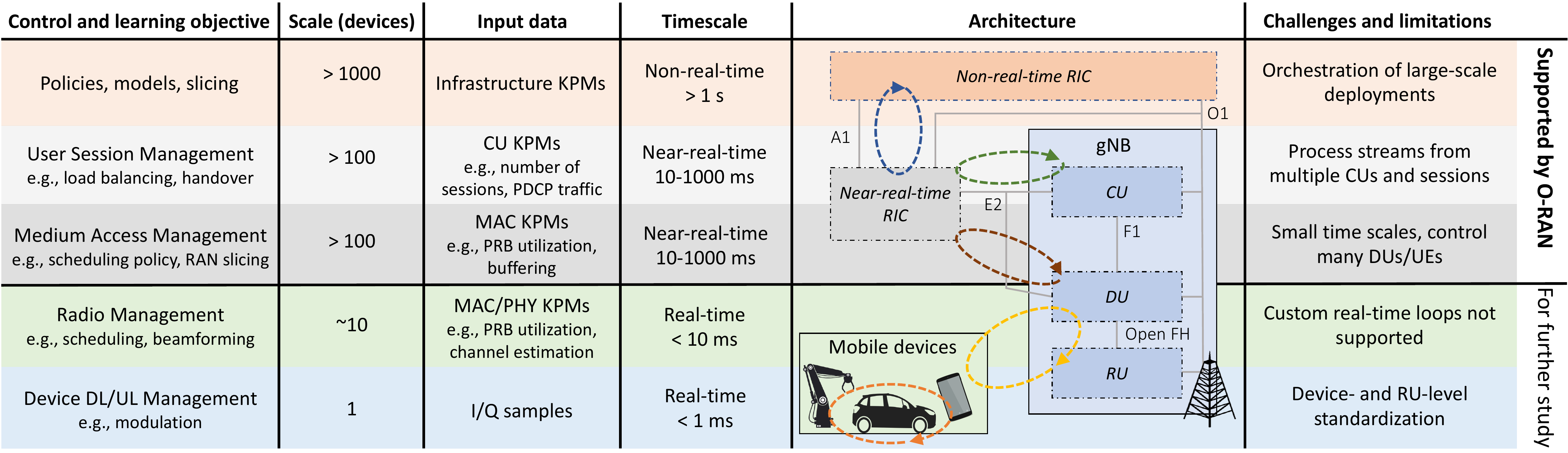}
    \caption{Closed-loop control enabled by the O-RAN architecture, and possible extensions, adapted from~\cite{bonati2021intelligence}. The control loops are represented by the dashed arrows over the architectural diagram.}
    \label{fig:loops}
\end{figure*}

\glsreset{cu}
\glsreset{du}
\glsreset{ru}

\section{O-RAN Key Architectural Principles}
\label{sec:arch}


The Open RAN vision is based on years of research on open and programmable networks. These principles have been at the center of the \gls{sdn} transformation in wired networks~\cite{mckeown2008openflow} in the past 15 years, and have started moving into the wireless domain more recently. 
For example, the xRAN Forum---an initiative led by operators---has proposed a standardized fronthaul interface, and introduced the idea of open, standardized interfaces for the integration of external controllers in the \gls{ran}~\cite{xran2021story}. 
In parallel, the \gls{cran} architecture (promoted, among others, by the operator-led C-RAN Alliance~\cite{xran2018pr}) has emerged as a solution to centralize most of the baseband processing for the \gls{ran} in virtualized cloud data centers~\cite{chih2014recent,checko2015cloud}, connected to remote radio units through high speed fronthaul interfaces. 
\gls{cran} enabled more refined signal processing and load balancing techniques by leveraging centralized data and control paths, while reducing costs by multiplexing computational resources. 
In 2018, these two initiatives joined forces to launch the O-RAN Alliance with the overall goal of standardizing an architecture and a set of interfaces to realize 
an Open \gls{ran}~\cite{xran2018pr}. 
In just four years, the O-RAN Alliance has 
scaled up to 
more than~300 members and contributors.
Its specifications are expected to drive~$50\%$ of \gls{ran}-based revenues by 2028~\cite{deloitte2021oran}.

Overall, it is possible to identify four foundational principles for the Open RAN in the literature and in the O-RAN specifications, as discussed next. These include disaggregation; intelligent, data-driven control with the \glspl{ric}; virtualization; and open interfaces~\cite{oran-wg1-arch-spec}.

\subsection{Disaggregation} 

As shown in Figure~\ref{fig:gnb-split}, \gls{ran} disaggregation splits base stations into different functional units, thus effectively embracing and extending the functional disaggregation paradigm proposed by \gls{3gpp} for the NR \glspl{gnb}~\cite{3gpp.38.401}. 
The \gls{gnb} is split into a \gls{cu}, a \gls{du}, and a \gls{ru} (called O-CU, O-DU, and O-RU in O-RAN specifications). The \gls{cu} is further split into two logical components, one for the \gls{cp}, and one for the \gls{up}. This logical split allows different functionalities to be deployed at different locations of the network, as well as on different hardware platforms. For example, \glspl{cu} and  \glspl{du} can be virtualized on white box servers at the edge (with hardware acceleration for some of the physical layer functionalities)~\cite{murti2021optimal,bonati2020open}, while the \glspl{ru} are generally implemented on \glspl{fpga} and \glspl{asic} boards and deployed close to RF antennas.

The O-RAN Alliance has evaluated the different \gls{ru}/\gls{du} split options proposed by the \gls{3gpp}, with specific interest in alternatives for physical layer split across the \gls{ru} and the \gls{du}~\cite{3gpp.38.801}.
The selected 7.2x~split strikes a balance between simplicity of the \gls{ru} and the data rates and latency required on the interface between the \gls{ru} and \gls{du}.
In split 7.2x, the \gls{ru} only performs \gls{fft} and cyclic prefix addition/removal operations, which makes the \gls{ru} inexpensive and easy to deploy.
The \gls{du} then takes care of the remaining functionalities of the physical layer, and of the \gls{mac} and \gls{rlc} layers~\cite{3gpp.38.322,3gpp.38.321,3gpp.38.201}. The operations of these three layers are generally tightly synchronized, as the \gls{mac} layer generates \glspl{tb} for the physical layer using data buffered at the \gls{rlc} layer. Finally, the \gls{cu} units (CP and UP) implement the higher layers of the \gls{3gpp} stack, i.e., the \gls{rrc} layer, which manages the life cycle of the connection~\cite{3gpp.38.331}; the \gls{sdap} layer, which manages the \gls{qos} of the traffic flows (also known as bearers)~\cite{3gpp.37.324}; and the \gls{pdcp} layer, which takes care of reordering, packet duplication, and encryption for the air interface, among others~\cite{3gpp.38.323}.

\subsection{\acrlongpl{ric} and Closed-Loop Control}

The second innovation is represented by the \glspl{ric}, which introduce programmable components that can run optimization routines with closed-loop control and orchestrate the \gls{ran}.
Specifically, the O-RAN vision includes two logical controllers that have an abstract and centralized point of view on the network, thanks to data pipelines that stream and aggregate hundreds of \glspl{kpm} on the status of the network infrastructure (e.g., number of users, load, throughput, resource utilization), as well as additional context information from sources outside of the \gls{ran}. The two \glspl{ric} process this data and leverage \gls{ai} and \gls{ml} algorithms to determine and apply control policies and actions on the \gls{ran}. Effectively, this introduces data-driven, closed-loop control that can automatically optimize, for example, network and \gls{ran} slicing, load balancing, handovers, scheduling policies, among others~\cite{bonati2021intelligence}. 
The O-RAN Alliance has drafted specifications for a non-real-time \gls{ric}, which integrates with the network orchestrator and operates on a time scale longer than 1 s, and a near-real-time \gls{ric}, which drives control loops with \gls{ran} nodes with a time scale between 10 ms and 1 s.
Figure~\ref{fig:loops} provides an overview of the closed-loop control that the \glspl{ric} enable throughout the disaggregated O-RAN infrastructure, together with real-time extensions that are considered for future work. In the next paragraphs, we will discuss the role of each \gls{ric} and related control loops.
%
%

\textbf{Non-real-time \gls{ric} and Control Loop.} 
The non-real-time (or non-RT) \gls{ric} is a component of the \gls{smo} framework, as illustrated in Figure~\ref{fig:oran-architecture}, and complements the near-RT \gls{ric} for intelligent RAN operation and optimization on a time scale larger than 1 second~\cite{oran-wg2-non-rt-ric-architecture,oran-wg2-non-rt-ric-functional-architecture,oran-wg2-non-rt-ric-use-cases}. Using the non-real-time control loop, the non-RT \gls{ric} provides guidance, enrichment information, and management of ML models for the near-RT \gls{ric}~\cite{oran-wg2-ml}. Additionally, the non-RT \gls{ric} can influence \gls{smo} operations, which gives the non-RT \gls{ric} the ability to indirectly govern all the components of the O-RAN architecture connected to the \gls{smo}, thus making decisions and applying policies that influence thousands of devices. This presents scalability challenges, as shown in Figure~\ref{fig:loops}, which need to be addressed through efficient process and software design.
Further details on the non-RT \gls{ric} and \gls{smo} will be given in Section~\ref{sec:orchestration}.

\textbf{Near-real-time \gls{ric} and Control Loop.}
The near-real-time (or near-RT) \gls{ric} is deployed at the edge of the network and operates control loops with a periodicity between 10\:ms and 1\:s~\cite{oran-wg3-ricarch}. As shown in Figure~\ref{fig:loops} and Figure~\ref{fig:oran-architecture}, the near-RT \gls{ric} interacts with \glspl{du} and \glspl{cu} in the \gls{ran}, as well as with legacy O-RAN-compliant \gls{lte} \glspl{enb}. The near-RT \gls{ric} is usually associated to multiple \gls{ran} nodes, thus the near-RT closed-loop control can affect the \gls{qos} of hundreds or thousands of \glspl{ue}.

The near-RT \gls{ric} consists of multiple applications supporting custom logic, called xApps, and of the services that are required to support the execution of the xApps. An xApp is a microservice that can be used to perform radio resource management through standardized interfaces and service models. It receives data from the \gls{ran} (e.g., user, cell, or slice \glspl{kpm}, as shown in Figure~\ref{fig:loops}) and (if necessary) computes and sends back control actions. 
To support xApps, the near-RT \gls{ric} includes (i) a database containing information on the \gls{ran} (e.g., list of connected \gls{ran} nodes, users, etc.) and serving as a shared data layer among xApps; (ii) messaging infrastructure across the different components of the platform, also supporting the subscription of \gls{ran} elements to xApps; (iii) terminations for open interfaces and \glspl{api}, and (iv) conflict resolution mechanisms to orchestrate control of the same \gls{ran} function by multiple xApps. We will further discuss characteristics and functionalities of the xApps in Section~\ref{sec:xapps}.

\textbf{Future Extensions to Real-Time Control Loops.} Figure~\ref{fig:loops} also includes loops that operate in the real-time domain, i.e., below 10 ms, for radio resource management at the \gls{ran} node level, or even below 1 ms, for device management and optimization. Typical examples of real-time control include scheduling, beam management, and feedback-less detection of physical layer parameters (e.g., modulation and coding scheme, interference recognition)~\cite{oshea2017introduction}. These loops, which have a limited scale in terms of devices being optimized, are not part of the current O-RAN architecture, but are mentioned in some specifications~\cite{oran-wg2-ml} as for further study.

\subsection{Virtualization} 

\begin{figure}[t]
        \centering
    \includegraphics[width=\columnwidth]{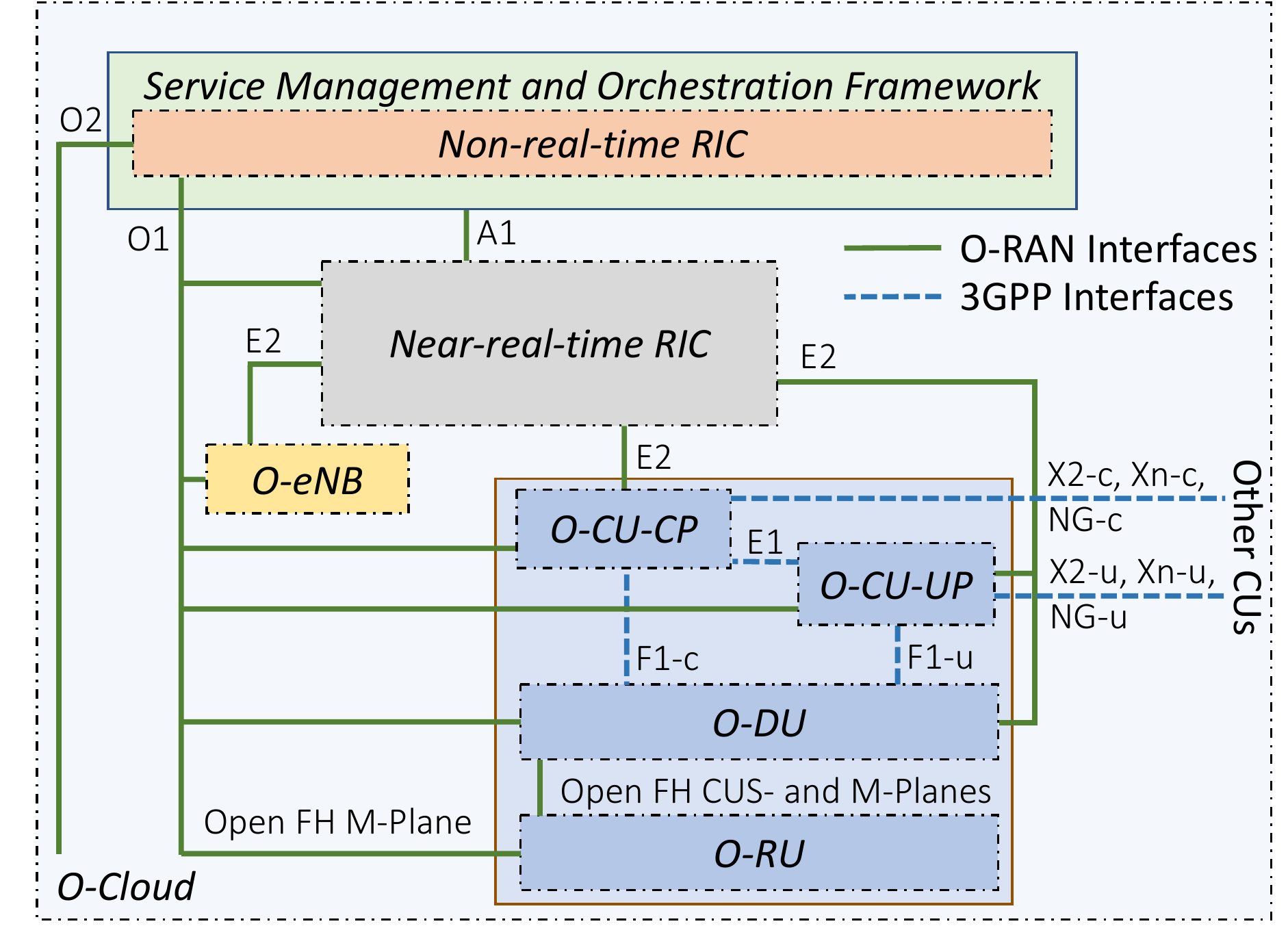}
    \caption{O-RAN architecture, with components and interfaces from O-RAN and 3GPP. O-RAN interfaces are drawn as solid lines, 3GPP ones as dashed lines.}
    \label{fig:oran-architecture}
\end{figure}

The third principle of the O-RAN architecture is the introduction of additional components for the management and optimization of the network infrastructure and operations, spanning from edge systems to virtualization platforms. According to~\cite{oran-wg1-arch-spec}, all the components of the O-RAN architecture shown in Figure~\ref{fig:oran-architecture} can be deployed on a hybrid cloud computing platform called O-Cloud. Specifically, the O-Cloud is a set of computing resources and virtualization infrastructure that are pooled together in one or multiple physical datacenters.
This platform combines physical nodes, software components (e.g., the operating system, virtual machine hypervisors, etc.), and management and orchestration functionalities~\cite{oran-wg6-o-cloud}, and
specializes the virtualization paradigm for O-RAN~\cite{jain2013network}.
It enables (i)~decoupling between hardware and software components; (ii)~standardization of the hardware capabilities for the O-RAN infrastructure; (iii)~sharing of the hardware among different tenants, and (iv)~automated deployment and instantiation of \gls{ran} functionalities. 

The O-RAN Alliance \gls{wg} 6 is also developing standardized hardware acceleration abstractions (called \glspl{aal}) that define common \glspl{api} between dedicated hardware-based logical processors and the O-RAN softwarized infrastructure, e.g., for channel coding/decoding and \gls{fec}~\cite{oran-wg6-aal-fec,oran-wg6-aal-gap}. These efforts also reflect into commercial hardware-accelerated, virtualized \gls{ran} implementations that can support the requirements of 3GPP NR use cases (e.g., \gls{urllc} flows~\cite{nasrallah2019ultra}) also on commercial hardware (e.g., the NVIDIA Aerial platform~\cite{kelkar2021nvidia}, NEC Nuberu~\cite{garcia2021nuberu}, and~\cite{panchal2021enabling} from Intel). The authors of~\cite{papatheofanous2021ldpc} discuss \gls{fpga}-based acceleration of the physical layer decoding with a prototype based on OpenAirInterface. 

In parallel, \gls{wg} 7 is defining the characteristics that white box hardware needs to satisfy to implement an O-RAN-compliant piece of equipment, e.g., indoor picocells, outdoor microcells and macrocells (all at sub-6 GHz and mmWaves), integrated access and backhaul nodes, and fronthaul gateways. These cover different architectural elements from Figure~\ref{fig:gnb-split}, including the RAN nodes (\gls{cu}, \gls{du}, \gls{ru}) and enablers of the fronthaul interface. The specifications clarify the functional parameters corresponding to the scenarios of interest (e.g., frequency bands, bandwidth, inter-site distance, \gls{mimo} configurations), and the hardware characteristics (e.g., accelerators, compute, connectivity) of the nodes.

The virtualization for the RAN components and of the O-RAN compute elements is expected to introduce savings and optimization of the power consumption related to the \gls{ran}~\cite{chih2014toward}. Virtualization makes it possible to easily and dynamically scale up or scale down the compute resources required to support user requirements, thus limiting the power consumption to the actual network functions that are needed~\cite{sabella2014energy,garcia2021nuberu}. In this sense, the closed-loop control capabilities described above, together with the virtualization in the RAN, also enable more refined and dynamic sleep cycles for the base stations and the RF components~\cite{pamuklu2021reinforcement,luo2018reducing}, which generally are the cause of most of the power consumption in cellular networks~\cite{lopez2022survey}. 

\subsection{Open Interfaces}

Finally, the O-RAN Alliance has introduced technical specifications that describe open interfaces connecting a number of different components of the O-RAN architecture. Figure~\ref{fig:oran-architecture} reports the new, open interfaces defined by O-RAN, as well as the intra-RAN interfaces from the 3GPP specifications. The latter is a partial enabler of the \gls{gnb} disaggregated architecture, which, however, is complemented by the O-RAN Open Fronthaul between the \gls{du} and the \gls{ru}. The O-RAN interfaces, instead, help overcoming the traditional RAN black box approach, as they expose data analytics and telemetry to the \glspl{ric}, and enable different kinds of control and automation actions, from \gls{ran} control to virtualization and deployment optimization. 

Without O-RAN, radio resource management and virtual/physical network functions optimization would be closed and inflexible, i.e., the operators would not have the same level of access to the equipment in their \gls{ran}, or it would be performed through a custom, piecemeal approach.
Standardization of these interfaces is thus a key step toward breaking the vendor lock-in in the \gls{ran}, e.g., allowing a near-RT \gls{ric} of one vendor to interact with the base stations of another vendor, or again enabling the interoperability of \glspl{cu}, \glspl{du} and \glspl{ru} from different manufacturers. This also fosters market competitiveness, innovation, faster update/upgrade cycles, and eases the design and introduction of new softwarized components in the \gls{ran} ecosystem~\cite{bonati2021intelligence}. 

Among the O-RAN-specific interfaces, the E2 interface connects the near-RT \gls{ric} to the \gls{ran} nodes. E2 enables the near-real-time loops shown in Figure~\ref{fig:loops} through the streaming of telemetry from the \gls{ran} and the feedback with control from the near-RT \gls{ric}. The near-RT \gls{ric} is connected to the non-RT \gls{ric} through the A1 interface, which enables a non-real-time control loop and the deployment of policy, guidance, and intelligent models in the near-RT \gls{ric}. The non-RT \gls{ric} also terminates the O1 interface, which connects to every other \gls{ran} component for management and orchestration of network functionalities. Finally, the non-RT \gls{ric} and the \gls{smo} also connect to the O-RAN O-Cloud through the O2 interface, and the O-RAN Fronthaul interface connects \glspl{du} and \glspl{ru}. The O-RAN Alliance has also defined a set of standardized tests to promote interoperability across different interface implementations, with an initial focus on the fronthaul interface and E2.
We will provide details on each interface in Section~\ref{sec:interfaces}.

Thanks to the open interfaces, the O-RAN architecture described in Figure~\ref{fig:oran-architecture} can be deployed by selecting different network locations (cloud, edge, cell sites) for different pieces of equipment, with multiple configurations described in~\cite{bonati2020open}. An example of deployment (i.e., \emph{Scenario B}~\cite{bonati2020open}) is shown in Figure~\ref{fig:gnb-split}, with the \glspl{ric} deployed in the cloud, the \glspl{cu} and \glspl{du} at the edge, and the \gls{ru} cell sites. Other deployment strategies, with the \glspl{ric} and the \gls{ran} nodes co-located are also possible, for example to support local and private 5G networks.

\section{Near-RT RIC, xApps, and Control of the RAN}
\label{sec:xapps}

The near-RT \gls{ric} is the core of the control and optimization of the \gls{ran}, thanks to the capabilities offered by the E2 interface. In this section, we will discuss the functionalities of the near-RT \gls{ric} and the near-RT \gls{ric} implementations available in the open source domain.


As discussed in Sections~\ref{sec:arch}, the near-RT \gls{ric} platform hosts the terminations of three interfaces (O1, A1, and E2), the xApps, and the components required to execute and manage the xApps. The O-RAN specifications describe the requirements and functionalities of the different components of the \gls{ric}, so that different standard-compliant implementations can be expected to provide the same set of services, but they do not introduce implementation requirements~\cite{oran-wg3-ricarch}. However, the O-RAN Alliance, through the \gls{osc}, also provides a reference implementation of a functional near-RT \gls{ric} that follows the specifications in~\cite{oran-wg3-ricarch} and can be used to prototype O-RAN solutions. The \gls{osc} near-RT \gls{ric} is based on multiple components running as microservices on a Kubernetes cluster~\cite{d_release}.

\begin{figure}[t]
    \centering
    \includegraphics[width=.95\columnwidth]{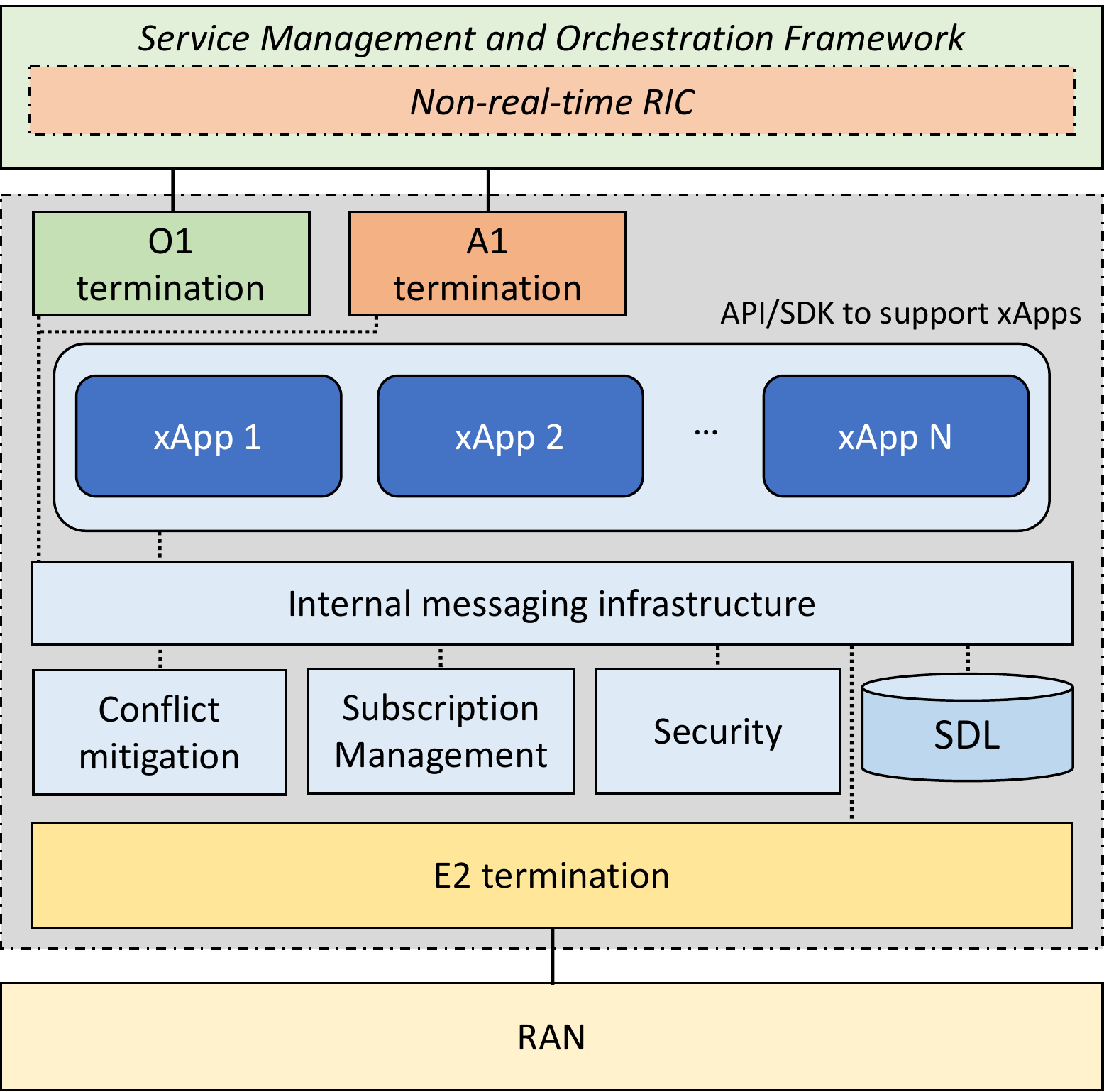}
    \caption{Near-RT \gls{ric} architecture. The near-RT RIC connects to the RAN through the E2 interface, at the bottom of the figure (yellow), and to the non-RT RIC/SMO through the A1 and O1 interfaces, at the top of the figures (orange and green, respectively). The communication among the RIC components (in light blue) is mediated by an internal messaging infrastructure. The near-RT RIC can onboard custom logic as xApps (dark blue).}
    \label{fig:near-rt-ric}
\end{figure}

\subsection{Near-RT RIC Internal Components}

Figure~\ref{fig:near-rt-ric} provides an overview of the architecture of a typical near-RT \gls{ric}. The main platform components include:
\begin{itemize}
    \item \textbf{Internal messaging infrastructure.} The internal messaging infrastructure connects xApps, platform services, and interface terminations to each other. The specifications do not mandate any specific technology (the \gls{osc} uses a custom library called \gls{rmr}~\cite{oran_rmr}), but they list the requirements and functionalities this sub-system needs to provide. The internal messaging infrastructure needs to support registration, discovery, and deletion of endpoints (i.e., internal \gls{ric} components and xApps), and provides \glspl{api} for sending and receiving messages, either through point-to-point communications or publish/subscribe mechanisms. It also provides routing and robustness to avoid internal data loss;
    
    \item \textbf{Conflict mitigation.} This component addresses possible conflicts emerging among different xApps. This is required because different, independent xApps may apply conflicting configurations while trying to achieve independent optimization goals, eventually resulting in performance degradation. The domain of the conflict may be a user, a bearer, or a cell, and can be related to any control action performed by the \gls{ric}. The O-RAN specifications highlight three different classes of conflicts. Direct conflicts can be directly detected by this internal component. For example, multiple xApps can apply different settings for the same parameters in the same control target, or xApps can request more resources than those available. This can be solved by the conflict mitigation component that decides which xApp prevails or limits the scope of a control action (pre-action resolution). Indirect and implicit conflicts, instead, cannot be observed directly and may or may not depend on the relationships among different xApps. For example, configurations that optimize the performance of certain classes of users may degrade others in non-obvious ways. 
    These conflicts may be detected and mitigated through post-action verification, i.e., by monitoring the performance of the system after the application of different control policies. Overall, conflict mitigation is a key component of the \gls{ric} but, at the time of this writing, it is not included in the \gls{osc} near-RT \gls{ric};
    
    \item \textbf{Subscription manager.} The subscription management functionality allows xApps to connect to functions exposed over the E2 interface. It also controls the access that individual xApps can have to E2 messages, and can merge multiple, identical subscription requests to the same E2 node into a single one;
    
    \item \textbf{Security sub-system.} According to~\cite{oran-wg3-ricarch}, this component has the high-level goal to prevent malicious xApps from leaking sensitive \gls{ran} data or from affecting the \gls{ran} performance. The details of this component are still left for further studies;
    
    \item \textbf{\gls{nib} Database and \acrlong{sdl} \acrshort{api}.} The \gls{rnib} database stores information on the E2 nodes, and the \gls{ue}-\gls{nib} contains entries for the \glspl{ue} and their identity. The UE identity (i.e., the UE-ID) is a key and sensitive piece of information in the \gls{ric}, as it allows UE-specific control, but at the same time it can expose sensitive information on the users. The UE-\gls{nib} makes it possible to track and correlate the identity of the same user in different E2 nodes. The database can be queried by the different components of the \gls{ric} platform (including the xApps) through the \gls{sdl} \glspl{api}. The \gls{osc} \gls{ric} provides an implementation of a \gls{sdl} library that can be compiled  inside xApps, as well as a Redis-based database~\cite{carlson2013redis};
    \item \textbf{xApp management.} The near-RT \gls{ric} features services and \glspl{api} for the automated life-cycle management of the xApps, from onboarding, to deployment and termination (triggered by the \gls{smo}), as well as tracing and logging for \gls{fcaps}. In the \gls{osc} \gls{ric}, this is done through wrappers on the Kubernetes infrastructure.
\end{itemize}

\subsection{Near-RT RIC xApps}

The main components of the near-RT \gls{ric} are the xApps. As previously discussed, an xApp is a plug-and-play component that implements custom logic, for example for \gls{ran} data analysis and \gls{ran} control. xApps can receive data and telemetry from the \gls{ran} and send back control using the E2 interface, as we described in Section~\ref{sec:e2}.

According to the O-RAN specifications~\cite{oran-wg3-ricarch}, an xApp is defined by a descriptor and by the xApp software image (i.e., the set of files needed to deploy the fully-functional xApp). The xApp descriptor (e.g., a YAML or JSON file) includes information on parameters needed to manage the xApp, such as, for example, autoscaling policies, deployment, deletion, and upgrade information. Additionally, it can describe the data types consumed by the xApp as well as its control capabilities.
Specifically, in the \gls{osc} \gls{ric}, the xApp is defined by a Docker image that can be deployed on a Kubernetes infrastructure by applying the descriptor schema, which is a file that specifies the attributes of the container.

At the time of this writing, the O-RAN specifications only mandate a limited set of \glspl{api} that the near-RT \gls{ric} platform needs to provide to xApps (including the \gls{sdl} \glspl{api} and the registration/discovery/subscription \glspl{api}). The definition of a broader set of \glspl{api} into a software development kit, however, would foster the development of xApps that can be seamlessly ported across different near-RT \gls{ric} implementations. Efforts in this direction have been promoted by the Telecom Infra Project (TIP) RAN Intelligence and Automation (RIA) subgroup~\cite{openranwp-ria,openranwp}.

\subsection{Open Source Near-RT RIC Implementations}

Besides the one provided by the \gls{osc}, the open source community includes third-party \gls{ric} implementations that enrich the Open RAN ecosystem. An example is ColO-RAN, which is an implementation focused on O-RAN experimentation based on the \gls{osc} near-RT RIC~\cite{polese2021coloran}, described in Section~\ref{sec:testbeds}.
%
The SD-RAN project by the \gls{onf} is developing an open source and cloud-native implementation of O-RAN near-RT \gls{ric}, together with xApps to control the \gls{ran}, and an \gls{sdk} to facilitate the design of new xApps~\cite{sdranwp,sdran_website}.
These xApps leverage both standard-compliant \glspl{e2sm}, and custom service models developed by the SD-RAN community.
The microservices of this \gls{ric}, which is based on the \gls{onos} controller, include xApp subscription services, network- and user-based information services, distributed data store services for high availability and operator services.
%
FlexRIC, instead, provides a monolithic near-RT \gls{ric} and a \gls{ran} agent to interface the OpenAirInterface radio stack with the \gls{ric}~\cite{schmidt2021flexric}.
It includes \glspl{sm} for monitoring and slicing programmability use cases, and an \gls{sdk} to build specialized service-oriented controllers.
%
Finally, 5G-EmPOWER is a near-RT \gls{ric} for heterogeneous \glspl{ran}~\cite{5gempower_website}.
It includes non-standard-compliant functionalities like mobility management for Wi-Fi and cellular networks, multi-tenant support, and deployment of custom resource allocation schema within network slices.

\section{Non-RT \gls{ric} and Orchestration Framework}
\label{sec:orchestration}

The second key element of the O-RAN architecture is the \gls{smo} framework. This component is in charge of handling all orchestration, management and automation procedures to monitor and control \gls{ran} components.\footnote{It is worth mentioning that although \gls{smo} functionalities are usually referred to as network-wide orchestration and management procedures (e.g., spanning both core and \gls{ran} portions of the network), the O-RAN specifications describe \gls{smo} operations and functionalities pertaining to \gls{ran} components only.}
Primarily, the \gls{smo} hosts the non-RT \gls{ric} and provides a set of interfaces (described in detail in Section~\ref{sec:interfaces}) that support the interaction between the different network components as well as data collection capabilities to facilitate network monitoring and control via \gls{ai}/\gls{ml}~\cite{oran-wg2-non-rt-ric-architecture,oran-wg2-ml}. 

\begin{figure}[t]
    \centering
    \includegraphics[width=.95\columnwidth]{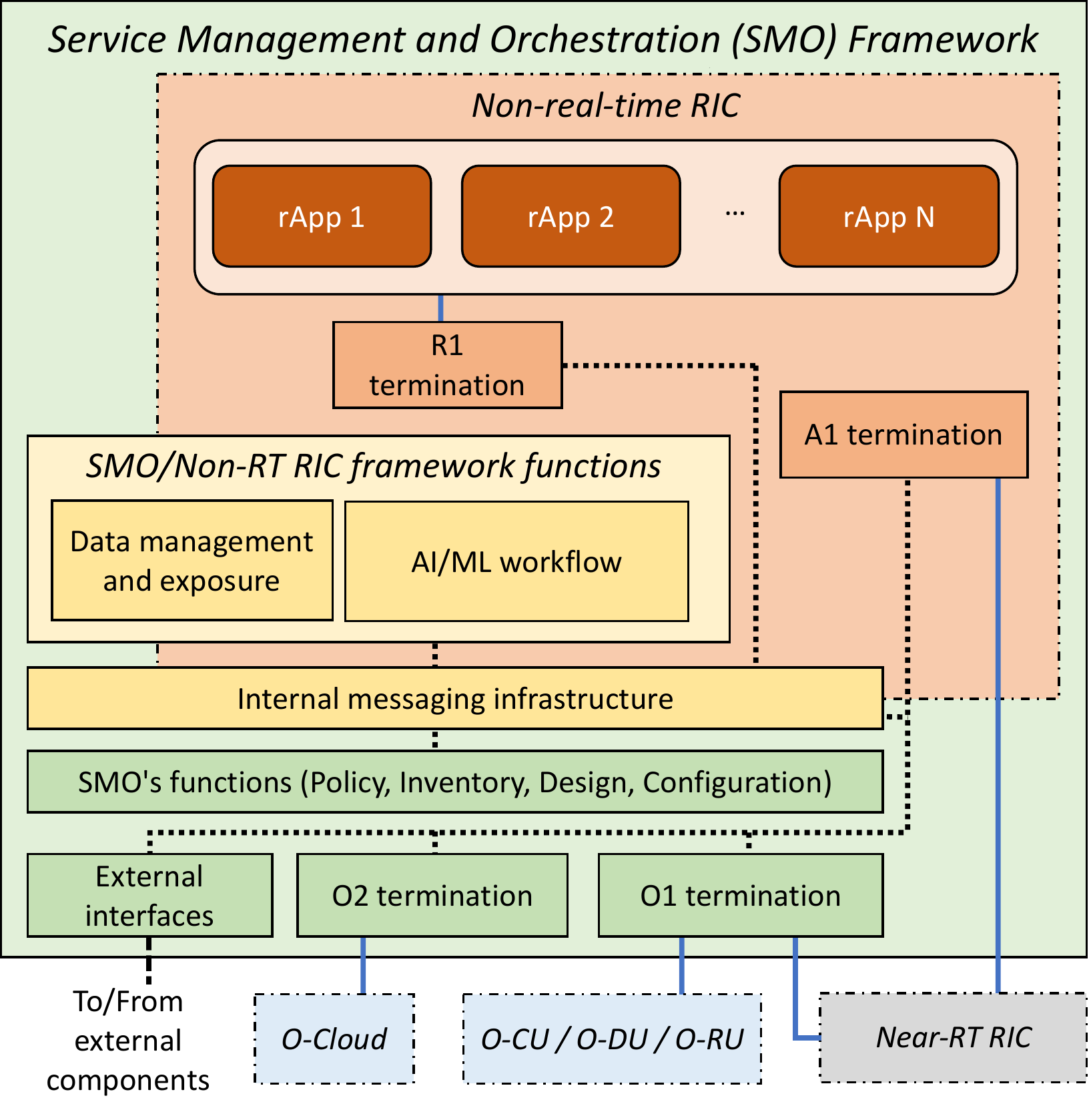}
    \caption{Non-RT \gls{ric} and \gls{smo} architecture. The SMO functionalities (in green) enable connectivity to the O-Cloud (through the O2 interface) and the other RAN components (through O1) for management and orchestration. The non-RT RIC features custom logic (rApps, in red), and a termination of the A1 interface to the near-RT RIC (orange). Shared functionalities between the non-RT RIC and the SMO are in yellow.}
    \label{fig:nonrtric}
\end{figure}

The high-level architecture of the \gls{smo} is illustrated in Figure~\ref{fig:nonrtric}. Its building blocks their main functionalities will be detailed in the remainder of this section. It is worth mentioning that at the time of writing, the O-RAN specifications do not provide strict guidelines regarding the split between \gls{smo} and non-RT \gls{ric} functionalities. However, the specifications group such functionalities into three distinct sets~\cite{oran-wg2-non-rt-ric-architecture}. The first set (orange-shaded blocks in Figure~\ref{fig:nonrtric}) identifies those functionalities and interfaces that are anchored to the non-RT \gls{ric}. A second set (green-shaded blocks) identifies functionalities anchored outside the non-RT \gls{ric}, while the functionalities from the remaining set (yellow-shaded blocks) are either not yet anchored to any specific \gls{smo} component or they span multiple components. Similarly to Figure~\ref{fig:near-rt-ric}, the non-RT \gls{ric} architecture embeds implementation-specific interfaces that interconnect and regulate the interactions between functionalities and components within the non-RT \gls{ric} and the \gls{smo} domains.
This infrastructure is depicted as the \textit{internal messaging infrastructure} in Figure~\ref{fig:nonrtric}.

The goal of the next sections is to describe these functionalities and interfaces, as well as to highlight their relevance to O-RAN systems and operations.


\subsection{Non-real-time RIC}

The non-RT \gls{ric} is one of the core components of the O-RAN architecture.
%
Similarly to the near-RT \gls{ric}, it enables closed-loop control of the \gls{ran} with timescales larger than 1\:s. Moreover, it also supports the execution of third-party applications, i.e., the \textit{rApps}, which are used to provide value-added services to support and facilitate RAN optimization and operations, including policy guidance, enrichment information, configuration management and data analytics. 

As shown in Figure~\ref{fig:nonrtric}, the non-RT \gls{ric} hosts the R1 termination, which allows rApps to interface with the non-RT \gls{ric}.
This allows them to obtain access to data management and exposure services, \gls{ai}/\gls{ml} functionalities, as well as A1, O1 and O2 interfaces
through the internal messaging infrastructure. 
It is worth mentioning that although rApps can support
the same control functionalities provided by xApps  (e.g., traffic steering, scheduling control, handover management) at larges timescales, they have been standardized to derive control policies that operate at a higher level and affect a larger number of users and network nodes. Relevant examples of rApps for non-RT \gls{ran} control applications include frequency and interference management, \gls{ran} sharing, performance diagnostics, end-to-end \gls{sla} assurance and network slicing~\cite{oran-wg2-non-rt-ric-use-cases}. 

To provide a flexible architecture in which the behavior of each and every network component and functionality can be adjusted in real time to meet the intents and goals of the operators, the non-RT \gls{ric} offers the following two high-level management and orchestration services~\cite{oran-wg2-non-rt-ric-functional-architecture}: (i)~intent-based network management, and (ii)~intelligent orchestration.

\textbf{Intent-based Network Management.} This functionality allows operators to specify their intent via a high-level language through a human-machine interface (it is expected that intents will follow practices used in currently available \glspl{smo}, and formalized as YAML or XML configuration files). Intents are then automatically parsed by the non-RT \gls{ric}, which determines the policies and the set of rApps and xApps that need to be deployed and executed to satisfy them.

In this context, the efforts of the \gls{osc}, which is working toward a non-RT \gls{ric} implementation that hosts a human-machine interface (i.e., the Intent Interface~\cite{nonrtric-osc}), are worth mentioning.
%

\textbf{Intelligence Orchestration.} Indeed, the O-RAN architecture enables and facilitates the development, deployment, execution and maintenance of network intelligence. However, this inevitably makes network control more complex due to the increasing number of xApps and rApps that execute at different \glspl{ric} and locations of the network. This calls for solutions that are capable of coordinating and orchestrating these applications. Specifically, the non-RT \gls{ric} is in charge of orchestrating network intelligence~\cite{oran-wg2-ml} to make sure that selected xApps and rApps are: (i) well-suited to satisfy operator intents and meet their requirements; (ii) instantiated at the appropriate \gls{ric} location to ensure control over the specific RAN elements specified in the intent; (iii) fed with relevant data; and (iv) robust enough not to generate conflicts due to multiple applications controlling the same functionalities and/or parameters simultaneously. For example, if the operator has instantiated multiple network slices and wants to control and optimize scheduling policies for each slice in near real time for a selected set of base stations close to a landmark of interest, the non-RT RIC must be able to determine automatically that only xApps executing at a specific near-RT \gls{ric} can satisfy the timing requirement. Moreover, the non-RT \gls{ric} must select only those xApps that are able to control scheduling decisions, and eventually dispatch them to the near-RT \glspl{ric} that control the base stations deployed in the area of interest, and are thus capable of generating data to be fed to the xApps.

\subsection{Other SMO/Non-RT RIC functionalities}

In this section, we describe the functionalities and architectural components that can reside both at the \gls{smo} and at the non-RT \gls{ric}~\cite{oran-wg2-non-rt-ric-architecture,oran-wg2-non-rt-ric-functional-architecture}. 

The internal messaging infrastructure is a composite of several \gls{smo} functions that allow all components within the \gls{smo} (even those included in the non-RT \gls{ric}) to access and utilize interfaces, data and functionalities offered by both the \gls{smo} and the non-RT \gls{ric}. For example, all interface terminations are tied to interface-specific functions included in the internal messaging infrastructure that are designed to facilitate the exchange of messages between terminations.
In this way, policies computed by rApps can reach the non-RT \gls{ric} through the R1 termination, and eventually reach the near-RT \gls{ric} through the A1 interface.

\textbf{Data Management and Exposure Services.} The O-RAN specifications also include data management and exposure services pertaining to the \gls{smo}/non-RT \gls{ric} framework. To this purpose, O-RAN follows a consumer/producer protocol in which data producers in the \gls{smo}/non-RT \gls{ric} can advertise and publish data (e.g., performance reports or AI-based prediction of \glspl{kpm} and network load). On the other hand, data consumers (e.g., rApps that determine high-level control policies) can discover, subscribe, receive and consume relevant data types from a selected number of nodes in the \gls{smo}/non-RT \gls{ric} domain. In order to fully support AI/ML solutions, the \gls{smo}/non-RT \gls{ric} can also perform collection of all data being produced, as well as relevant AI/ML pre-processing operations involving data analytics (e.g., correlation analysis), labeling, and normalization. 

\textbf{AI/ML Workflow.} Another important capability offered by the non-RT \gls{ric} is the possibility to oversee the entire \gls{ai} life cycle, and cover all aspects of AI/ML development, including data collection, training, validation, deployment and execution. This AI/ML workflow, which will be detailed in Section~\ref{sec:ai-ml-workflow}, is illustrated in Figure~\ref{fig:ai}.

\subsection{SMO Framework and Open Source SMO Implementations}

Besides hosting the non-RT \gls{ric}, the \gls{smo} also offers additional functionalities and interfaces, summarized in Figure~\ref{fig:nonrtric}. These include management and interactions with the O-Cloud via the O2 interface (see Section~\ref{sec:other-int}) as well as other O-RAN components via the O1 interface. The \gls{smo} takes care of \gls{fcaps} management procedures as well as service and resource inventory, topology and network configuration, as well as policy management for network orchestration services.

Although several members of the O-RAN Alliance have announced the development and availability of proprietary \glspl{smo} compliant with the latest O-RAN releases, these \glspl{smo} are closed solutions not open to the general public and whose implementation details and offered functionalities are not generally available. For this reason, we focus on two open-source solutions---with publicly-available code and functionalities---that are currently being integrated with the O-RAN architecture~\cite{bonati2020open}: \gls{onap}~\cite{onap_5g,onap_architecture} and \gls{osm}~\cite{osm_architecture}. 

Both \gls{onap} and \gls{osm} are comprehensive platforms that enable automation and orchestration in virtualized and softwarized networks. \gls{onap} is one of the main projects being developed and maintained by the Linux Foundation, while \gls{osm} is hosted by \gls{etsi} and follows ETSI \gls{vnf} standard specifications~\cite{osm_architecture}.
Being maintained by the Linux Foundation, \gls{onap} provides native integration with other major projects such as Kubernetes, Akraino, Acumos and OpenDaylight~\cite{onap_website}. This makes \gls{onap} quite a complex environment compared to \gls{osm} which, instead, offers similar services in a more lightweight framework. 
Although the development of \gls{smo} functionalities and the integration of O-RAN components with the \gls{smo} is still at its early stages, \gls{onap} is already being used by the \gls{osc} (which is a joint effort between the Linux Foundation and the O-RAN Alliance) as the preferred \gls{smo} platform for open-source O-RAN code releases~\cite{nonrtric-osc}. 
It is also worth mentioning that in May 2021 \gls{etsi} signed a cooperation agreement with the O-RAN Alliance~\cite{etsi-oran}. Because of this, the integration efforts between \gls{osm} and the O-RAN architecture are still at an early stage~\cite{osm2020oran}.

\section{The O-RAN Open Interfaces}
\label{sec:interfaces}

As discussed in Section~\ref{sec:arch}, the control loops supported by near-RT and non-RT \glspl{ric} are enabled by a set of standardized interfaces. Each interface enables services (e.g., reporting of telemetry from the \gls{ran}) through the combination of well-defined procedures (e.g., the subscription and indication procedures for E2). Procedures involve the exchange of messages between the endpoints of an interface (e.g., the indication message for E2). This section reviews the logical abstractions and procedures that define such interfaces, providing insights on their role in the Open \gls{ran} ecosystem. Specifically, the E2 interface is described in Section~\ref{sec:e2}, the O1 interface in Section~\ref{sec:o1}, the A1 interface in Section~\ref{sec:a1}, and the fronthaul interface in Section~\ref{sec:fh}. Finally, Section~\ref{sec:other-int} reviews the remaining O-RAN and 3GPP interfaces.

\subsection{E2 Interface}
\label{sec:e2}

The E2 interface is an open interface between two endpoints, i.e., the near-RT \gls{ric} and the so-called E2 nodes, i.e., \glspl{du}, \glspl{cu}, and O-RAN-compliant LTE \glspl{enb}~\cite{oran-wg3-e2-gap}. The E2 allows the \gls{ric} to control radio resource management and other functionalities of the E2 nodes. Moreover, this interface also enables the collection of metrics from the \gls{ran} to the near-RT \gls{ric}, either periodically or after pre-defined trigger events. Both control and data collection procedures can pertain to one or more cells, slices, \gls{qos} classes, or specific \glspl{ue}.

To support the above operations, the O-RAN Alliance uses a variety of unique identifiers. Specifically, O-RAN uses identifiers based on \gls{3gpp} specifications for the \gls{gnb}, slice, and \gls{qos} class~\cite{3gpp.23.501}.
Regarding specific \glspl{ue}, the O-RAN Alliance 
has instead introduced a common user identifier (i.e., the UE-ID) across its specifications. This provides a consistent and uniform user identity across the system without exposing sensitive information related to the user. 

\begin{figure}[t]
    \centering
    \includegraphics[width=\columnwidth]{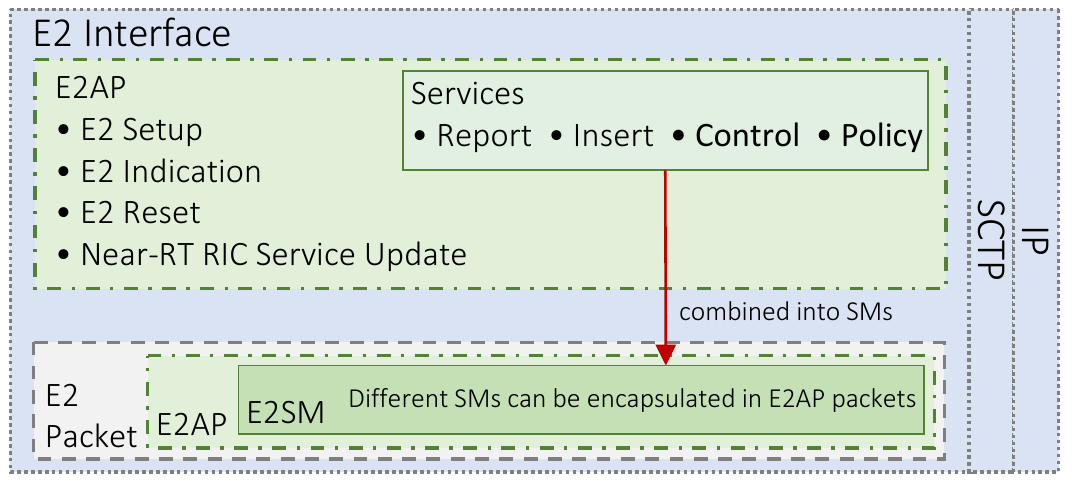}
    \caption{Representation of an O-RAN E2AP packet (bottom left), which includes an E2SM payload (top left). The E2 payload is then encapsulated in SCTP and IP headers (right). The top part of the figure also summarizes the services provided by the E2 interface.}
    \label{fig:e2}
\end{figure}

The E2 interface has been logically organized in two protocols: E2 \gls{ap} and E2 \gls{sm}. The E2 \gls{ap}~\cite{oran-wg3-e2-ap} is a basic procedural protocol that coordinates how the near-RT \gls{ric} and the E2 nodes communicate with each other, and provides a basic set of services, as shown in Figure~\ref{fig:e2}. E2AP messages can embed different E2 \glspl{sm}~\cite{oran-wg3-e2-sm}, which implement specific functionalities (i.e., the reporting of \gls{ran} metrics or the control of \gls{ran} parameters). The E2 interface runs on top of the \gls{sctp} protocol~\cite{stewart2001sctp}. 

\begin{figure*}[t]
\begin{subfigure}[t]{0.48\textwidth}
    \centering
    \includegraphics[width=0.9\columnwidth]{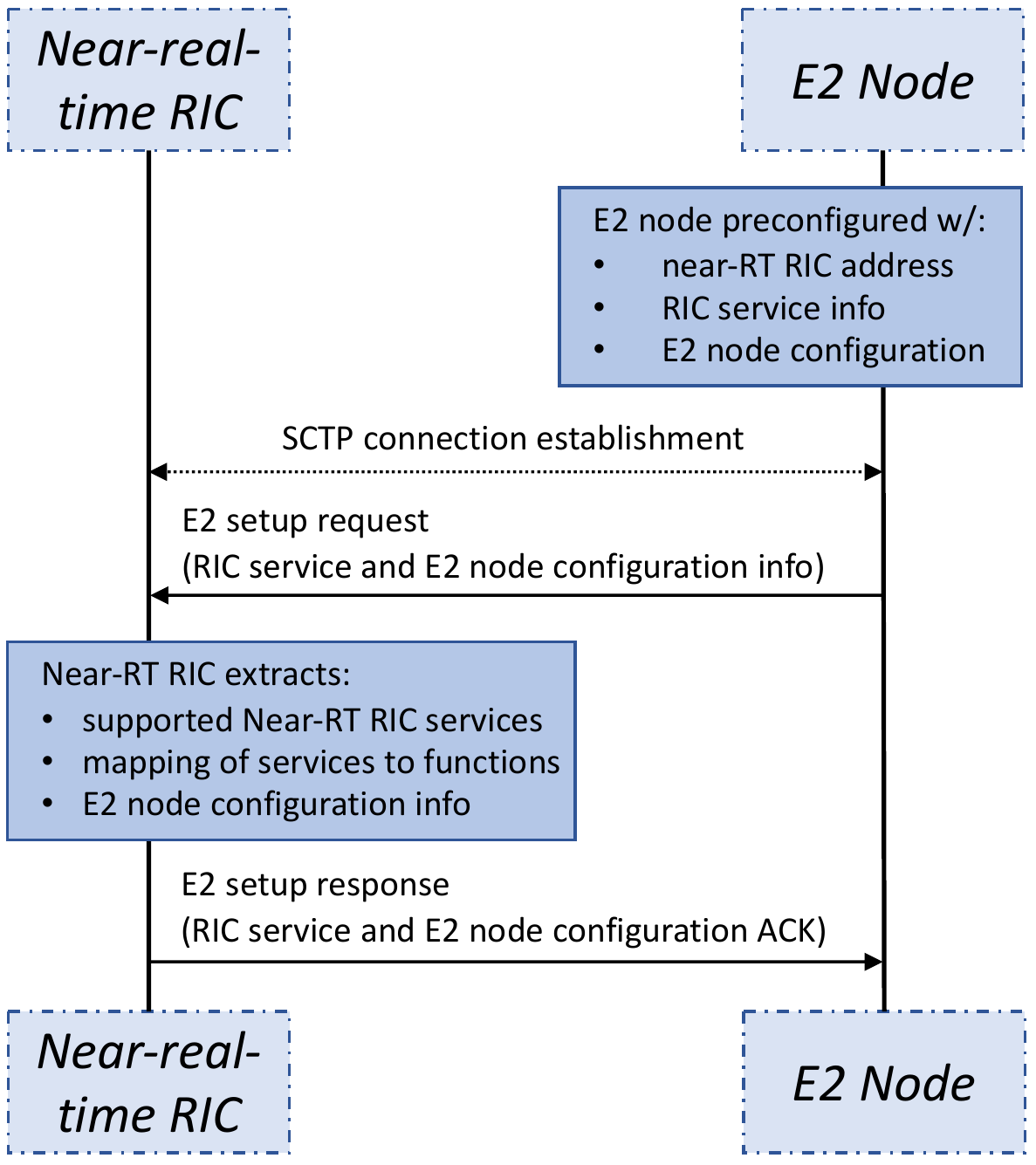}
    \caption{Procedure for the setup of an E2 session between the near-RT RIC and an E2 node. The procedure is initiated by the E2 node which interacts with the near-RT RIC.}
    \label{fig:e2-setup}
\end{subfigure}%
\hfill%
\begin{subfigure}[t]{0.48\textwidth}
    \centering
    \includegraphics[width=0.9\columnwidth]{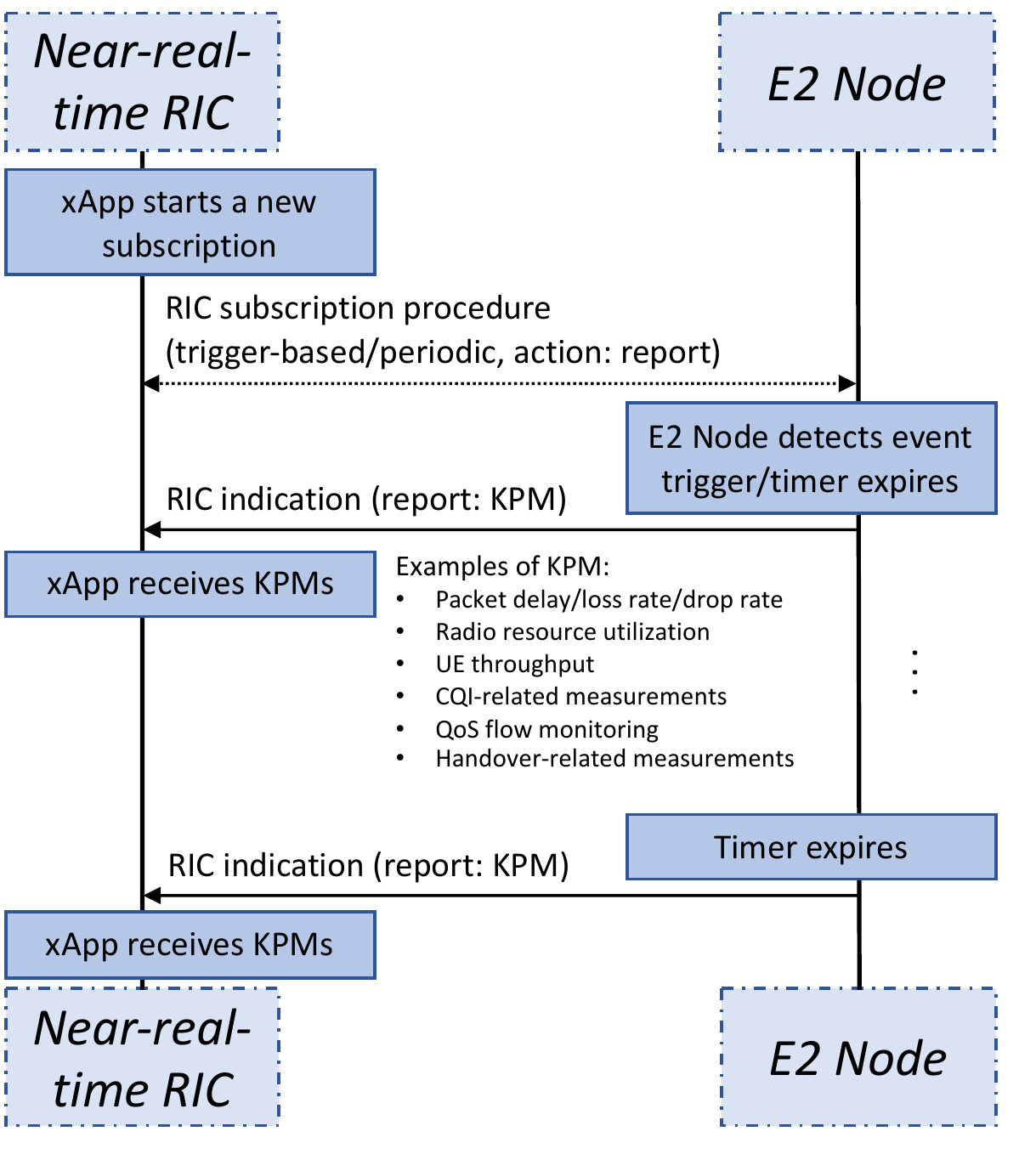}
    \caption{Procedures related to the streaming of \glspl{kpm} from the E2 node to the near-RT RIC. The subscription procedure is started by an xApp on the near-RT RIC, which then receives the reports.}
    \label{fig:e2-sm-kpm}
\end{subfigure}
\caption{Procedures for E2 setup and E2SM KPM. The vertical lines represent the temporal evolution of the process, while horizontal lines are the messages exchanged by the near-RT RIC and the E2 node.}
\label{fig:e2-setup-e2sm-kpm}
\end{figure*}

Each E2 node exposes a number of \textit{\gls{ran} functions}, i.e., the services and capabilities it supports. For example, \glspl{du} from different vendors may expose different control knobs depending on which parameters and functionalities can be tuned, as well as their capability in collecting and reporting different performance metrics. By using publish-subscribe mechanics, E2 nodes can publish their data and the xApps on the near-RT \gls{ric} can subscribe to one or more of these \gls{ran} functions through the E2 interface. This makes it possible to clearly separate the capabilities of each node and to define how the xApps interact with the \gls{ran}.

At the lowest level, the E2AP handles interface management (setup, reset, reporting of errors for the E2 interface itself) and near-RT \gls{ric} service updates (i.e., the exchange of the list of the RAN functions supported by the E2 node). As an example, the E2 setup procedure is shown in Figure~\ref{fig:e2-setup}. At first, the \gls{sctp} connection is established between the near-RT \gls{ric} and the E2 node (which is aware of the IP address and port of the E2 termination of the near-RT \gls{ric}). Then, the E2 node transmits an E2 setup request, in which it lists the \gls{ran} functions and configuration it supports, together with the identifiers for the node. The near-RT \gls{ric} processes this information and replies with an E2 setup response. 

After the connection is established, the E2AP provides four services which can be combined in different ways to implement an \gls{e2sm}~\cite{oran-wg3-e2-sm}. These services, also shown in Figure~\ref{fig:e2}, are~\cite{oran-wg3-e2-ap}:
\begin{itemize}
    \item \textit{E2 Report}. The report service involves \textit{E2 RIC Indication} messages that contain data and telemetry from an E2 node. The E2 report service is activated upon subscription from an xApp to a function offered by the E2 node. During the subscription negotiation, the xApp in the near-RT \gls{ric} can specify trigger events or the periodicity with which the E2 node should send report messages. Based on this periodicity, a timer is set in the E2 node and a report is sent whenever the timer expires.
    The \gls{ric} Indication message is of type \textit{report}. 
    
    \item \textit{E2 Insert}. Similarly, the insert service involves messages sent from an E2 node to an xApp in the near-RT \gls{ric} to notify the xApp about a specific event in the E2 node (e.g., a \gls{ue} signaling the possibility to perform a handover).
    It is activated upon subscription from an xApp and involves a RIC Indication message (of type \textit{insert}). In this case, the trigger is associated to a RAN radio resource management procedure which is suspended when the insert message is sent. A wait timer is also started, and, if the \gls{ric} does not reply before the timer expires, the procedure in the E2 node can be resumed or definitely halted.
    
    \item \textit{E2 Control}. The control service can be autonomously initiated by the \gls{ric}, or it can be the consequence of the reception of an \textit{insert} message at the near-RT \gls{ric}. This service is based on a procedure with two messages, a \textit{RIC Control Request} from the \gls{ric} to the E2 node, and a \textit{RIC Control Acknowledge} in the opposite direction. The control services can influence parameters exposed by the RAN functions of the E2 node.
    
    \item \textit{E2 Policy}. This service involves a subscription procedure that specifies (i) an event trigger, and (ii) a policy that the E2 node should autonomously follow to perform radio resource management.
\end{itemize}

These services are then combined to create a service model. The service model message is inserted as payload in one of the E2AP messages, as shown in Figure~\ref{fig:e2}. The actual content is encoded using ASN.1 notation, i.e., through well-defined types for numbers and key-value pairs~\cite{larmouth2000asn}. We provide examples of an E2 Subscription Request message and of an E2 Indication message (of type report) in Appendices~\ref{app:e2sub} and~\ref{app:e2rep}, respectively.

\textbf{E2 Service Models.} 
At the time of this writing, the O-RAN Alliance \gls{wg}3 has standardized three service models: (i)~the \gls{e2sm} \gls{kpm}~\cite{oran-wg3-e2-sm-kpm}; (ii)~the \gls{e2sm} \gls{ni}~\cite{oran-wg3-e2-sm-ni}, and (iii)~the \gls{e2sm} \gls{rc}~\cite{oran-wg3-e2-sm-rc}.

The \gls{e2sm} \gls{kpm}~\cite{oran-wg3-e2-sm-kpm} reports performance metrics from the \gls{ran}, using the E2 report service. The procedures associated to the \gls{kpm} service model are shown in Figure~\ref{fig:e2-sm-kpm}. During the E2 setup procedures, the E2 node advertises the metrics it can expose. An xApp in the near-RT \gls{ric} can then send a subscription message specifying which \glspl{kpm} are of interest, and whether the reporting is periodic or trigger-based. Finally, the E2 node uses E2 Indication messages of type \emph{report} to stream the selected \glspl{kpm}.
Different \gls{kpm} messages are generated from different E2 nodes, i.e., the specifications defines performance metrics \textit{containers} with different fields to be populated for \gls{du}, \gls{cu}-\gls{up}, and \gls{cu}-\gls{cp}. Additionally, the recent Version 2 of the \gls{kpm} specifications~\cite{oran-wg3-e2-sm-kpm} has introduced cell-specific and user-specific performance containers, which are based on the metrics defined in \gls{3gpp} technical specifications for LTE and NR~\cite{3gpp.28.552,3gpp.32.425}. 

The \gls{e2sm} \gls{ni}~\cite{oran-wg3-e2-sm-ni} is used to take the messages received by the E2 node on specific network interfaces and forward them to the near-RT \gls{ric} domain via E2 \emph{report} messages over the E2 interface.
The E2 node advertises which interfaces it supports during the subscription procedure, and they include X2 (which connects LTE \glspl{enb}), Xn (which connects different NR \glspl{gnb}), and F1, which connects \glspl{du} and \glspl{cu}). 


\textbf{Closed-loop control with E2SM RAN Control (RC).} One of the goals of the xApps is to optimize the radio resource management in the E2 nodes. The \gls{e2sm} \gls{rc}, introduced in July 2021, implements control functionalities through E2 control services. 
%
%
The first release of the specifications for \gls{e2sm} includes a broad set of control domains:
\begin{itemize}
    \item radio bearer control, to modify parameters for the bearers of an E2 node or of a specific UE, for example related to \gls{qos} parameters, bearer admission control, split bearer and \gls{pdcp} duplication control;
    \item radio resource allocation control, to modify, among others, discontinuous reception, semi-persistent scheduling, for slice-level \glspl{prb} for an E2 node, a cell, a slice, a UE, or \gls{qos} classes;
    \item connected mode mobility control, to initiate a mobility procedure for \glspl{ue} in \gls{rrc} connected state, i.e., a handover to a specific cell, or a conditional handover\footnote{The conditional handover is a feature in 3GPP NR Release 16 that mandates conditions for which the \gls{ue} should handover to a target cell, but does trigger an immediate handover~\cite{3gpp.38.331}.} to a set of cells;
    \item radio access control, to set parameters for random access backoff, \gls{ue} admission to a cell, among others;
    \item \gls{dc} control, to configure and trigger the handover of a UE to selected target secondary cells, secondary cell updates, or \gls{dc} release;
    \item \gls{ca} control, to initiate \gls{ca} and modify the component carriers for a specific UE;
    \item idle mobility control, to modify functions for mobility procedures of \glspl{ue} in a \gls{rrc} idle state, including cell re-selection priorities and idle timers.
\end{itemize}

\gls{e2sm} \gls{rc} also provides capabilities for UE identification and UE information reporting. The control actions and policies that \gls{e2sm} specifies relate to specific parameters standardized by the \gls{3gpp}, i.e., the control action will usually carry the value for a \gls{3gpp} \gls{ie} defined in the \gls{e2sm} specifications~\cite{oran-wg3-e2-sm-rc}. For example, the handover control \gls{ie} includes (among other things) a target cell ID for the handover encoded as a \gls{3gpp} \emph{Target Cell Global ID} \gls{ie} in~\cite[Section 9.2.3.25]{3gpp.38.423}.

\begin{figure}[t]
    \centering
    \includegraphics[width=.95\columnwidth]{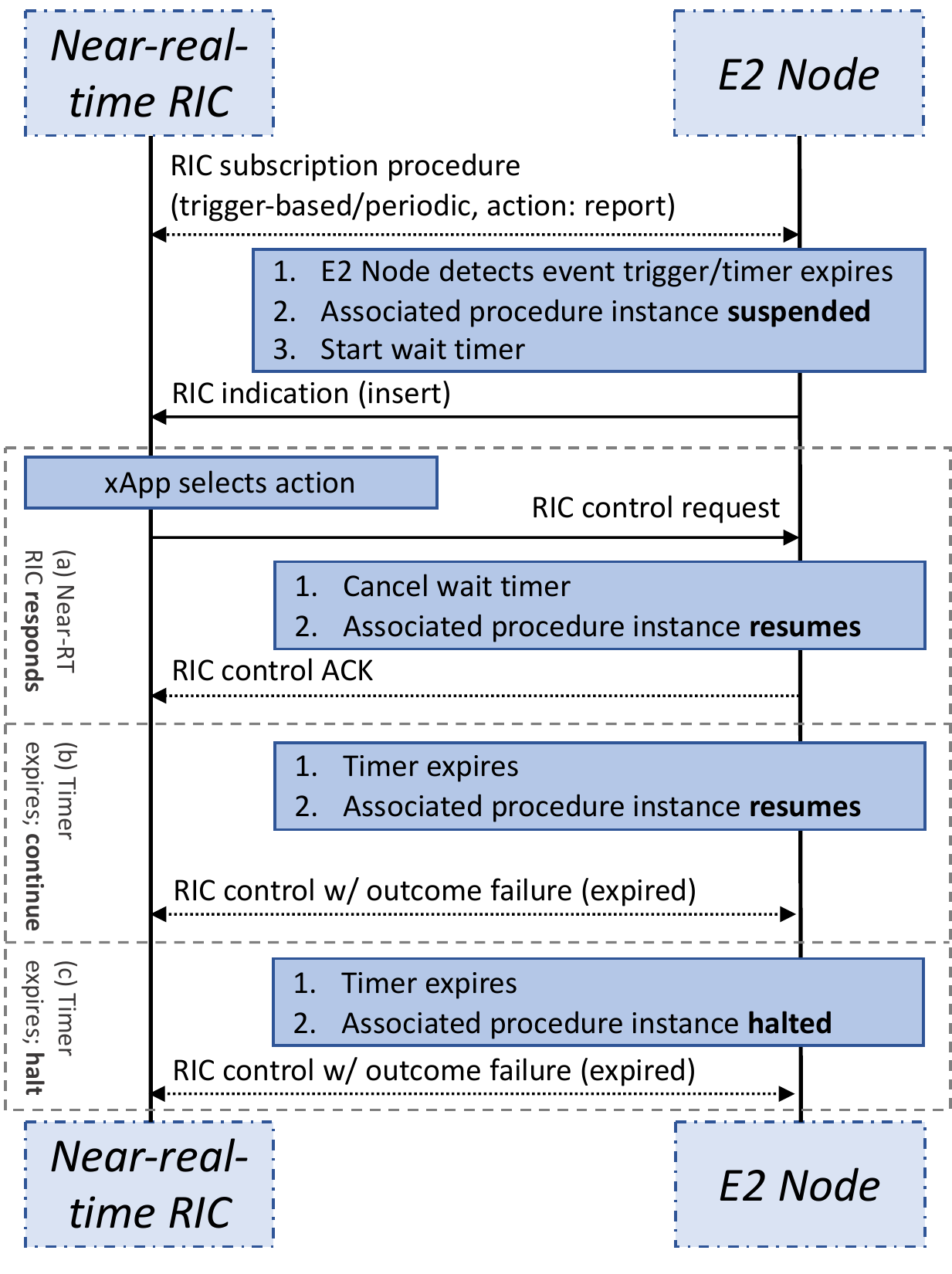}
    \caption{E2 insert service with subsequent E2 control service response. The vertical lines represent the temporal evolution of the process, while horizontal lines are the messages exchanged by the near-RT RIC and the E2 node.}
    \label{fig:e2-insert-control}
\end{figure}

To effectively implement control actions and/or enforce policies, the \gls{e2sm} leverages a combination of the E2 services described in Section~\ref{sec:e2}. Figure~\ref{fig:e2-insert-control} shows an example of the E2 messages that an E2 node and an xApp in the near-RT \gls{ric} can exchange during a typical control procedure. First, the xApp subscribes to the E2 node, which needs to expose a specific \gls{ran} function called \emph{\gls{ran} control}. During the subscription phase, the xApp can specify a triggering event or a timer. Then, if the timer expires or the condition defined in the trigger is verified, the E2 node sends an E2 Insert message to the \gls{ric}. For example, this may happen when a metric that the \gls{gnb} monitors exceeds a certain threshold, e.g., when the \gls{rsrp} for a neighboring cell (or target cell) becomes better than that of the current cell plus an offset (also called \emph{A3 event} in the \gls{3gpp} specifications~\cite{3gpp.38.331}). In this case, the E2 node sends a handover control request insert message (in which it can also specify the target cell defined by the E2 node itself), suspends the radio resource management procedure, and starts a timer. Three different outcomes can follow this event. The near-RT \gls{ric} can reply with an E2 control message that either denies the control request, or accepts it and replies with a control action (for example, a target cell ID selected by the xApp). Alternatively, if the E2 node timer expires before the reception of a message from the \gls{ric}, the procedure may continue autonomously, or it may be halted, according to the specific procedure and configuration of the E2 node. 

The control action sent by the xApp can also be asynchronous, i.e., it does not depend on the reception of an insert message from the E2 node. Additionally, the \gls{e2sm} can also be used to specify policies, i.e., to alter pre-defined behaviors in the E2 nodes. Policies can be of two different types: (i) control policies that allow the E2 node to perform radio resource management actions without the interaction with the \gls{ric}, when certain conditions are satisfied, and (ii) offset policies, which change \gls{3gpp}- or vendor-defined thresholds by adding or removing offsets and thus modifying how the E2 node performs specific functions.

\subsection{O1 Interface}
\label{sec:o1}

Besides E2, the other interface that connects O-RAN specific components with \gls{ran} nodes is the O1 interface~\cite{oran-wg1-o1}. In general, O-RAN-managed elements (including the near-RT \gls{ric}, \gls{ran} nodes) are connected via O1 to the \gls{smo} and the non-RT \gls{ric}. The O1, thus, is an open interface which adopts and extends standardized practices for \textit{operations and maintenance}.

\begin{figure}[t]
    \centering
    \includegraphics[width=\columnwidth]{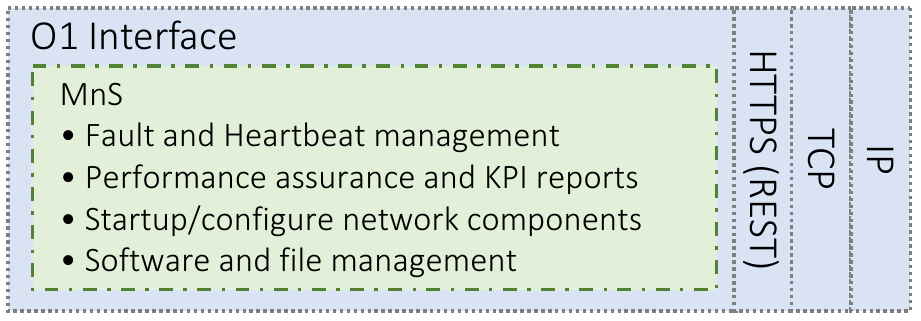}
    \caption{O-RAN O1 interface and \acrlong{mns} payload (left), which is encapsulated in HTTPS/TCP/IP packets (right).}
    \label{fig:o1}
\end{figure}

The O1 interface supports \textit{\gls{mns}}, which include the management of the life-cycle of O-RAN components (from startup and configuration to fault tolerance and heartbeat services~\cite{3gpp.28.622}), performance assurance and trace collection through \glspl{kpi} reports, and software and file management (see Figure~\ref{fig:o1}).
The O1 interface generally connects one \gls{mns} provider (i.e., generally the node managed by the \gls{smo}) to one \gls{mns} consumer (i.e., the \gls{smo}).

The \emph{Provisioning Management Services} allow the \gls{smo} to push configurations to the managed nodes, and the reporting of external configuration updates from managed nodes to the \gls{smo}.
For this, O1 uses a combination of REST/HTTPS \glspl{api} and NETCONF~\cite{RFC6241}, which is a protocol standardized by the \gls{ietf} for the life-cycle management of networked functions. The supported provisioning services include those defined in 3GPP technical specifications~\cite{3gpp.28.531,3gpp.28.532}. An additional \emph{Fault Supervision \gls{mns}} is used to report errors and events to the \gls{smo}. It is also based on 3GPP-defined fault events~\cite{3gpp.28.545,3gpp.28.532,3gpp.28.622,3gpp.28.623}, and can be used by the \gls{ran} nodes to report errors (through standardized JSON payloads) using REST APIs. For each node, the \gls{smo} can also query a list of alarms (i.e., probes that monitor the status of specific elements and components in the node), and, in case, acknowledge or clear them. Finally, the \gls{smo} can provision heartbeats on the devices it manages, through the \emph{Heartbeat \gls{mns}}, and manage not only virtual but also \glspl{pnf}. Heartbeat messages are used to monitor the status and availability of services and nodes.

The \textit{Performance Assurance \gls{mns}} can be used to stream (in real time) or report in bulk (through file transfer) performance data to the \gls{smo}, to enable, for example, data analytics and data collection for AI/ML. The \gls{smo} can select the \glspl{kpi} (also referred to as counters) to be reported and the frequency of reporting. It relies on use cases and formats for \gls{kpi} reporting defined by the 3GPP~\cite{3gpp.28.550,3gpp.28.532,3gpp.28.622,3gpp.28.623} or by the \gls{ves} project. The performance metrics are also either based on 3GPP documents~\cite{3gpp.28.552}, vendor specific, or standardized by the different \glspl{wg} of the O-RAN Alliance.
For bulk transfer, a file-ready notification is first sent from the \gls{mns} provider (e.g., the specific node of the \gls{ran}) to the \gls{smo} through HTTP APIs. Then, a file transfer through SFTP is performed. The \gls{smo} can also download files for which the file-ready notification was received in previous instants. A WebSocket is instead used for real-time streaming, following a handshake. The \gls{smo} can also monitor trace-based events through the Trace \gls{mns}, e.g., to profile calls, \gls{rrc} connection establishment or radio link failures.

The O1 interface can also be used to push and/or download files on the nodes managed by the \gls{smo}. This enables, for example, software updates, new beamforming configuration files for the \glspl{ru}, and deployment of \gls{ml} models and security certificates.

\subsection{A1 Interface}
\label{sec:a1}

The A1 interface connects two O-RAN-specific components, i.e., the non-RT \gls{ric} (or \gls{smo}) and the near-RT \gls{ric}, as shown in Figure~\ref{fig:oran-architecture}~\cite{oran-wg2-a1}. It allows the non-RT \gls{ric} to deploy policy-based guidance for the near-RT \gls{ric} (e.g., to set high-level optimization goals), to manage \gls{ml} models used, for example, in xApps, and to negotiate and orchestrate the transfer of enrichment information for the near-RT \gls{ric}. This is done through standard-defined mechanisms that are based on a specific syntax (based on JSON schema) which can express policies and high-level intent. The policies, \gls{ml} models, and enrichment information can refer to a group of \glspl{ue}, or even to a specific \gls{ue}. Notice that, at the time of writing, the A1-based \gls{ml} model management is still considered for further study~\cite{oran-wg2-a1,oran-wg2-ml}.

\begin{figure}[t]
    \centering
    \includegraphics[width=\columnwidth]{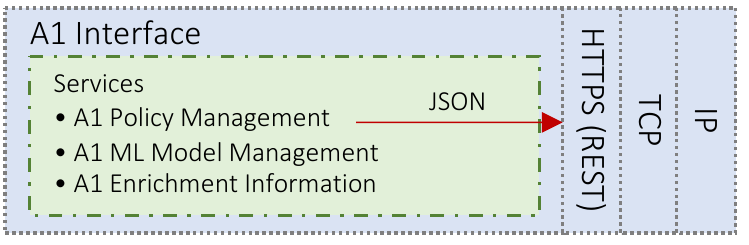}
    \caption{O-RAN A1 JSON payload (left), which is encapsulated in HTTPS/TCP/IP packets (right). The figure also summarizes the services that can be provided over A1.}
    \label{fig:a1}
\end{figure}

The A1 interface, illustrated in Figure~\ref{fig:a1}, relies on the A1AP application protocol, whose functionalities are then further specified for each service it supports~\cite{oran-wg2-a1-ap}. The A1AP is based on a 3GPP framework for policy deployment for network functions~\cite{3gpp.23.501}, which combines REST APIs over HTTP for the transfer of JSON objects. For each service, both the non-RT and the near-RT \glspl{ric} feature a pair of HTTP clients and servers, which are used alternatively for service management and for the actual data transfer and/or notifications.

The \textit{A1 Policy management} is used by the non-RT \gls{ric} to drive the functionalities of the near-RT \gls{ric} to achieve high-level intent for the \gls{ran}, as we will discuss in Section~\ref{sec:orchestration}. This intent is generally defined through \gls{qos} or \gls{kpi} goals for all users or subsets of users (e.g., a slice) and monitored using the reporting functionalities of the O1 interface and feedback over A1.\footnote{This can only notify if the policy is enforced or not.} The policies are defined by the non-RT \gls{ric} and then deployed over A1. The non-RT \gls{ric} is also tasked with monitoring and managing the life cycle of the policies, thanks to APIs for deleting, updating, and querying policies in the near-RT \gls{ric}. 

Each policy is based on specific JSON schema which are grouped according to different policy types~\cite{oran-wg2-a1-types}. All JSON schema have in common a policy identifier, which is unique for the non-RT \gls{ric}, a scope identifier, and one or more policy statements. The scope can be a single \gls{ue}, a group of \glspl{ue}, slices, cells, bearers, and application classes. The policy itself is then expressed through a sequence of policy statements, which can cover policy resources (i.e., the conditions for resource usage for a policy) and policy objectives (i.e., the goal of the policy in terms of \gls{qos} or \gls{kpi} targets). The O-RAN technical specification~\cite{oran-wg2-a1-types} lists several standardized types for policy statements, which depend on specific use cases (e.g., throughput maximization, traffic steering preferences, \gls{qos} targets, among others).

Finally, the \textit{A1 \gls{ei}} service aims at improving \gls{ran} performance by providing information that is generally not available to the \gls{ran} itself (e.g., capacity forecasts, information elements from sources outside the \gls{ran}, aggregate analytics). The non-RT \gls{ric} and the \gls{smo} have indeed a global perspective on the network and access to external sources of information, and can relay it to the xApps in the near-RT \gls{ric} using A1 \gls{ei}. The flow of information can also bypass the non-RT \gls{ric}, which can instruct the near-RT \gls{ric} (over A1) to connect directly to sources of \gls{ei}. 


\subsection{O-RAN Fronthaul}
\label{sec:fh}

\begin{figure}[t]
    \centering
    \includegraphics[width=\columnwidth]{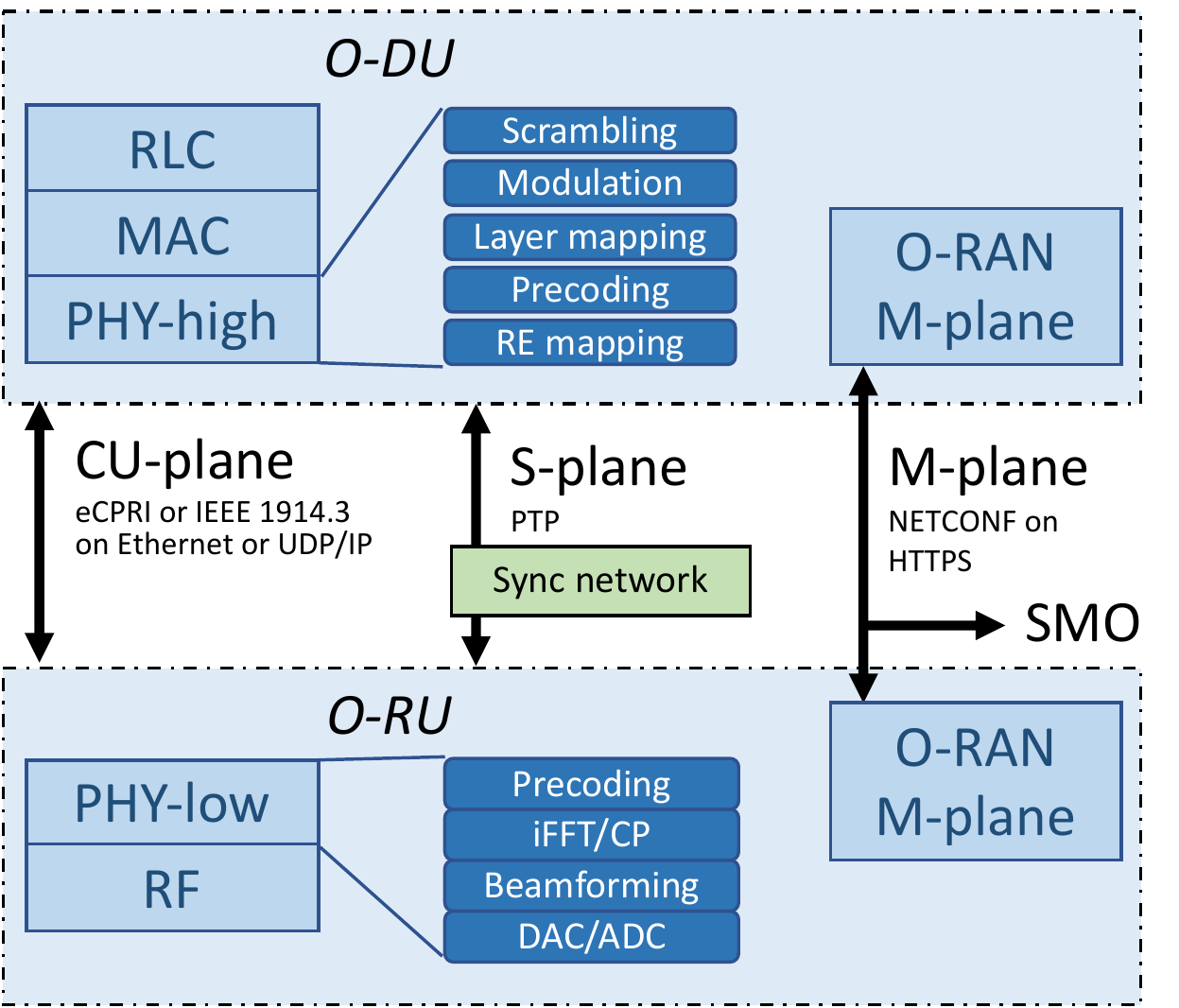}
    \caption{O-RAN fronthaul interface and fronthaul planes. This interface enables the 7.2x split of the physical layer functionalities across DU and RU, with data and control between the PHY-high and PHY-low transported over the Control and User- (CU-) planes. The S-plane provides synchronization, and the M-plane management and orchestration functionalities.}
    \label{fig:oran-fh-interface}
\end{figure}

The O-RAN \gls{fh} interface connects a \gls{du} to one or multiple \glspl{ru} inside the same \gls{gnb}~\cite{oran-wg4-fronthaul-cus,oran-wg4-fronthaul-m}. The O-RAN \gls{fh} interface makes it possible to distribute the physical layer functionalities between the \gls{ru} and the \gls{du}, and to control \gls{ru} operations from the \gls{du}. As discussed in Section~\ref{sec:arch}, the O-RAN Alliance has selected a specific configuration (split 7.2x) for the splitting of the physical layer among those proposed by the 3GPP~\cite{3gpp.38.801}. As shown in Figure~\ref{fig:oran-fh-interface}, the lower part of the physical layer (\emph{low PHY}) resides in the \gls{ru} and performs \gls{ofdm} phase compensation~\cite{petrovic2007effects}, the inverse \gls{fft} and \gls{cyp} insertion for frequency-to-time conversion in downlink, and \gls{fft} and \gls{cyp} removal in uplink. More capable \glspl{ru} (i.e., category B \glspl{ru}) can also perform precoding. This functionality is implemented at the \gls{du} for less capable \glspl{ru} (i.e., category A). To complement the \gls{du} capabilities, category A \glspl{ru} need to support low PHY processing for at least 8 data streams. The physical layer in the \gls{du} (\emph{high PHY}) performs scrambling, modulation, layer mapping, and resource element mapping. 

According to O-RAN specifications~\cite{oran-wg4-fronthaul-cus}, the 7.2x split strikes a balanced trade-off among the simplicity of the interface and of the \gls{ru} design, the potential for interoperability (fewer parameters to configure than higher layer splits), and the data rate required for fronthaul transport (lower with respect to configurations that split the physical layer even further). The latter can be based on Ethernet or UPD/IP encapsulation, carrying either an eCPRI~\cite{ecpri} or an IEEE 1914.3~\cite{ieee-1914.3} payload. Note that one \gls{du} can support more than one \gls{ru}, e.g., to serve carriers of the same cells from different \glspl{ru}, or to process multiple cells with one \gls{du} and multiple \glspl{ru}. To do this, the O-RAN \gls{fh} specifications foresee an additional component which multiplexes a fronthaul stream to multiple \glspl{ru}, or the daisy-chaining of \glspl{ru}. In addition, the O-RAN \gls{fh} interface has been designed to support reliable, low-latency communications between \glspl{du} and \glspl{ru} with timing that matches the requirements of \gls{urllc} flows. For example, the fronthaul interface includes different modulation compression techniques, to reduce the load on the fronthaul network~\cite{lagen2021modulation}, and fronthaul networks can be designed to support URLLC flows with minimal jitter~\cite{Chitimalla:17}.

The O-RAN \gls{fh} protocol includes four different functionalities (or \emph{planes}). Besides the user (U-) and control (C-) planes (for transport of data and PHY-layer control commands)~\cite{oran-wg4-fronthaul-cus}, the O-RAN \gls{fh} also features a synchronization plane (S-plane), for timing management among \glspl{du} and \glspl{ru}~\cite{oran-wg4-fronthaul-cus}, and a management plane (M-plane), for the configuration of the \gls{ru} functionalities from the \gls{du} itself~\cite{oran-wg4-fronthaul-m-yang,oran-wg4-fronthaul-m}. In the following, we describe the functionalities of the different \gls{fh} planes.

\textbf{C-plane.} The C-plane takes care of transferring commands from the high-PHY in the \gls{du} to the low-PHY of the \gls{ru}, including scheduling and beamforming configurations, management of different NR numerologies in different subframes, downlink precoding configuration, and spectrum sharing control. For the latter, the specifications include \gls{laa} procedures, such as the possibility to perform listen-before-talk in the \gls{ru} and \gls{du} pair, and dynamic spectrum sharing operations, with the possibility to specify which \glspl{prb} are dedicated to spectrum sharing between LTE and NR.

The C-plane messages are encapsulated in eCPRI or IEEE 1914.3 headers and payloads, with specific fields and commands for the different control procedures. The O-RAN \gls{fh} specifications also provide details on how specific C-plane directives (e.g., related to the usage of a specific beamforming vector) can be coupled with specific U-plane packets (and thus symbols to be transmitted).

The combination of C-plane and M-plane can be used to configure and manage beamforming capabilities of the \gls{ru}, a key feature in 5G networks, especially at FR2~\cite{giordani2018tutorial}. In particular, each \gls{ru} can support multiple antenna panels, each with multiple \gls{tx} or \gls{rx} arrays~\cite{oran-wg4-fronthaul-m}. Each array can be mapped to one or more data flows. The fronthaul interface allows the control of amplitude and phase of the radiating elements in the phased arrays of the \gls{ru} (for beamforming in the time domain), or the selection of digital precoding weights (for beamforming in the frequency domain), with four different beamforming options. With \emph{predefined-beam beamforming}, the \gls{du} dynamically selects (through the C-plane) time and/or frequency beamforming vectors\footnote{The combination of time and frequency beamforming enables hybrid beamforming strategies~\cite{gomez2021hybrid}.} that the \gls{ru} advertises as available at startup (through the M-plane). Alternatively, with \emph{attributed-based beamforming}, the \gls{du} can select beams based on specific attributes, e.g., azimuth and elevation angles. With \emph{weight-based beamforming}, the \gls{du} also specifies the weights for generic time and/or frequency domain beamforming vectors. The last option is \textit{channel-information-based beamforming}, in which the \gls{du} provides the \gls{ru} with \gls{csi} for a specific user and the \gls{ru} computes the beamforming weights. The O-RAN \gls{fh} supports a well-defined model for the \gls{ru} antennas, so that the \gls{du} can unambiguously identify antenna elements, their polarization, position, and orientation of the panel.

\textbf{U-plane.} The main functionality of the U-plane is transferring I/Q samples in the frequency domain between the \gls{ru} and the \gls{du}. Typically, a C-plane message specifies scheduling and beamforming configuration, and is followed by one or more U-plane messages with the I/Q samples to be transmitted in the corresponding transmission opportunities. The U-plane also takes care of timing the transmission of its messages so that they are received at the \gls{ru} with enough time for processing before transmission. Additionally, the U-plane specifies the digital gain of the I/Q samples, and can compress them for more efficient data transfer. 

\textbf{S-plane.} The S-plane takes care of time, frequency, and phase synchronization between the clocks of the \glspl{du} and of the \glspl{ru}. This is key to a correct functioning of a time- and frequency-slotted system distributed across multiple units. Thanks to the shared clock reference, the \gls{du} and \gls{ru} can properly align time and frequency resources for the transmission and reception of the different LTE and NR data and control channels.

The O-RAN \gls{fh} S-plane can be deployed with different topologies, specified in~\cite{oran-wg4-fronthaul-cus}, which differ according to whether a direct \gls{du}-\gls{ru} link exists, or if the two elements are connected through a fabric of Ethernet switches. Additionally, the \gls{fh} specifications include different synchronization profiles, based on different protocols, such as \gls{plfs} or \gls{ptp}, which can achieve sub-microsecond time accuracy~\cite{scheiterer2009synch}. 

\textbf{M-plane.} The O-RAN \gls{fh} M-plane is a protocol that runs in parallel to the C-, U-, and S-planes, with dedicated endpoints in the \gls{du} and \gls{ru} that establish an IPv4 or IPv6 tunnel~\cite{oran-wg4-fronthaul-m}. It enables the initialization and the management of the connection between the \gls{du} and the \gls{ru}, and the configuration of the \gls{ru}. In this context, the specifications foresee two architectural options, i.e., hierarchical, in which the \gls{smo} manages the \gls{du} and the \gls{du} manages the \gls{ru}, and hybrid, in which the \gls{smo} can also interact directly with the \gls{ru}. The M-plane of the O-RAN \gls{fh} can thus function as the O1 interface of the \gls{ru}. As for O1, the management directives are based on NETCONF. Finally, contrary to the C-, U-, and S-planes, the M-plane is end-to-end encrypted through SSH and/or TLS.

The M-plane takes care of several operations related to the life cycle of the \gls{ru}. First, it manages the start-up, during which the \gls{ru} establishes the management with the \gls{du} and/or the \gls{smo} thanks to pre-defined IP addresses or DHCP configurations. Then, it enables software updates, configuration management, performance and fault monitoring, and file management for bulk transfer of data. Among others, the M-plane manages the registration of the \gls{ru} as \gls{pnf}, the parameters of the \gls{ru}-to-\gls{du} link (including timing), and the update of beamforming vectors (from the deployment of new codebooks, to the tilting of existing ones, and calibration of the antennas).

Besides standardizing the \gls{fh} interface, the O-RAN Alliance is also developing a set of specifications to characterize the transport and synchronization capabilities of an open fronthaul or crosshaul network that supports the connectivity between \glspl{du} and \glspl{ru}. For example,~\cite{oran-wg9-sync} reviews network-enabled synchronization, by discussing \gls{ptp} profiles, support required by the Ethernet substrate, points of failures, among others. Other areas of interest are related to the management of the open fronthaul network, wave-division-multiplexing-based networks, and packet-switched architectures with modern features such as, for example, slicing.

\subsection{Other Interfaces}
\label{sec:other-int}

The O2 interface connects the \gls{smo} to the O-RAN O-Cloud, enabling the management and provisioning of network functions in a programmatic manner~\cite{oran-wg6-o2}. It allows the definition of an inventory of the facilities controlled by the O-Cloud, monitoring, provisioning, fault tolerance and updates. For this interface, the O-RAN Alliance \gls{wg}6 is considering the adoption of well-known standards and open-source solutions, e.g., relevant \gls{etsi} \gls{nfv} standards, 3GPP service-based interfaces, and the Kubernets, Open Stack, and \gls{onap}/\gls{osm} projects. 

Finally, as shown in Figure~\ref{fig:oran-architecture}, the O-RAN disaggregated architecture also leverages additional interfaces defined by the 3GPP. Notably, the E1 interface connects the \gls{cu} control and user functions~\cite{3gpp.38.460}. The F1 interface connects the \gls{cu} to the \gls{du}, with dedicated sub-interfaces for user and control planes~\cite{3gpp.38.473}. The Xn (X2) interface connects different \glspl{gnb} (\glspl{enb}), for example to perform handovers and to enable dual connectivity~\cite{3gpp.36.423,3gpp.38.423}. The Uu interface exists between an \gls{ue} and the \gls{gnb}~\cite{3gpp.38.401}, and the NG interface connects the \gls{gnb} to the 5G core, i.e., to the \gls{upf} for the user plane and the \gls{amf} for the control plane~\cite{3gpp.38.401}.

\begin{figure*}
    \centering
    \includegraphics[width=\textwidth]{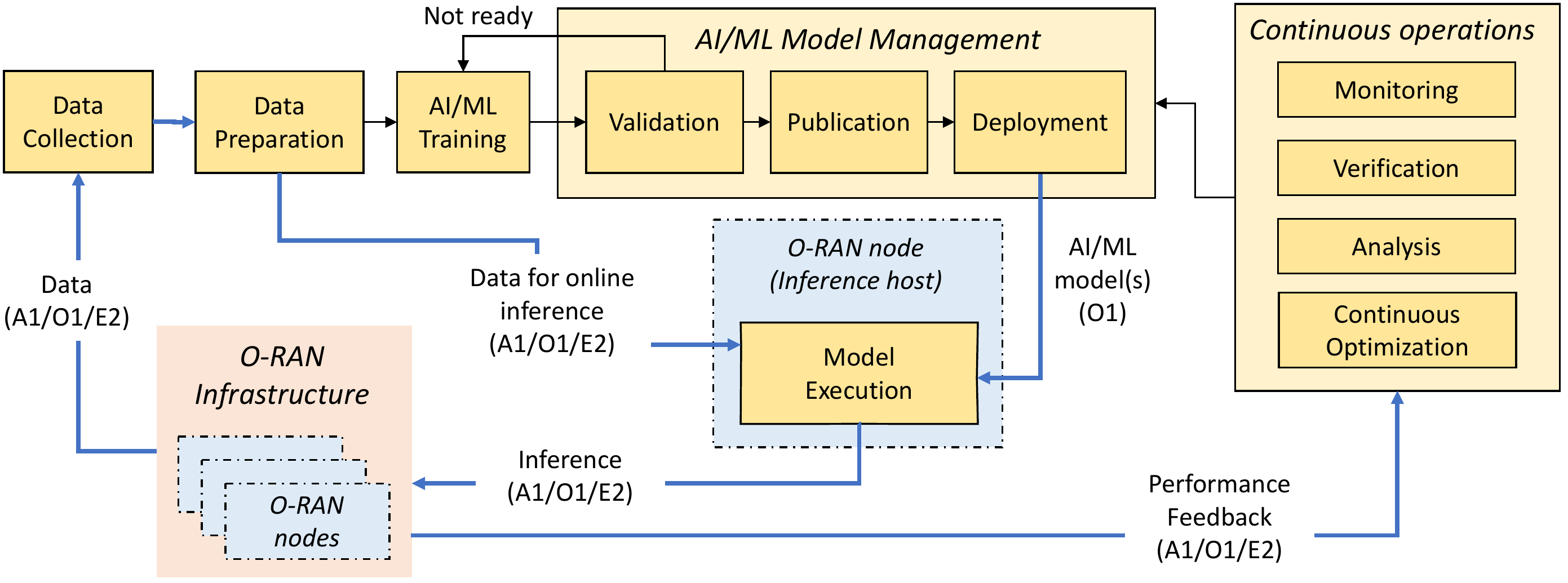}
    \caption{AI/ML workflow in the O-RAN architecture. The RAN infrastructure provides data through O-RAN interfaces to the data collection and preparation logical blocks. The AI/ML models are then trained, validated, and deployed, and execute on O-RAN nodes (e.g., the RICs). Models can be further refined through monitoring and continuous optimization based on performance feedback from the RAN.}
    \label{fig:ai}
\end{figure*}

\section{AI/ML Workflows}
\label{sec:ai-ml-workflow}

The goal of this section is to provide a detailed overview of the procedures and operational steps that regulate the AI/ML workflow in the O-RAN architecture. In the following, we provide a step-by-step guide on the life cycle of this workflow---from data collection to the actual deployment and execution of network intelligence in real time---that
relies on the architectural components described in Sections~\ref{sec:arch}, \ref{sec:xapps}, and~\ref{sec:orchestration}, and on the interfaces discussed in Section~\ref{sec:interfaces}.

The AI/ML workflow is being standardized by O-RAN WG2, with its specifications described in~\cite{oran-wg2-ml}. However, not all the procedures, features and functionalities have been finalized yet, with some of them left for further studies.
This workflow is composed of six main steps, which we will describe next: (i)~data collection and processing, (ii)~training; (iii)~validation and publishing; (iv)~deployment; (v)~AI/ML execution and inference, and (vi)~continuous operations.

For the sake of illustration and clarity, in the following we will describe the procedures involved in the AI/ML life-cycle within O-RAN systems. We take as an example the case where an operator aims at developing, training and deploying an xApp that controls RAN slicing policies by adapting them in near real time according to current network load and traffic demand via AI-based algorithms. In this example, each base station hosts three slices, namely a \gls{urllc} slice for ultra-low latency services, an \gls{embb} slice for high-throughput traffic (e.g., video streaming and file transfer), and a \gls{mmtc} slice to handle traffic generated, for example, by small sensors and \gls{iot} devices. The goal of the xApp is to control RAN slicing policies by assigning the available \glspl{prb} to each slice so that the diverse performance requirements of each slice are satisfied.

\textbf{Data Collection and Processing.}
First, data is collected over the O1, A1 and E2 interfaces and stored in large datasets (e.g., \textit{data lake} centralized repositories~\cite{MILOSLAVSKAYA2016300}) where it can be extracted upon request. Since different AI/ML solutions might use different \gls{kpm} types collected over different time periods and with different granularity (e.g., throughput, latency, \gls{mcs}, \gls{cqi}, data demand, jitter, to name a few), the O-RAN specifications consider a preliminary data pre-processing (or preparation) step. In this step, data
for both training and online inference is shaped and formatted according to the input size of the specific AI/ML model being considered~\cite{garcia2015data}. \textit{Example:} with respect to the xApp controlling RAN slicing policies, this step involves collecting data and performance metrics over the O1 interface to generate a training dataset to be used in the next step (i.e., training phase). For example, since the xApp must be able to adapt RAN slicing policies for the different slices according to current data demand and required minimum performance levels, the collected data must include how many \glspl{prb} are necessary to transmit the data requested by each user of the three slices, as well as throughput (\gls{embb}), number of transmitted packets (\gls{mmtc}) and latency (\gls{urllc}) measurements. 
In this, data processing might include the use of autoencoders for dimensionality reduction~\cite{ng2011sparse,bonati2021intelligence}, as well as well-established AI data processing procedures such as normalization, scaling and reshaping.

\textbf{Training.} The O-RAN specifications do not allow the deployment of any untrained data-driven solution~\cite{oran-wg2-ml}. All the AI/ML models are required to go through an offline training phase to ensure the reliability of the intelligence and avoid inaccurate predictions, classifications and/or actions that might result in outages or inefficiencies in the network~\cite{levine2020offline,pmlr-v119-agarwal20c,nair2020accelerating}. However, this does not preclude online training, which is still supported by O-RAN provided that it is only used to fine-tune and update a model previously trained offline~\cite{oran-wg2-ml,kading2016fine}. \textit{Example:} In our example, the operator can train a variety of AI algorithms all controlling the number of \glspl{prb} allocated to each slice but differing one from another with respect to their implementation details. For example, the operator can train a set of \gls{drl} agents and decision trees and explore different combination and input formats (e.g., the specific subset of \glspl{kpm} and their amount), different architectures (e.g., depth and width of a \gls{drl} agent, number of neurons, among others). The goal of this procedure is to train a large number of AI algorithms and identify which ones are the most suitable to accomplish a specific task.

\textbf{Validation and Publishing.} Once models are trained, they go through a validation phase to make sure they are reliable, robust and effective in performing classification, prediction or control tasks. If the validation is successful---and the models are deemed ready for deployment---they are published and stored in an AI/ML catalog on the \gls{smo}/non-RT \gls{ric}. Otherwise, they are required to go through additional re-design and re-training phases until the validation tests are successful~\cite{zappone2019wireless}. \textit{Example:} Once training has been completed, the different AI algorithms are compared one another and against diverse validation datasets including previously unseen data to identify which models are the most effective in controlling RAN slicing policies. For example, a typical validation test includes evaluating how well diverse AI solutions perform under diverse traffic patterns and demand, number and distribution of users, available bandwidth and operational frequencies. This procedure can either point out AI solutions that are not performing well and need to be retrained, as well as determine the subset of AI algorithms that can be published to the AI/ML catalog as well as provide side information on the ideal network conditions (e.g., network load, mobility pattern, size of deployment) under which the specific AI solution delivers the best performance so that the operator can deploy the AI solution that is best suited to a specific deployment.

\textbf{Deployment.} Models stored in the AI/ML catalog can be downloaded, deployed and executed following two different options, namely \textit{image-based} and \textit{file-based deployments}. In both cases, the deployment of the model is performed by using the O1 interface, and the node where the model executes is referred to as \textit{inference host}.

In the image-based deployment, the AI/ML model executes as a containerized image in the form of an O-RAN application (e.g., xApps and rApps) deployed at the O-RAN nodes, where it is executed to perform online inference.
At the time of this writing, these nodes are limited to the \glspl{ric} and the execution of \gls{ai} at the \glspl{cu}/\glspl{du} is left for further studies.
The file-based deployment, instead, considers the case where the AI/ML model is downloaded as a standalone file that executes within an inference environment---outside the O-RAN application domain---that forwards the inference output of the model to one or more O-RAN applications.
\textit{Example:} In our case, the operator will select the pre-trained AI-based RAN slicing models from the AI/ML catalog and deploy them as xApps that will be executed in the near-RT \gls{ric}.

\textbf{AI/ML Execution and Inference.} Once models are deployed on the inference host, they are fed with data to perform diverse online inference tasks. 
These include classification and prediction tasks, deriving policies at both \glspl{ric} (transmitted over the A1 and E2 interfaces), and taking management and control actions (over the O1 and E2 interfaces, respectively). \textit{Example:} Once the xApp has been deployed on the near-RT RIC, RAN slicing control is performed by executing the operations  described in Fig.~\ref{fig:e2-insert-control} where the xApp (i) is fed with \glspl{kpm} (e.g., requested \glspl{prb}, latency and throughput measurements) collected over the E2 interface; and (ii) computes control actions that are used to pilot the \gls{du} and assign the available \glspl{prb} to the different slices in near-RT.

\textbf{Continuous Operations.} An important aspect of the AI/ML workflow is the ability to monitor and analyze the intelligence deployed throughout the network to verify that the inference outputs of AI/ML models are effective, accurate and do not negatively affect the performance of the network. Continuous operations ensure that models that perform poorly online can be refined and re-trained to improve their functionalities~\cite{ebert2016devops,makinen2021who,li2021rlops}. \textit{Example:}  In our case, the operator can monitor constantly the performance of the RAN slicing xApp and, if any anomalies or inefficiencies are detected, it can decide to re-train the AI/ML model embedded in the xApp over new data collected through the O1 and E2 interfaces.

\begin{figure}[t]
    \centering
    \includegraphics[width=\columnwidth]{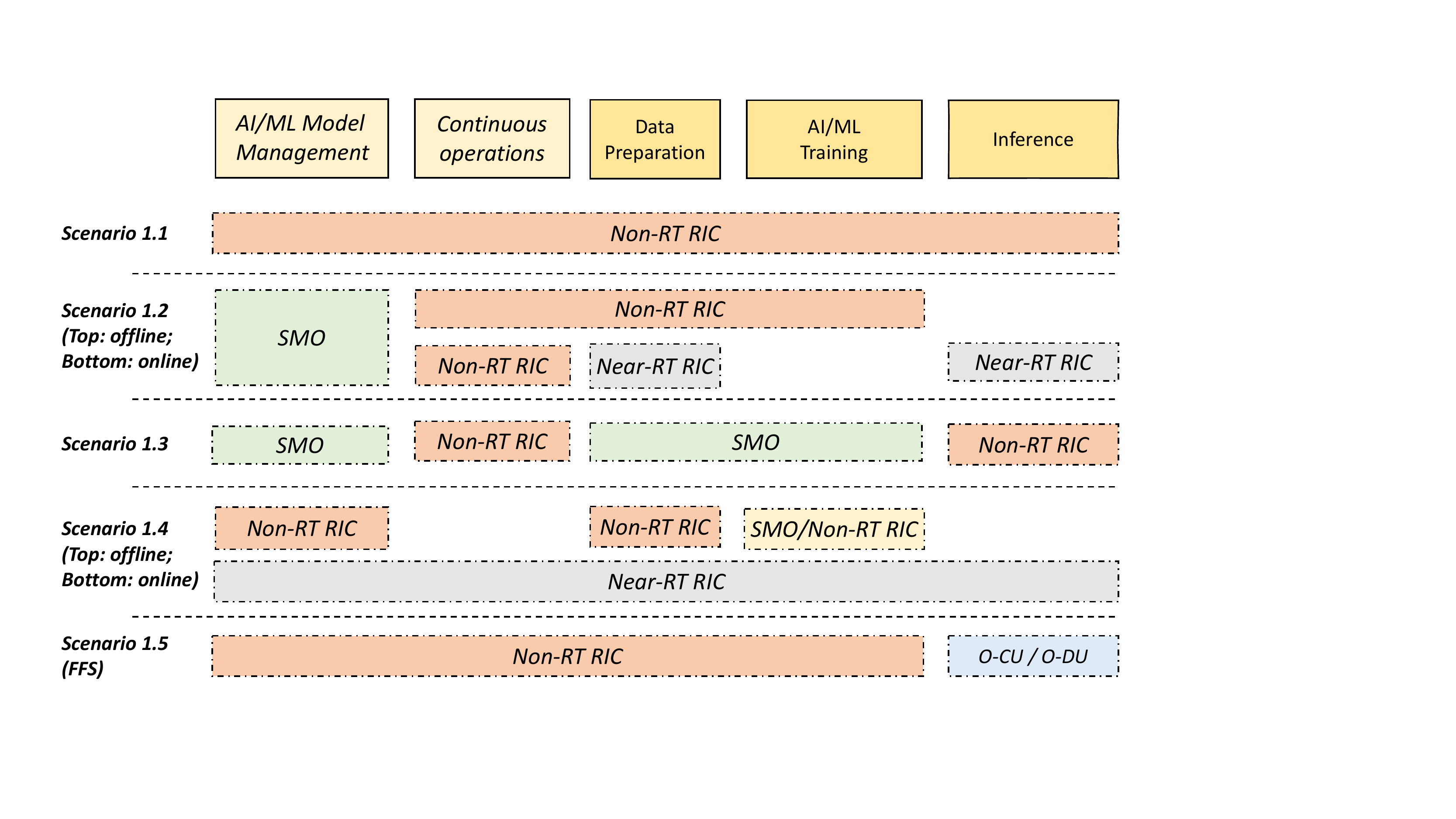}
    \caption{O-RAN AI/ML deployment scenarios, adapted from~\cite{oran-wg2-ml}. Different scenarios embed different components of the O-RAN AI/ML workflows in different components of the RAN, from the SMO/non-RT RIC to the near-RT RIC or the RAN nodes.}
    \label{fig:ai_deployments}
\end{figure}

\subsection{AI/ML deployment scenarios}

One of the main features of 5G networks is the ability to support a large variety of use-cases and applications. Given the diversity of the 5G ecosystem, it is clear that a one-size-fits-all solution in deploying and controlling network intelligence is unlikely to exist. For this reason, the O-RAN Alliance has specified five different deployment scenarios that define the location where the different components of the AI/ML workflow of Figure~\ref{fig:ai} are instantiated and executed~\cite{oran-wg2-ml}.
%
%
Although these deployment scenarios, which are shown in Figure~\ref{fig:ai_deployments}, cover a large set of real-world use cases, practical deployments might deviate from them to accommodate operator- and application-specific requirements.\footnote{Although the O-RAN specifications consider the case in which \gls{ai} is executed at the \glspl{cu}/\glspl{du} (Scenario 1.5 in Figure~\ref{fig:ai_deployments}), in practice this is left for further studies.}

As mentioned before, O-RAN specifications are specifically designed for the \gls{ran} portion of the network and its functionalities. However, it is worth mentioning that O-RAN and its \glspl{ric} can influence decisions regarding the core network and the \gls{mec} infrastructure. Indeed, the \gls{smo} can act as a gateway between the \gls{ran} and other network components, as it has the capability of orchestrating functionalities across the whole network. In this way, xApps and rApps executing within the O-RAN environment can be leverage to gather information on the \gls{ran} (e.g., traffic load forecast, mobility prediction, network state dynamics) that can be used by the \gls{smo} (e.g., ONAP or OSM) to take informed decisions about \gls{mec} service instantiation and delivery as well as network slicing policies in the core network~\cite{doro2020sledge}.

\subsection{Gathering inputs for online inference}

Data for online inference can be collected from multiple data producers over the O1, A1 and E2 interfaces, which are designed to support O-RAN control loops operating at different time scales.
The O1 interface allows components in the \gls{smo}/non-RT \gls{ric} domain to gather data from any O-RAN management component and perform non-real-time optimization.
The A1 interface can be used by the non-RT \gls{ric} to send enrichment information from the \gls{smo}/non-RT \gls{ric} domain to the near-RT \gls{ric} and its applications. For example, an rApp in the non-RT \gls{ric} can send enrichment information to the near-RT \gls{ric} on the predicted KPI evolution over the next few seconds.
Finally, the E2 interface allows the near-RT \gls{ric} and its xApps to collect data from E2 nodes (e.g., \glspl{kpm}) for near-RT control of the RAN. 
\begin{figure}[t]
    \centering
    \includegraphics[width=\columnwidth]{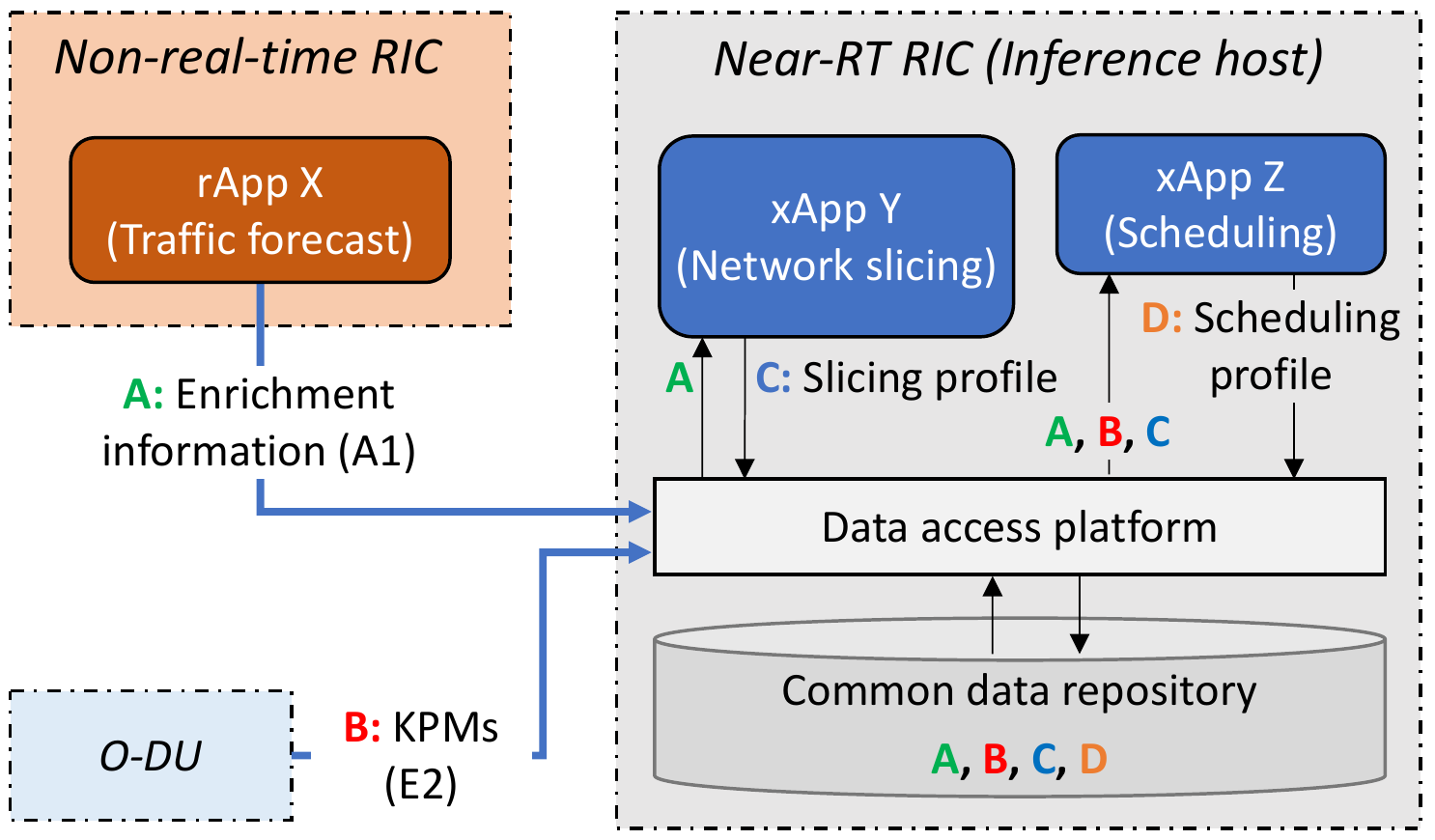}
    \caption{An example of chained AI/ML models with diverse input types and data producers. Letter \emph{A} indicates the traffic forecast generated by rApp X. Letter \emph{B} represents the \glspl{kpm} from the \gls{ran}. xApp Y uses data \emph{A} to generate control action \emph{C} (i.e., a slicing profile). xApp Z, instead, using data \emph{A}, \emph{B} and control \emph{C} as input to generated a new control action \emph{D} (i.e., a scheduling profile).}
    \label{fig:ai_chaining}
\end{figure}
It is worth mentioning that the O-RAN specifications consider the case where input data might be generated within the same inference host by different data producers~\cite{oran-wg2-ml}. This is the case of chained AI/ML models, where one control task consists of the execution of several sub-tasks, each involving a different AI/ML model and requiring diverse input data and types.

To regulate data production and consumption between applications hosted in the same node (e.g., the near-RT \gls{ric}), the O-RAN specifications lay the basis for a \textit{data access platform} that regulates data production, sharing and access. This platform acts as a middleware between the applications and a common data repository where data is stored and shared.
To provide a high-level example of this, in Figure~\ref{fig:ai_chaining} we show how data produced by different sources can be consumed by O-RAN applications performing heterogeneous tasks. 
We consider the case of an rApp~X forecasting traffic load (data type~A). Two xApps (xApp~Y and xApp~Z) leverage \gls{ai} to control network slicing and scheduling strategies, respectively. xApp~Y consumes data type~A received from the data access platform and produces a slicing profile (data type~C), while xApp~Z consumes data types~A, B and~C coming from the data access platform (with type~B consisting of \glspl{kpm} sent from a \gls{du} over the E2 interface) and selects a scheduling profile D to be used by the controlled \gls{du}.

\section{Open RAN Use Cases and Research}
\label{sec:results}

The capabilities introduced by the \glspl{ric}, the open interfaces, and the AI/ML workflow described in Sections~\ref{sec:xapps}-\ref{sec:ai-ml-workflow} make it possible to support advanced use cases and scenarios for the RAN control and deployment self-optimization.  

Recent years have seen an increase of O-RAN-driven research on applications and use cases, e.g., on the design of xApps and rApps or on the optimal configuration of O-RAN networks. 
This section describes the areas where the application of the Open RAN paradigm is the most promising, and recent research results that show how O-RAN-based solutions can be used to optimize the RAN. 

The O-RAN Alliance has collected an extensive list of 19 exemplary use cases for Open RAN deployments in~\cite{oran-wg1-use-cases,oran-wg1-use-cases-analysis}, and the literature further discusses at a high level some of these in~\cite{rimedo2021wp,nolle2021wp,ntt2021wp,comcores2020wp,dell2021wp,deloitte2021oran,viavi2021wp,altiostar2021wp,ericsson2021securitywp,security2022,xran2021story,openranwp-ria,openranwp,lee2020hosting,abdalla2021generation,garciasaavedra2021oran,brik2022deep}. At a high level, the scenarios and use cases can be classified in different ways, e.g., by considering the control knob or inference target, or the domain that is being controlled or optimized (e.g., a UE, a slice, a RAN node, or the whole network).
In terms of control knobs, we discussed E2-specific targets in Section~\ref{sec:e2}. More generally, it is possible to identify several areas of interest, as follows.

\textbf{Mobility.} Open RAN networks can influence the mobility management or the performance of mobile users by tuning handover, load balancing, multi-connectivity, access barring, and beamforming parameters in the \gls{ran}. Differently than in traditional 3GPP networks, this can be done in a closed-loop fashion by exploiting knowledge on the state of multiple base stations, and predictions on the user mobility based on information from the RAN or external enrichment information. For example,~\cite{polese2018machine} presents an Open-RAN-based mechanism for the prediction of the load of multiple base stations in a cellular network, with application to the dynamic routing of autonomous vehicles to avoid introducing congestion. The O-RAN documents~\cite{oran-wg1-use-cases,oran-wg1-use-cases-analysis} also include context-based handover for vehicular scenarios, in which xApps exploiting enrichment information from the non-RT RIC and inference on AI/ML RAN data to manage handovers, and applications related to \glspl{uav}, with configuration of RAN parameters based on the expected trajectory of the UAV. 

\textbf{Resource allocation.} Control in this area spans network slicing, scheduling, and provisioning of services and network functions. As for mobility control, the advantages of Open RAN compared to traditional cellular networks lie in the possibility of adapting to dynamic, evolving contexts, to new user requirements (e.g., for different slices), and to external events that alter the state and configurations of the RAN. 

In this sense, the research on intelligent O-RAN control has adopted network slicing as one of the most interesting and promising areas for ML-based optimization. Bonati et al.\ analyze data-driven approaches in O-RAN, and provide the first demonstration of closed-loop control of a softwarized cellular network instantiated on a large-scale experimental platform~\cite{bonati2021intelligence}. The performance of the \gls{ran}---implemented on the Colosseum testbed (see Section~\ref{sec:testbeds})---is optimized through xApps that control the scheduling policies of various network slices on the base stations. The slices have different optimization targets, e.g., for the URLLC slice the reward minimizes the buffer occupancy as a proxy for the end-to-end latency. 
Polese et al.\ propose ColO-RAN, a pipeline for the design, training, testing and experimental evaluation of \gls{drl}-based control loops in O-RAN~\cite{polese2021coloran}. The capabilities of ColO-RAN---prototyped on the Colosseum and Arena testbeds---are showcased through xApps to control the \gls{ran} slicing allocation and scheduling policies, and for the online training of \gls{ml} models.
Johnson et al.\ propose NexRAN, an xApp to perform the closed-loop control of the slicing resources of softwarized base stations, and demonstrate it on the POWDER platform of the U.S.\ PAWR program~\cite{johnson2021nextran,johnson2021open}.
Sarikaya and Onur consider the placement of \gls{ran} slices in a multi-tier 5G Open \gls{ran} and formulate it mathematically, showing the benefits of flexible functional splits compared to fixed split options~\cite{sarikaya2021placement}. Niknam et al.\ analyze the principles and requirements of the O-RAN specifications, and propose an intelligent scheme for the management of radio resources and traffic congestion. The effectiveness of this solution has been proved on real-world operator data~\cite{niknam2020intelligent}.
Similarly, Mungari assesses the performance of \gls{ml}-driven radio resource management in O-RAN-managed networks through an xApp deployed on the near-RT \gls{ric}, and evaluates it in a small laboratory setup~\cite{mungari2021rl}. The authors of~\cite{coronado2022roadrunner} focus on the cell selection process, showing how O-RAN can help improve the allocation of users to specific cells based on forecasted throughput metrics rather than simply signal level metrics. 

Lien et al.~\cite{lien2021session} and Filali et al.~\cite{filali2022communication} explicitly consider xApp-based optimization of radio resources for URLLC users. In particular, the first paper shows how a reinforcement learning agent running in the near-RT RIC can effectively control the instantiation of new URLLC guaranteed bitrate sessions and configure the session-level parameters to increase the probability of successfully onboarding new URLLC users~\cite{lien2021session}. The paper by Filali et al., instead, studies an O-RAN-based slicing mechanism that controls the resource allocation for URLLC users and manages to achieve quality of service targets~\cite{filali2022communication}.

Additional examples can be found in~\cite{oran-wg1-use-cases,oran-wg1-use-cases-analysis}, e.g., resource allocation for mobile users (including UAVs) with anticipatory mobility prediction, and QoS-based resource allocation. The latter aims at dynamically provisioning the set of resources that selected users require to satisfy their \gls{qos}, for example through ad hoc slicing and subsequent allocation of \glspl{prb} to the \gls{qos}-driven slices. Finally, the O-RAN infrastructure can also be used to predict the emergence of congestion and apply appropriate remediations by adding more resources (e.g., carriers, MIMO layers, cells).

\textbf{QoE/QoS-based control.} The optimization of \gls{ran} resources to meet specific \gls{qos} and \gls{qoe} requirements also extends beyond resource allocation. For example, Bertizzolo et al.\ consider drone-enabled video streaming applications and propose a control system for the Open \gls{ran} for the joint optimization of transmission directionality and the location of the drone~\cite{bertizzolo2021streaming}. This solution is evaluated both experimentally, in a multi-cell outdoor \gls{ran} testbed, and through numerical simulations.

The O-RAN Alliance use cases also include a \gls{qoe} optimization scenario, where inputs from external systems, monitoring of application performance, and multi-dimensional data are combined to identify the best \gls{ran} configuration for the optimization of the \gls{qoe} on a user basis~\cite{oran-wg1-use-cases}. Here, O-RAN interfaces play a unique role, as they make it possible to combine and fuse data input from different, heterogeneous sources (RAN, external) in a way that is not usually possible in closed, traditional \gls{ran} deployments. 

\textbf{RAN sharing.} This use case covers different scenarios, from spectrum sharing to infrastructure sharing on a neutral host architecture, which are expected to increase the spectral and energy efficiencies, and to reduce operational and capital expenditures. This scenario combines the flexibility of O-RAN softwarization and virtualization with the dynamicity and reconfigurability of closed-loop control (including with external information). The O-RAN specifications include multiple mechanisms for spectrum sharing, including slicing and control of dynamic spectrum access at the \gls{du}/\gls{ru}.

In this regard,~\cite{smith2021oran} studies a sharing scenario between a 5G \gls{ran} and a low-Earth orbit satellite constellation (in particular, the uplink between a ground station and a satellite), managed by a \gls{ric}. The study includes input from a RAN-independent spectrum-sensing framework and a sharing mechanism applied on a UE-basis by the RIC. Paper~\cite{baldesi2022charm} also proposes an inter-technology spectrum sharing solution enabled by O-RAN, but, in this case, the sensing is performed inside the \gls{du} and \gls{ru} themselves, with I/Q-based deep learning analysis of the received signals. The proposed sharing scheme detects Wi-Fi users (or unknown users occupying the spectrum of interest) and adapts the configuration of the \gls{ran} to avoid interference.~\cite{kulacz2022dynamic} proposes a framework based on the aggregation of contextual information from multiple sources to create spectrum maps that are then used by xApps for dynamic spectrum access control. An outlook to spectrum sharing enabled by O-RAN controllers in future 6G applications is provided in~\cite{polese2022dynamic}, where a backhaul link carrier frequency is changed by a centralized controller based on external information on incumbents that may suffer from interference from the communication links.

Blockchain-based approaches are also considered in~\cite{giupponi2021blockchain,enayati2022blockchain}. Notably,~\cite{giupponi2021blockchain} embeds a blockchain framework on top of the O-RAN infrastructure to enable secure and trusted exchange of RAN resources (as for example \glspl{du}, \glspl{ru}, etc.) among multiple operators.~\cite{enayati2022blockchain} combines Open-RAN-enabled neutral infrastructure and dynamic allocation based on blockchain as a practical way to bridge the digital divide gap. Finally,~\cite{suzuki2022implementation} explores the usage of smart contracts to activate and deactivate carrier aggregation across different operators on a shared O-RAN infrastructure.

\textbf{Massive MIMO.}
This technology represents a key enabler of 5G networks. Through an Open RAN architecture, it is possible to embed dynamic control and adaptability to the configuration of the \gls{mimo} codebook (or group of beams\footnote{The group of beam is a set of beams on which the base stations transmits synchronization signals for initial access, to improve the initial access performance especially at higher frequencies~\cite{oran-wg1-use-cases-analysis,giordani2018tutorial}.}) or of the beam selection process, and to make mobile experience more reliable and robust. 

From the signal processing point of view, there is an extensive body on research that benefits from \gls{cran} and virtualized, centralized \glspl{cu} and \glspl{du}~\cite{checko2015cloud,parsaeefard2017dynamic}. Paper~\cite{hewavithana2022overcoming} studies how channel state information available at the \gls{ru} or \gls{du} may need to be shared across the two nodes. The authors consider specifically the capabilities provided by the O-RAN FH. In this context,~\cite{lagen2021modulation,lagen2021modulationtmc} discuss different compressions schemes for the O-RAN FH interface, also considering multi-stream capabilities typical of \gls{mimo} setups. Finally, the authors of~\cite{electronics10172162} analyze different beamforming options based on the O-RAN 7.2x split of the physical layer.

When it comes to the \glspl{ric} introduced by O-RAN, the control and optimization is generally related to beam parameters and to the codebook or group of beams in the \gls{du} and \gls{ru}. For example, the O-RAN technical document on use cases investigates two different data-driven solutions~\cite{oran-wg1-use-cases}. The first adapts the group of beams based on telemetry collected at the non-RT RIC, e.g., user activity and reports, measurement reports, GPS coordinates, and reconfiguration through the O1 and O-RAN FH interfaces. The second is an optimization on the near-RT RIC of the mobility configuration, e.g., the beam-specific offsets that will determine whether a user should change beam or not. Another scenario of interest is the grouping of the users into multi- and single-user \gls{mimo} groups, which would then benefit from capacity enhancement or diversity. This can be done through policy guidance and enrichment information from the non-RT \gls{ric}, and with the actual control being performed by the near-RT \gls{ric}.
The dynamic, data-driven reconfiguration of beamforming with the \glspl{ric} is relatively unexplored in the Open RAN literature. 

\textbf{Security.} We will discuss security in details in Section~\ref{sec:security}.

\textbf{New applications.} The next generations of wireless cellular networks also embed and expand to other use cases. An example is the support for \glspl{uav} and vehicular communications, which we discussed as part of the mobility and resource allocation use cases. 

Another example is represented by industrial \gls{iot} scenarios, which require high reliability and precise timing and synchronization achieved with data duplication, multi-connectivity, dedicated \gls{qos} and packet compression techniques~\cite{godor2020look}. Considering the number of parameters that can be tuned in this context, the closed-loop control enabled by the O-RAN \glspl{ric} can provide elastic configurations that adapt to the evolving conditions on the factory floor.

Moving away from communication scenarios, 5G networks from Release 16 also support positioning through dedicated signals on the air interfaces (both LTE and NR) and a location management function in the core network~\cite{keating2019overview}. Location services will be used, for example, in indoor scenarios (e.g., factories, malls), to provide value-added services, or to implement location-based safety information broadcasting. However, relaying the location information to the core for analysis and processing may be affected by delays or jitter, making precise and timely estimation more difficult. In this case, the O-RAN specifications~\cite{oran-wg1-use-cases} see the near-RT \gls{ric} with a dedicated positioning xApp as an alternative to the 5G core location management function, leveraging the deployment of the near-RT \gls{ric} at the edge of the network.












\textbf{O-RAN deployments optimization.} Besides optimization of the \gls{ran} through O-RAN, several papers also study how to optimize the deployment of the Open RAN components themselves (e.g., \glspl{ric}, xApps and rApps, \gls{ran} nodes). This leverages the management functionalities of the \gls{smo}, as well as its centralized point of view and the telemetry and statics from the \gls{ran}.

D'Oro et al.\ design a zero-touch orchestration framework that optimizes network intelligent placement in O-RAN-managed networks, and prototype it at scale on Colosseum using open source \gls{ric} and \gls{ran} components~\cite{doro2022orchestran}. In this context, the paper~\cite{li2021rlops} introduces the concept of RLOps, or reinforcement learning operations, i.e., a framework to manage the life-cycle of intelligence (specifically, reinforcement learning) for Open RAN. RLOps cover the whole end-to-end workflow of AI/ML for the \gls{ran} (Section~\ref{sec:orchestration}), from the design and development to the operations (i.e., deployment, updates), and the management of safety and security during the overall intelligence life-cycle.
Huff et al., instead, develop a library, namely RFT, to make xApps fault-tolerant while preserving high scalability~\cite{huff2021rft}. This is achieved through techniques such as state partitioning, partial replication and fast re-route with role awareness.

%
Other papers focus on the virtualized components and of the disaggregated base stations.
Tamim et al.\ maximize the network availability by proposing deployment strategies for the virtualized O-RAN units in the O-Cloud~\cite{tamim2021downtime}.
Pamuklu et al.\ propose a function split technique for green Open \glspl{ran}~\cite{pamuklu2021reinforcement}. The proposed solution, which is based on \gls{drl}, is evaluated on a real-world dataset.
The authors of \cite{WANG2022108682} develop a matching scheme between \glspl{du} and \glspl{ru} with a 2D bin packing problem. Finally, the O-RAN specifications consider a similar use case, with data-driven pooling of the \glspl{cu} and \glspl{du} on shared, virtualized resources~\cite{oran-wg1-use-cases}, orchestrated through the non-RT \gls{ric}.


\textbf{O-RAN white papers and surveys.} Finally, overviews of O-RAN and of its components are given by Lee et al.\ in~\cite{lee2020hosting}, which implements \gls{ai}/\gls{ml} workflows through open-source software frameworks; by Abdalla et al.\ in~\cite{abdalla2021generation}, which reviews O-RAN capabilities and shortcomings; by Garcia-Saavedra and Costa-Perez in~\cite{garciasaavedra2021oran}, which gives a succinct overview of O-RAN building blocks, interfaces and services; and by Brik et al.\ in~\cite{brik2022deep} and Arnaz et al.\ in~\cite{arnaz2022toward}, which discuss deep learning and artificial intelligence applications for the Open RAN.~\cite{bonati2020open} discusses open source software that can be used to deploy 5G and Open RAN networks. Several white papers~\cite{rimedo2021wp,nolle2021wp,ntt2021wp,comcores2020wp,dell2021wp,deloitte2021oran,viavi2021wp,altiostar2021wp,ericsson2021securitywp,security2022,xran2021story,openranwp-ria,openranwp} provide high-level overviews of the Open RAN vision and of the O-RAN architecture.
Differently from the above works, in this paper, we present a comprehensive overview of the O-RAN specifications, deep-diving into the standardized interfaces, protocols and services, and discussing in detail use cases, \gls{ai}/\gls{ml} workflows, deployment scenarios, and open platforms for O-RAN-enabled experimental research.

\section{Security in the Open RAN}
\label{sec:security}

It is undeniable that the introduction of new architectural components, open interfaces, disaggregation, and the integration of custom and possibly data-driven control logic will make next-generation cellular networks more efficient and flexible. On the other hand, however, this revolution comes with unprecedented security challenges that primarily stem from the fact that the distributed and disaggregated nature of the O-RAN infrastructure effectively extends the attack surface for malicious users, thus posing severe threats to the network and its users~\cite{ericsson2021securitywp}. 
At the same time, the unparalleled monitoring capabilities, the intelligence and the cloud-native deployment that characterize O-RAN architectures truly add insights on the state of the network and provide the necessary tools to implement advanced solutions to monitor, detect, prevent and counteract threats~\cite{altiostar2021wp}. In this regard, the O-RAN Alliance has created a dedicated working group for the analysis and definition of threat models for O-RAN networks, as well as for the definition of security measures and policies for the components of the O-RAN architecture toward a zero-trust model~\cite{oran-sfg-threat,oran-sfg-protocols,oran-sfg-requirements}. 

\subsection{Security Stakeholders}

The O-RAN \gls{sfg} (recently promoted to a full \gls{wg}, i.e., WG11) has defined a list of stakeholders that need to proactively secure the \gls{ran}. This extends the interested parties beyond those considered in traditional 4G and 5G networks, e.g., vendors, operators, and system integrators. As also discussed in~\cite{altiostar2021wp}, operators will assume a more predominant role in securing the infrastructure, as the openness of the platform and the usage of multivendor components allows them to customize the build (and thus the security) of the infrastructure. This also means that operators can assess and vet the security standard of the open components introduced in the network, which is often not possible in close architectures that are fully vendor-driven.~\cite{altiostar2021wp} also identifies network functions and virtualization platform vendors as new stakeholders (e.g., third-party xApp and rApps developers, O-Cloud providers), along with administrator profiles that manage virtualized and disaggregated components. In addition, the orchestrator (e.g., the entity that manages the \gls{smo}) also has a role in securing the operations of the network.

\subsection{Extended Threat Surface} 

According to the threat analysis in~\cite{oran-sfg-threat}, the O-RAN Alliance plans to complement the \gls{3gpp} security requirements and specifications~\cite{3gpp.33.117,3gpp.33.501,3gpp.33.511,3gpp.33.818,3gpp.33.848} to address security issues specific to the O-RAN architecture. \gls{sfg} has identified seven threat categories:

\begin{itemize}
	\item \textbf{Threats against the O-RAN system.} The Open RAN architecture introduces new architectural elements and interfaces (from the fronthaul to control and management interfaces), which become part of an extended threat surface. These components can be subject to different attacks, which may compromise (i) the availability of the infrastructure (e.g., unauthorized access to disaggregated \gls{ran} components aiming at deteriorating the performance of the network, or the malicious deployment of xApps that intentionally introduce conflicts with other xApps); (ii) data and infrastructure integrity (e.g., compromised software trust chains or the misconfiguration of interfaces), and (iii) data confidentiality (e.g., through attacks that disable over-the-air encryption, or facilitate unregulated access of user data from xApps and rApps). 

	Data-related threats encompass (i) information transported over O-RAN interfaces for control, management, and configuration of the \gls{ran}; (ii) data used for training and testing of the \gls{ml} models; (iii) sensitive data on the users, e.g., their identities; and (iv) the cryptographic keys deployed on the elements of the network. The threat surface also expands to the new logical components of the architecture, i.e., the \glspl{ric}, the \gls{smo}, and the software frameworks (e.g., xApps, rApps) that execute on the \glspl{ric}.
	
	Finally, attacks on the Open Fronthaul 7.2x split interface are also often mentioned as a potential vulnerability~\cite{ericsson2021securitywp}. As of today, the 7.2x interface is not encrypted on the control plane, because of the challenging timing requirements that encryption would introduce. This introduces man-in-the-middle attacks, in which the attacker impersonates the \gls{du} (or \gls{ru}), and compromises user data or configurations in either of the two endpoints. Another attack can be carried out against the S-Plane, with a malicious actor compromising the synchronization infrastructure and thus causing performance degradation.

	\item \textbf{Threats against the O-Cloud.} The O-Cloud provides a virtualization environment that encompasses RAN elements as well as O-RAN components. To this end, threats identified for the O-Cloud relate to attacks in virtualized environments. Possible attacks include (i) compromising virtual network functions, either being executed or their snapshots or images (e.g., to leak embedded cryptographic secrets); (ii) exploitation of the O2 interface between the O-Cloud and the SMO to gain access and escape isolation; (iii) misuse of containers or virtual machines for network functions to attack other entities in the network; and (iv) spoofing or compromising the underlying networking or auxiliary services.

	\item \textbf{Threats against open-source code.} The softwarization of the RAN and of O-RAN components opens new vulnerabilities because of backdoors in the O-RAN code by (i) trusted developers, which intentionally compromise software components, or by (ii) upstream libraries that are not in control of the O-RAN developers.

	\item \textbf{Physical threats.} The additional hardware introduced to support the \gls{gnb} split and the O-RAN infrastructure can be compromised by attackers that gain physical access to the infrastructure. The attacks can range from power availability attacks, to cabling reconfigurations, or addition of hardware backdoors.

	\item \textbf{Threats against the wireless functionalities.} Attacks on the \gls{ru} or the O-RAN fronthaul interface between \glspl{ru} and \glspl{du} can lead to performance degradation on the air interface, with typical attacks related to jamming of data or synchronization signals.

	\item \textbf{Threats against the AI/ML components.} Finally, the O-RAN specification~\cite{oran-sfg-threat} and the literature~\cite{altiostar2021wp,ericsson2021securitywp} also describe a new class of threats, i.e., attacks against \gls{ai}/\gls{ml} models used for inference and control in xApps and rApps. A practical instances of such an attack is that of poisoning attacks, in which an attacker exploits unregulated access to the data stored in the \gls{smo}/non-RT \gls{ric} to inject altered and misleading data into the datasets used for offline training of \gls{ai}/\gls{ml} algorithms. Another example is that of an adversary gaining unrestricted control over one or more O-RAN nodes to produce synthetic data fed in real time to \gls{ai}/\gls{ml} solutions being fine-tuned online, or being used to perform online inference. These attacks are extremely relevant as they might result in \gls{ai}/\gls{ml} solutions that output wrong predictions, or make wrong control decisions that result in performance degradation or---even worse---outages. Similar attacks can also target the \gls{ml} model directly (e.g., by modifying the weights or configurations of the model)~\cite{mcgraw2019security}.

\end{itemize}
Based on this security analysis, the O-RAN \gls{sfg} has identified 30 O-RAN-specific critical assets related to interfaces and data, and 12 related to logical components, also partially discussed in~\cite{altiostar2021wp,ericsson2021securitywp,mimran2022evaluating}.

\subsection{O-RAN Security Principles and Opportunities}

While the new architecture and its interfaces introduce new threats and opportunities for attackers, they also come with the opportunity to re-think the security principles and best practices for designing, deploying and operating cellular networks and align them to the best practices of cloud-native deployments~\cite{altiostar2021wp}. In general, openness is associated with increased visibility into the processes and operations of the \gls{ran}, which puts operators in control of their network. The O-Cloud and the virtualized nature of the O-RAN platforms enable quick deployment of security patches and updates, automated testing and deployment, with full control over the entire end-to-end process including information on vendors and software components being used at any moment. The disaggregation also makes it possible to deploy simpler network functions, with more atomic components that are easier to test and profile. Finally, the virtualized \gls{cu} is generally deployed in a centralized data center, which makes it easier to physically secure the \gls{ran} cryptographic keys. 
In this sense, the O-RAN \gls{sfg} and WG11 have published a number of technical specifications that mandate authentication and encryption procedures across the different elements of the O-RAN architecture~\cite{oran-sfg-protocols,oran-sfg-requirements}. 

Finally, the availability of data and the insights on the \gls{ran} that the different interfaces (E2, O1) provide can also be leveraged to increase the security of the \gls{ran} itself. This is due to the intelligent, data-driven self-monitoring of the \gls{ran} performance, which can automatically trigger warnings and alarms in case unintended behaviors are detected~\cite{jin2019rogue,savic2021deep}. 

\section{O-RAN Standardization}
\label{sec:standardization}

The O-RAN Alliance is a consortium of operators, vendors, research institutions, and industry partners that focuses on reshaping the \gls{ran} ecosystem toward an intelligent, open, virtualized and interoperable architecture.
To this aim, the efforts of the Alliance cover three macro-areas~\cite{oran2018building}: (i)~\textit{specification}, aimed at extending \gls{ran} standards to include openness and intelligence; (ii)~\textit{software development}, focused on developing and contributing open source software for the \gls{ran} components of the O-RAN architecture, and (iii)~\textit{testing and integration}, in which it provides guidance to members of the Alliance willing to perform testing and integration of the O-RAN-compliant solutions they develop.

\begin{table}[t]
    \centering
    \footnotesize
    \renewcommand{\arraystretch}{1.2}
    \setlength{\tabcolsep}{2pt}
    \caption{O-RAN working groups.}
    \label{tab:oran-wg}
    \begin{tabularx}{0.75\columnwidth}{
        >{\raggedright\arraybackslash\hsize=0.75\hsize}X 
        >{\raggedright\arraybackslash\hsize=1.25\hsize}X }
        \toprule
        \acrlong{wg} & Focus \\
        \midrule
        WG1 & Architecture and use cases \\
        WG2 & Non-RT RIC and A1 interface \\
        WG3 & Near-RT RIC and E2 interface \\
        WG4 & Open Fronthaul interfaces \\
        WG5 & Open F1/W1/E1/X2/Xn interfaces \\
        WG6 & Cloudification and orchestration \\
        WG7 & White-box hardware \\
        WG8 & Stack reference design \\
        WG9 & Open X-haul transport \\
        WG10 & OAM and O1 interface \\
        WG11 & Security\\
        \bottomrule
    \end{tabularx}
\end{table}

\begin{table}[t]
    \centering
    \footnotesize
    \renewcommand{\arraystretch}{1.2}
    \setlength{\tabcolsep}{2pt}
    \caption{O-RAN focus groups.}
    \label{tab:oran-fg}
    \begin{tabularx}{\columnwidth}{
        >{\raggedright\arraybackslash\hsize=0.6\hsize}X
        >{\raggedright\arraybackslash\hsize=1.95\hsize}X 
        >{\raggedright\arraybackslash\hsize=0.45\hsize}X }
        \toprule
        \acrlong{fg} & Focus & State \\
        \midrule
        OSFG & Open source issues, establishment of the \acrshort{osc} & Dormant \\
        SDFG & Standardization strategies & Active \\
        TIFG & Testing and integration, PlugFests & Active \\
        \bottomrule
    \end{tabularx}
\end{table}

The specification tasks of the O-RAN Alliance are divided among 10~\glspl{wg}, each responsible for specific parts of the O-RAN architecture. The main focus of each of these \glspl{wg} is summarized in Table~\ref{tab:oran-wg}:
\begin{itemize}
    \item \textit{WG1: Use Cases and Overall Architecture}. This \gls{wg} identifies key O-RAN use cases, deployment scenarios and development tasks of the overall O-RAN architecture. It includes three task groups: (i)~Architecture Task Group, focused on specifying the overall O-RAN architecture, on describing its functions and interfaces, on illustrating relevant implementation options, and on facilitating cross-\gls{wg} architectural discussions; (ii)~Network Slicing Task Group, focused on studying network slicing in O-RAN and on defining its use cases, requirements and extensions to the O-RAN interfaces, and (iii)~Use Case Task Group, focused on identifying, defining and disseminating use cases enabled by O-RAN. We discussed WG1 activities primarily in Sections~\ref{sec:arch} and~\ref{sec:results}.
    
    \item \textit{WG2: Non-RT \gls{ric} and A1 Interface}. This \gls{wg} specifies the architecture and functionalities of the non-RT \gls{ric} and of the A1 interface, which are discussed in Sections~\ref{sec:orchestration} and~\ref{sec:a1} of this paper.
    
    \item \textit{WG3: Near-RT \gls{ric} and E2 Interface}. This \gls{wg} specifies the architecture and functionalities of the near-RT \gls{ric} and of the E2 interface. It also takes care of providing support to \gls{ai}/\gls{ml} and data analytics model design to train models and enhance radio resource management and allocation. WG3 content is discussed in Sections~\ref{sec:xapps} and~\ref{sec:e2} of this paper.
    
    \item \textit{WG4: Open Fronthaul Interfaces}. This \gls{wg} focuses on defining an Open Fronthaul interface that supports interoperability of \glspl{du} and \glspl{ru} manufactured by different vendors, as discussed in Section~\ref{sec:fh}. 
    
    \item \textit{WG5: Open F1/W1/E1/X2/Xn Interface}. This \gls{wg} provides multi vendor profile specifications for the F1/W1/E1/X2/Xn interfaces that are compliant with the \gls{3gpp} specifications. These interfaces are briefly discussed in Sections~\ref{sec:arch} and~\ref{sec:interfaces} of this paper.
    
    \item \textit{WG6: Cloudification and Orchestration}. This \gls{wg} identifies use cases to demonstrate the benefits of the software/hardware decoupling of the O-RAN elements (e.g., \glspl{ric}, \gls{cu}, \gls{du}, \gls{ru}) and deployment scenarios, and develops requirements and reference designs for the cloud platform. This \gls{wg} also develops life-cycle flows and commonalities of O2 interface \glspl{api} between the \gls{smo} and the O-Cloud. This is primarily described in Section~\ref{sec:arch}.
    
    \item \textit{WG7: White-box Hardware}. This \gls{wg} specifies and releases a reference design toward a decoupled software/hardware platform, as discussed in Section~\ref{sec:arch}.
    
    \item \textit{WG8: Stack Reference Design}. This \gls{wg} develops software architecture, design, and plans for the \gls{cu} and \gls{du} compliant with the \gls{3gpp} NR specifications, and a set of tests that promote interoperability across different implementations of the O-RAN interfaces. We reviewed this in Section~\ref{sec:arch}.
    
    \item \textit{WG9: Open X-haul Transport}. This \gls{wg} focuses on the network transport, including transport equipment, physical media and protocols, as discussed in Section~\ref{sec:fh}.
    
    \item \textit{WG10: \acrshort{oam}}. This \gls{wg} focuses on the O1 interface \gls{oam} specifications (e.g., unified O1 operation and notification) and on creating \gls{oam} architecture and requirements for the O-RAN architecture and use cases identified by WG1. O1 and related topics are presented in Section~\ref{sec:o1}.

    \item \textit{WG11: \acrfull{swg}}. This \gls{wg} focuses on the security aspects of the O-RAN ecosystem, as discussed in Section~\ref{sec:security}.

\end{itemize}

\begin{figure*}[b]
    \centering
    \includegraphics[width=\textwidth]{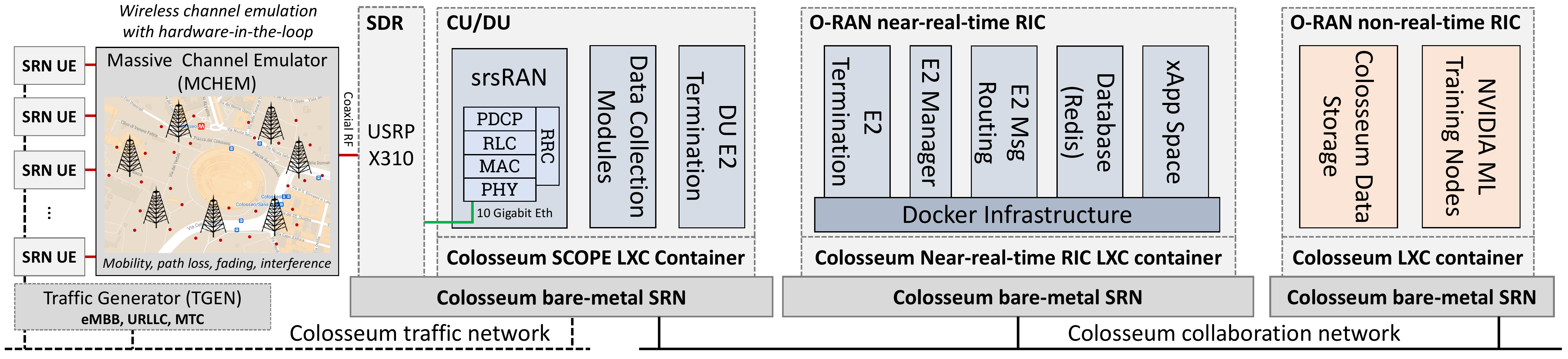}
    \caption{OpenRAN Gym components~\cite{bonati2022openrangym} deployed on the Colosseum experimental testbed~\cite{bonati2021colosseum}, adapted from~\cite{polese2021coloran}. The non-RT RIC, near-RT RIC, and RAN nodes (either base stations or users) are deployed as containers on different Colosseum \acrfullpl{srn}. Different Colosseum networks can be used to provide application traffic and support for the O-RAN interfaces, while the wireless communications among RAN components is emulated by the Massive Channel Emulator (MCHEM).}
    \label{fig:ric-colosseum}
\end{figure*}

\noindent Besides \glspl{wg}, O-RAN also includes groups that focus on topics that are relevant to the whole Alliance. These are named \glspl{fg} and they are summarized in Table~\ref{tab:oran-fg}:

\begin{itemize}
    \item \textit{\gls{osfg}}. This \gls{fg} used to deal with open source-related issues in O-RAN. These included the planning, preparation, and establishment of the \gls{osc}, and the coordination with other open source communities. Since the launch of the \gls{osc} in 2019, most of the activities of the \gls{osfg} have been taken care directly by the \gls{osc}. As a result, this \gls{fg} is currently dormant~\cite{ghadialy2021oran}.
    
    \item \textit{\gls{sdfg}}. This \gls{fg} programs the standardization strategies of the O-RAN Alliance and interfaces with other standard development organizations (e.g., 3GPP Service and System Aspects, 3GPP \gls{ran}, ETSI, ITU-T, Small Cell Forum, IEEE~1914, NGMN Alliance). The \gls{sdfg} also collects requirements and suggestions from the other entities of the Alliance (e.g., the \glspl{wg}), and provides them guidance on standardization use cases.
    
    \item \textit{\gls{tifg}}. This \gls{fg} defines the overall approach of O-RAN for testing and integration, including the coordination of tests across the \glspl{wg}. Examples are the test and integration of specifications, the creation of profiles to facilitate O-RAN commercialization, and the specification of processes for O-RAN integration and solution verification. The \gls{tifg} also plans and coordinates the O-RAN PlugFests, and sets guidelines for third-party open test and integration centers.
    
\end{itemize}

Finally, the O-RAN Alliance also features a \gls{tsc}, which guides the activities of the Alliance, and features four sub-committees. These aim at (i) defining a minimum viable plan for a fully-compliant O-RAN network; (ii) define the procedures for the O-RAN Alliance; (iii) promote industry engagement; and (iv) develop research activities toward next generation or 6G networks.

\section{Experimental Wireless Platforms for O-RAN}
\label{sec:testbeds}

The recent adoption of the softwarization paradigm in next generation cellular networking, and the emergence of open frameworks and interfaces for the Open \gls{ran}, such as the ones described in this work, bring unprecedented opportunities to the field of experimental research on cellular networks. In this regard, experimental research will be fundamental in developing and validating the use cases described in Section~\ref{sec:results}.
Once hard to experiment upon---due to the complex, hardware-based and closed implementations of \gls{ran} nodes, e.g., the base stations---in recent years, the cellular networking ecosystem has seen this task facilitated by open and open-source protocol stacks (e.g., srsRAN~\cite{gomez2016srslte} and OpenAirInterface~\cite{kaltenberger2020openairinterface}).
These software-based implementations enable the instantiation of 3GPP-compliant network elements on general-purpose, off-the-shelf devices, allowing virtually anyone to instantiate a complete and operational cellular network with multiple nodes, and to experimentally validate solutions for cellular applications.

Thanks to the open protocol stacks mentioned above, in the last few years, the wireless community has seen the creation and broader adoption of publicly-available platforms and frameworks open and available to the research community. The prominent ones are summarized in Table~\ref{tab:platforms}.
\begin{table}[t]
    \centering
    \footnotesize
    \renewcommand{\arraystretch}{1.2}
    \setlength{\tabcolsep}{2pt}
    \caption{Experimental wireless platforms and frameworks for O-RAN.}
    \label{tab:platforms}
    \begin{tabularx}{\columnwidth}{
        >{\raggedright\arraybackslash\hsize=.8\hsize}X 
        >{\raggedright\arraybackslash\hsize=1.14\hsize}X 
        >{\raggedright\arraybackslash\hsize=1.06\hsize}X }
        \toprule
        Platform & O-RAN-compatible & Deployment \\
        \midrule
        Arena~\cite{bertizzolo2020arena} & yes (container ready to use) & sub-6\:GHz indoors \\
        Colosseum~\cite{bonati2021colosseum} & yes (container ready to use) & sub-6\:GHz network emulator \\
        AERPAW~\cite{panicker2021aerpaw} & yes & aerial outdoors \\
        COSMOS~\cite{raychaudhuri2020cosmos} & yes & mmWave outdoors, sub-6\:GHZ indoors \\
        POWDER~\cite{breen2021powder} & yes (container ready to use) & sub-6\:GHz outdoors \& indoors \\
        5GENESIS~\cite{koumaras20185genesis} & will be (platform under development) & 5G outdoors \\
        \midrule
        Framework & O-RAN-compatible & Deployment \\
        \midrule
        OpenRAN Gym~\cite{bonati2022openrangym} & yes & end-to-end pipeline for AI/ML O-RAN research \\
        Open AI Cellular~\cite{openaicellular,upadhyaya2022prototyping} & yes & AI-enabled framework \\
        \bottomrule
    \end{tabularx}
\end{table}
These platforms play a vital role as they provide the means---and scale---to virtualize cellular stacks and controllers on their publicly-available infrastructure, and to design and prototype solutions in deployments as close as possible to those of commercial networks.
Moreover, when it comes to \gls{ai}/\gls{ml} solutions---which require large amounts of data for the training and testing processes---they operate as \textit{wireless data factories}, providing users with the tools to perform data collection at scale in controlled---yet realistic---wireless environments.

Open \gls{ran} experimentation relies on a combination of (i)~compute resources, to run the virtualized O-RAN components (e.g., the \glspl{ric}); and (ii)~radio resources, to host the over-the-air component of the \gls{ran}. For example, the base stations can be implemented through open and softwarized protocol stacks, such as srsRAN~\cite{gomez2016srslte} and OpenAirInterface~\cite{kaltenberger2020openairinterface}.
These stacks leverage \glspl{sdr} (e.g., NI USRPs) as radio front-ends, and serve users implemented through analogous protocol stacks, or via commercial smartphones. The O-RAN software can be either developed ad hoc, e.g., as FlexRIC~\cite{schmidt2021flexric}, or be based on the \gls{osc}, \gls{onf}, Linux foundation, or ETSI open source frameworks, as discussed in Sections~\ref{sec:xapps} and~\ref{sec:orchestration}. For example, when focusing on RAN control, the research pipeline generally involves a near-RT \gls{ric}, the \gls{ran}, the E2 interface, and custom xApps implementing the desired control on the \gls{ric}.

All the platforms in Table~\ref{tab:platforms} provide these two ingredients, at different extents, and are thus fit for Open RAN research. Some of them, as discussed next, are already equipped with software and pipelines for this, while others can be used in combination with other frameworks. 



\subsection{Experimental Open RAN research with OpenRAN Gym}

The Open RAN experimental workflow is enabled, for example, by OpenRAN Gym~\cite{bonati2022openrangym}. OpenRAN Gym combines software-defined cellular stacks with a lightweight \gls{ric}, which can be deployed on multiple experimental platforms from Table~\ref{tab:platforms}, E2 termination, and an end-to-end AI/ML pipeline for O-RAN.

OpenRAN Gym offers a ready-to-use \gls{lxc}-based implementation with the main components of the \gls{osc} near-RT \gls{ric}, as shown in Figure~\ref{fig:ric-colosseum}~\cite{polese2021coloran}.
This can be instantiated on top of any bare metal compute and includes \gls{ric} services implemented as Docker containers.
They include the O-RAN \textit{E2 termination}, \textit{manager} and \textit{message routing} containers, which are used to communicate with the E2 nodes, and \textit{Redis database} container, used to keep a record of the E2 nodes associated with the near-RT \gls{ric}.
Additionally, external xApps can be instantiated in what is shown in the figure as \textit{xApp space}. As part of OpenRAN Gym, we provide a sample xApp that manages the connectivity to and from the \gls{ric} platform. 

Through the E2 termination, the near-RT \gls{ric} can connect to the E2 nodes (e.g., \glspl{cu}/\glspl{du}) to implement softwarized control of the \gls{ran} (see Figure~\ref{fig:ric-colosseum}, left).
In OpenRAN Gym, the latter can be implemented through the SCOPE framework that extends srsRAN~\cite{gomez2016srslte} with additional slicing, MAC- and PHY-layer functionalities, control \glspl{api}, and data collection capabilities~\cite{bonati2021scope}.
This can be paired with the \gls{osc} RAN-side E2 termination to interface with the E2 termination on the near-RT \gls{ric}.

\subsection{Colosseum and Arena}

Two of the prominent platforms that allow users to instantiate an O-RAN-compliant network (e.g., with OpenRAN Gym) and components on a white-box infrastructure are Colosseum and Arena~\cite{bonati2021colosseum,bertizzolo2020arena}.
Colosseum is the world's largest wireless network emulator with hardware-in-the-loop. Through a first-of-its-kind \gls{fpga} fabric, Colosseum empowers researchers with the tools to capture and reproduce different conditions of the wireless channel, and to experiment at scale through 256~\gls{usrp} X310 \glspl{sdr}~\cite{bonati2021colosseum}.
Colosseum provides researchers with access to 128 \glspl{srn}, i.e., a combination of a server and of an \gls{usrp} X310 \gls{sdr} acting as a \gls{rf} front-end. \glspl{srn} can be used to instantiate the \glspl{ric} (without considering the radio component) or softwarized base stations and users.
Colosseum also provides data storage and NVIDIA \glspl{gpu} for the training of \gls{ml} models (see Figure~\ref{fig:ric-colosseum}, right). These resources can be used, for example, as a component of the \gls{smo}/non-RT RIC.

Finally, the \glspl{sdr} are connected through coaxial \gls{rf} cables to Colosseum \acrlong{mchem}---which takes care of emulating in \gls{fpga} different conditions of the wireless environment (Figure~\ref{fig:ric-colosseum}, left)---and through the traffic network to Colosseum \acrlong{tgen}---which leverages \acrlong{mgen}~\cite{mgen} to generate and stream IP traffic flows to the \glspl{srn}.

After prototyping O-RAN-powered solutions on Colosseum with the setup of Figure~\ref{fig:ric-colosseum}, users can port them to other experimental testbeds, such as Arena and the platforms of the U.S.\ PAWR program~\cite{bertizzolo2020arena,pawr}.
Arena is an indoor testbed with 24~\glspl{sdr} (16~\glspl{usrp} N210 and 8~\glspl{usrp} X310) connected to a ceiling grid with 64~antennas and controlled by a set of 12~high-performance servers.
The combination of servers, \glspl{sdr}, and antenna layout offer the ideal setup for testing of \gls{mec} capabilities~\cite{doro2020sledge} and private indoor cellular deployments~\cite{bonati2020cellos,bonati2021stealte}. An example of O-RAN-related research that combines Colosseum and Arena is described in~\cite{polese2021coloran}.

\subsection{Other Experimental Research Platforms}

Other publicly-available testbeds include the city-scale platforms of the U.S.\ National Science Foundation PAWR program~\cite{pawr}.
These consist of POWDER~\cite{breen2021powder}, focused on sub-6~GHz cellular deployments, COSMOS~\cite{kohli2021openaccess}, on mmWave communications, and AERPAW~\cite{panicker2021aerpaw}, on aerial cellular deployments.\footnote{A fourth platform, ARA, which will focus on rural broadband connectivity, has been announced~\cite{zhang2021ara}. However, this platform is not yet operational.} Namely, all of these platforms are compatible with the O-RAN paradigm, as they allow users to instantiate white-box base stations managed by the O-RAN \glspl{ric}. However, at the time of this writing, the only testbed that offers a ready-to-use O-RAN implementation (in the form of a pre-compiled container image) is POWDER~\cite{johnson2021nextran,powder_oran}. OpenRAN Gym has been tested on POWDER and COSMOS.

%
%

In Europe, the 5GENESIS consortium is working toward the implementation of various 5G components, and the validation of different use cases across its several testbeds~\cite{koumaras20185genesis}.
These include edge-computing \gls{nfv}-enabled heterogeneous radio infrastructure, orchestration and management frameworks, terrestrial and satellite communications, and ultra-dense network deployments.
Upon completion, these testbeds will be compatible with the O-RAN ecosystem.

Among the notable open-source initiatives, Open AI Cellular (OAIC) proposes a framework that is integrated with the O-RAN ecosystem.
This framework allows users to manage cellular networks through \gls{ai}-enabled controllers, and to interact with systems that locate implementation, system-level, and security flaws in the network itself~\cite{oaic_website,upadhyaya2022prototyping}.

\section{Future Research and Development Directions}
\label{sec:future}

While the foundational principles and the main specifications for O-RAN have been drafted, partially enabling the use cases described in Section~\ref{sec:results}, there are still several open issues for both standardization, development, and research. We identified some of them as follows:
\begin{itemize}
	\item \textbf{Identification of key O-RAN use cases.} While O-RAN provides the infrastructure to implement RAN closed-loop control, the identification of a key set use cases that leverage these extended capabilities is still ongoing. The O-RAN Alliance provides a list of relevant use cases for the RIC-enabled control, which include classic radio resource management optimization related to handover optimization, resource allocation, QoE optimization, traffic steering, among others~\cite{oran-wg1-use-cases}. Nonetheless, as the capabilities of the 3GPP RAN evolve toward, for example, non-terrestrial networks and support for \gls{ar}/\gls{vr} in the metaverse, it becomes necessary to further refine and evaluate future O-RAN use cases and the role intelligent, data-driven closed-loop control can have in future domains.

    \item \textbf{Open RAN beyond public cellular networks---the private cellular use case.} Finally, private cellular networks are quickly emerging as a key 5G deployment scenario, with applications in industrial automation, warehouses, healthcare industry, education, and entertainment. Greenfield private 5G deployments can easily embed Open RAN solutions for network control and optimization, as well as to reduce ownership and operation costs thanks to disaggregated and virtualized nodes. The design of O-RAN-enabled private networks introduces challenges in terms of domain-specific optimization, integration with edge systems and local breakouts, and support for connectivity in constrained environments.

    \item \textbf{Service models development and implementation.} As discussed in Section~\ref{sec:interfaces}, the E2 service models play a key role in the definition of what O-RAN and, specifically, the near-RT \gls{ric} can control in a \gls{3gpp}-defined \gls{ran}. As new use cases for O-RAN and xApp-driven control are identified, it is important to design service models that are comprehensive, and to identify profiles and basic set of functionalities that \gls{ran} equipment vendors need to implement to be O-RAN compliant. Indeed, the near-RT \gls{ric} effectiveness for \gls{ran} analysis and control ultimately depends on the E2 service models implemented in the \gls{ran}.
    
    \item \textbf{Interoperability and testing.} O-RAN has made significant steps toward enabling interoperability among vendors and networking components, thanks to the definition of open interfaces. This key principle needs to be realized in practice, with vendors that fully commit to implementing standard-compliant interfaces. In addition, a truly interoperable ecosystem will foster the development of xApps and rApps that can be ported across multiple near-RT and non-RT \glspl{ric}. In this sense, the standardization of \glspl{api} or of a \gls{sdk} for the \glspl{ric} is a key interoperability enabler. Finally, the fronthaul interface and the deployment of efficient, scalable fronthaul networks is one of the key challenges in the design and scaling of Open RAN deployments.

    \item \textbf{O-RAN Architecture and its evolution.} The foundational elements of the O-RAN architecture include the disaggregated RAN nodes (\glspl{cu}, \glspl{du}, \glspl{ru}) and the near-RT and non-RT \glspl{ric} hosting xApps and rApps, respectively. There are several open questions as of how this architecture can be effectively deployed, e.g., in terms of distribution of networking elements across the edge and cloud network, or ratio among RAN nodes and RIC elements. In addition, further research can help designing further extensions of the O-RAN architecture, which, for example, enable real-time control in the RAN nodes through what we define as \textit{dApps} in~\cite{doro2022dapps}. These elements can work together with xApps to leverage data that cannot be transferred for analysis from the \gls{ran} to the \gls{ric} (e.g., I/Q samples, or fine-grained channel estimation information).
    
    \item \textbf{Multi-time-scale control.}  When considering the full O-RAN architecture (and possible extensions as discussed above), different control loops will operate and have visibility on the system at different time scales. This opens challenges in terms of multi-scale control. Further research is required on the design of the multi-scale algorithms, on the identification of instability in the system as well as conflicts across the different control loops, and on the automated selection of the optimal control loops that can be used to reach specific high-level intents~\cite{doro2022orchestran}.
    
    \item \textbf{Effective \gls{ai}/\gls{ml} algorithm design, testing, and deployment.} The AI/ML workflow described in Section~\ref{sec:ai-ml-workflow} positions O-RAN to be a framework for the practical deployment of \gls{ml} solutions in the \gls{ran}. While this workflow is being standardized, several challenges still remain. These challenges are related to how to (i) collect training and testing datasets that are heterogeneous and representative of large-scale deployments; (ii) test and/or refine data-driven solutions through online training without compromising the production \gls{ran} performance, and (iii) design AI/ML algorithms that work with real, unreliable input, and can easily generalize to different deployment conditions~\cite{polese2021coloran}.

    \item \textbf{Spectrum sharing solutions enabled by Open RAN.} Network controllers and programmable RAN nodes open new opportunities for the development of spectrum sharing systems~\cite{baldesi2022charm}. The O-RAN specifications already include capabilities for LTE/NR dynamic spectrum sharing, but the design of algorithms to enable this is still an open issue. Future research can investigate how to practically enable O-RAN-based sensing, reactive and proactive spectrum adaptation solutions, considering 3GPP and non-3GPP systems, as well as sharing-related extensions of the O-RAN architecture. 
   
    \item \textbf{Security in O-RAN.} As discussed in Section~\ref{sec:security}, the openness of the RAN increases the threat surface but also enables new approaches to network security. For example, the improved visibility into the RAN performance and telemetry and the possibility of deploying plug-in xApps and rApps for security analysis and threat identification make it possible to explore novel approaches for securing wireless networks and make them more robust and resilient. 
    The research and development of security approaches that leverage O-RAN capabilities and improve the integrity, resiliency, and availability of its deployments is a key step toward making Open RAN approaches a viable and future-proof alternative to traditional \gls{ran} deployments.

    \item \textbf{Energy efficiency with Open RAN.} As discussed in Section~\ref{sec:arch}, virtualization and closed-loop control provide useful primitives for the dynamic network function allocation and thus for the energy efficiency maximization. Further research is required to develop orchestration routines at the non-RT RIC/SMO that embed energy efficiency in the optimization goal, as well as xApps and rApps that adopt control actions or policies that include energy efficiency targets.

\end{itemize}

\begin{lstlisting}[language=myxml,style=mystyle,float=*,
caption={Example of E2 Subscription Request message, compliant with E2AP V2.0~\cite{oran-wg3-e2-ap}. Generated using the E2 simulator library from~\cite{e2sim}.},
label={lst:e2-sub-request}]
<E2AP-PDU>
    <initiatingMessage>
        <procedureCode>8</procedureCode>
        <criticality><ignore/></criticality>
        <value>
            <RICsubscriptionRequest>
                <protocolIEs>
                    <RICsubscriptionRequest-IEs>
                        <id>29</id>
                        <criticality><reject/></criticality>
                        <value>
                            <RICrequestID>
                                <ricRequestorID>123</ricRequestorID>
                                <ricInstanceID>34</ricInstanceID>
                            </RICrequestID>
                        </value>
                    </RICsubscriptionRequest-IEs>
                    <RICsubscriptionRequest-IEs>
                        <id>5</id>
                        <criticality><reject/></criticality>
                        <value>
                            <RANfunctionID>1</RANfunctionID>
                        </value>
                    </RICsubscriptionRequest-IEs>
                    <RICsubscriptionRequest-IEs>
                        <id>30</id>
                        <criticality><reject/></criticality>
                        <value>
                            <RICsubscriptionDetails>
                                <ricEventTriggerDefinition>31 32 33 34</ricEventTriggerDefinition>
                                <ricAction-ToBeSetup-List>
                                    <ProtocolIE-SingleContainer>
                                        <id>19</id>
                                        <criticality><ignore/></criticality>
                                        <value>
                                            <RICaction-ToBeSetup-Item>
                                                <ricActionID>1</ricActionID>
                                                <ricActionType><report/></ricActionType>
                                                <ricActionDefinition>35 36 37 38</ricActionDefinition>
                                                <ricSubsequentAction>
                                                    <ricSubsequentActionType><continue/></ricSubsequentActionType>
                                                    <ricTimeToWait><w10ms/></ricTimeToWait>
                                                </ricSubsequentAction>
                                            </RICaction-ToBeSetup-Item>
                                        </value>
                                    </ProtocolIE-SingleContainer>
                                </ricAction-ToBeSetup-List>
                            </RICsubscriptionDetails>
                        </value>
                    </RICsubscriptionRequest-IEs>
                </protocolIEs>
            </RICsubscriptionRequest>
        </value>
    </initiatingMessage>
</E2AP-PDU>
\end{lstlisting}

\begin{lstlisting}[language=myxml,style=mystyle,float=*,
caption={Example of E2 Indication message of type report, compliant with E2AP V2.0~\cite{oran-wg3-e2-ap}. Generated using the E2 simulator library from~\cite{e2sim}.},
label={lst:e2-report}]
<E2AP-PDU>
    <initiatingMessage>
        <procedureCode>5</procedureCode>
        <criticality><ignore/></criticality>
        <value>
            <RICindication>
                <protocolIEs>
                    <RICindication-IEs>
                        <id>29</id>
                        <criticality><reject/></criticality>
                        <value>
                            <RICrequestID>
                                <ricRequestorID>123</ricRequestorID>
                                <ricInstanceID>26</ricInstanceID>
                            </RICrequestID>
                        </value>
                    </RICindication-IEs>
                    <RICindication-IEs>
                        <id>5</id>
                        <criticality><reject/></criticality>
                        <value>
                            <RANfunctionID>0</RANfunctionID>
                        </value>
                    </RICindication-IEs>
                    <RICindication-IEs>
                        <id>15</id>
                        <criticality><reject/></criticality>
                        <value>
                            <RICactionID>1</RICactionID>
                        </value>
                    </RICindication-IEs>
                    <RICindication-IEs>
                        <id>27</id>
                        <criticality><reject/></criticality>
                        <value>
                            <RICindicationSN>24</RICindicationSN>
                        </value>
                    </RICindication-IEs>
                    <RICindication-IEs>
                        <id>28</id>
                        <criticality><reject/></criticality>
                        <value>
                            <RICindicationType><report/></RICindicationType>
                        </value>
                    </RICindication-IEs>
                    <RICindication-IEs>
                        <id>25</id>
                        <criticality><reject/></criticality>
                        <value>
                            <RICindicationHeader>
                                <--- E2SM Header --->
                            </RICindicationHeader>
                        </value>
                    </RICindication-IEs>
                    <RICindication-IEs>
                        <id>26</id>
                        <criticality><reject/></criticality>
                        <value>
                            <RICindicationMessage>
                                <--- E2SM Payload --->
                            </RICindicationMessage>
                        </value>
                    </RICindication-IEs>
                    <RICindication-IEs>
                        <id>20</id>
                        <criticality><reject/></criticality>
                        <value>
                            <RICcallProcessID><--- E2SM process identifier ---></RICcallProcessID>
                        </value>
                    </RICindication-IEs>
                </protocolIEs>
            </RICindication>
        </value>
    </initiatingMessage>
</E2AP-PDU>
\end{lstlisting}

\section{Conclusions}
\label{sec:conclusions}

This paper presented a comprehensive overview of the O-RAN specifications, architectures and operations. We first introduced the main architectural building blocks and the principles behind the design of O-RAN networks. Then, we described the components of the near-RT and non-RT \glspl{ric} and of the \gls{smo}, and discussed the O-RAN interfaces, including E2, O1, A1, the fronthaul interface, and O2. The second part of this paper focused on topics spanning multiple components and interfaces in the O-RAN architecture. We provided details on the AI and ML workflow that O-RAN enables, on O-RAN use cases and research, and on the O-RAN security challenges and potential. Finally, we reviewed the structure and standardization efforts of the O-RAN Alliance, and discussed research platforms, and future research directions.

We believe that these insights, together with the deep dive on the O-RAN specifications, architecture, and interfaces, will foster and promote further efforts toward more open, programmable, virtualized, and efficient wireless networks.


\appendix

In the following appendices, we provide examples for messages exchanged over the E2 interface (the subscription message, in Appendix~\ref{app:e2sub}, and the Indication message of type report, in Appendix~\ref{app:e2rep}), and a list of acronyms used throughout this work (Appendix~\ref{app:acronyms}).

\subsection{Example of E2 Subscription Request Message}
\label{app:e2sub}

Listing~\ref{lst:e2-sub-request} shows an example of the fields of an E2AP message for a subscription request, which is generated in the near-RT RIC and sent to the E2 termination of an E2 node. The XML format is used to describe the set of fields and their entries (i.e., the \glspl{ie}), then the actual message is encoded in ASN.1 format (i.e., a sequence of bytes) before being encapsulated and transmitted on the SCTP socket. The first field is the message type, which contains a procedure code (8 for the subscription) and the actual type of message (an initiating message, as it begins a procedure on E2). Then, the message contains two IDs, one that uniquely identifies the RIC request, and the other the RAN function to which the RIC wants to subscribe. The core of the message is the set of \glspl{ie} with details on the subscription request. In particular, the message defines a list of actions (with only one entry in this example). Each action has a type (in this case, \texttt{report}), a definition (which is optional, and depends on what the specific RAN function supports), and, possibly, a subsequent action to perform once the first is completed (in this case, \texttt{continue}, after waiting for a timer of 10\:ms to expire).

\subsection{Example of E2 Indication (Report) Message. }
\label{app:e2rep}

Listing~\ref{lst:e2-report} features the XML for an E2AP Indication message, which is generated by the E2 node and sent to the near-RT RIC upon triggering of an event defined through the related subscription procedure. As for the E2 Subscription Request message, also this message is encoded using ASN.1, and features the message type and procedure code at the beginning of the message. Then, it lists \glspl{ie} related to the RIC request, the function identifier, the corresponding action identifier (which is a unique value for each RIC request), a sequence number (which is optional), and the type of indication, which in this case is \texttt{report} (the possible values are \texttt{insert} or \texttt{report}). The last three fields (indication header, indication message, and call process identifier) are encoded by the E2AP message but their semantics are defined by the specific E2SM used to populate the message (e.g., E2SM \gls{kpm}).

\subsection{Acronyms}
\label{app:acronyms}

\renewcommand{\arraystretch}{1}
\footnotesize
\renewcommand{\glossarysection}[2][]{}
\setlength{\glsdescwidth}{0.75\columnwidth}
\printglossary[style=index]

\balance
\footnotesize  
\bibliographystyle{IEEEtran}
\bibliography{biblio.bib}

\end{document}